\documentclass[11pt,letterpaper]{article}

\usepackage{jheppub}
\usepackage{color}
\usepackage{graphicx}
\usepackage{wrapfig}
\usepackage{epstopdf}

\usepackage{verbatim}
\usepackage{amsmath}
\usepackage{amssymb}
\usepackage{subfig}
\usepackage{url}
\usepackage{bbold}
\usepackage{xspace}
 \usepackage{setspace}

\usepackage{floatrow}
\newfloatcommand{capbtabbox}{table}[][\FBwidth]

\usepackage{slashed}
\usepackage{verbatim}

\newcommand{\met}{E\!\!\!/_T}

\newcommand{\be}{\begin{equation}}
\newcommand{\ee}{\end{equation}}

\DeclareRobustCommand{\Fig}[1]{figure~\ref{#1}}

\newcommand{\Dfbd}{\mathord{\buildrel{\lower3pt\hbox{$\scriptscriptstyle\leftrightarrow$}}\over {D}_{\mu}}}

\newcommand{\beq}{\begin{equation}}
\newcommand{\eeq}[1]{\label{#1}\end{equation}}
\def\beqa{\begin{eqnarray}}
\def\eeqa#1{\label{#1}\end{eqnarray}}
\newcommand{\bea}{\begin{eqnarray}}
\newcommand{\eea}{\end{eqnarray}}
\newcommand{\eeqn}{\end{equation}}

\newlength{\dhatheight}

\def\stacksymbols #1#2#3#4{\def\theguybelow{#2}
    \def\vp{\lower#3pt}
    \def\sp{\baselineskip0pt\lineskip#4pt}
    \mathrel{\mathpalette\intermediary#1}}

\def\intermediary#1#2{\vp\vbox{\sp
     \everycr={}\tabskip0pt
     \halign{$\mathsurround0pt#1\hfil##\hfil$\crcr#2\crcr
              \theguybelow\crcr}}}

\begin{document}

\title{The Hunt for the Rest of the Higgs Bosons}

\author[a]{Nathaniel Craig,}
\author[b,c]{Francesco D'Eramo,}
\author[a]{Patrick Draper,}
\author[d]{Scott Thomas,}
\author[a]{and \\ Hao Zhang}

\affiliation[a]{Department of Physics, University of California, Santa Barbara, CA 93106, USA}
\affiliation[b]{Department of Physics, University of California, Berkeley, CA 94720 USA}
\affiliation[c]{Theoretical Physics Group, Lawrence Berkeley National Laboratory, Berkeley, CA 94720, USA}
\affiliation[d]{Department of Physics and Astronomy, Rutgers University, Piscataway, NJ 08854, USA}

\emailAdd{ncraig@physics.ucsb.edu}
\emailAdd{fraderamo@berkeley.edu}
\emailAdd{pidraper@physics.ucsb.edu}
\emailAdd{scthomas@physics.rutgers.edu}
\emailAdd{zhanghao@physics.ucsb.edu}

\date{\today}

\abstract{We assess the current state of searches at the LHC for additional 
Higgs bosons in light of both direct limits and indirect bounds coming 
from coupling measurements of the Standard Model-like Higgs boson. 
Given current constraints, we 
identify and study three LHC searches that
are critical components of a comprehensive program 
to investigate extended electroweak symmetry breaking sectors:
production of a heavy scalar or pseudoscalar with decay to $t \bar t$; 
$b \bar b$ and $t \bar t$ associated production of a heavy scalar or 
pseudoscalar with decay to invisible final states; and $t \bar b$ associated 
production of a charged Higgs with decay to 
$\bar t b$. Systematic experimental searches in these channels 
would contribute to robust coverage of the possible single production modes 
of additional  heavy Higgs bosons.
}


\arxivnumber{15xx.xxxxx}

\preprint{}

\maketitle

\section{Introduction}
\label{sec:intro}

Following the discovery of a Standard Model-like Higgs boson 
at the LHC, the systematic search for 
additional weakly-coupled scalars near the electroweak scale 
is of paramount importance. A variety of experimental searches 
have been performed for such extended Higgs sectors to date, 
predominantly targeting new scalars with substantial couplings 
to the electroweak gauge bosons~\cite{Aad:2015wra,Aad:2015nfa,
ATLAS:2012ac,Chatrchyan:2013yoa,CMS:2014yra,Khachatryan:2015cwa,Khachatryan:2014jya} or to down-type fermions~\cite{
Aad:2014vgg,Khachatryan:2014wca}. However, these searches 
are only sensitive to a fraction of the interesting parameter space 
in general extended Higgs sectors. This raises the surprising 
possibility that a large number of additional Higgs bosons have 
been produced at the LHC without leaving signals in existing 
search channels. 

Moving forward into Run 2 at the LHC, a natural question is 
{\it how the search for additional Higgses should be organized}
in order to ensure systematic coverage of extended electroweak 
symmetry breaking sectors. Given the proliferation of potential signals,
it is useful to consider signatures broadly. Interesting topologies 
and searches for new Higgs bosons can be classified in terms of 
simplified models, much in the spirit of simplified model searches 
for supersymmetry~\cite{Dube:2008kf,Alwall:2008ag,Alves:2011wf}. These 
simplified models may then be combined to provide coverage of 
the parameter space of a given extended Higgs sector. 

A useful first step in organizing the search for additional states
is to begin with the signatures of a single additional Higgs boson, 
so that the available decay modes involve Standard Model (SM) 
bosons and fermions, the 125 GeV Higgs, and potentially additional 
invisible decays. In a general extended Higgs sector there may be 
numerous Higgs bosons beyond the SM Higgs. By focusing on one 
new state at a time, we can characterize the dominant signals of an 
extended Higgs sector if the additional Higgs bosons are well separated 
in mass, or if the additional Higgs bosons are approximately degenerate 
so that decays between heavy Higgs bosons are kinematically disfavored. 
Having comprehensively covered these signatures, it is then possible 
to systematically expand the picture to consider production and decay 
processes involving more than one heavy Higgs boson.
 
Within the space of signatures of a single new Higgs state, further 
powerful guidance is provided by the coupling measurements of the 
recently-discovered SM-like Higgs boson, which constrain the parameter space of 
extended Higgs sectors. These coupling measurements are currently 
consistent with SM predictions to within the $20-30\%$ level. Such 
agreement suggests that any extension of the Higgs sector must be 
near an {\it alignment limit} in its parameter space, wherein 
the SM-like Higgs 
boson is closely aligned with the vacuum expectation value (vev) that breaks 
electroweak symmetry and correspondingly exhibits the properties 
of a SM Higgs boson~\cite{Craig:2012vn, Craig:2013hca, Carena:2013ooa, 
Haber:2013mia}. In a given extended Higgs sector, this alignment limit 
may be approached either due to decoupling of additional Higgs states
\cite{Gunion:2002zf, Haber:2013mia}, or simply due to the organization 
of dimensionless couplings in the Higgs potential \cite{Craig:2012vn, 
Craig:2013hca, Carena:2013ooa, Dev:2014yca, Das:2015mwa, Dev:2015bta}. 
Proximity to the alignment limit then governs also the couplings of additional 
Higgs bosons, and may be used as a guide to searches for additional Higgses.

The precise properties of the SM-like Higgs boson and 
additional Higgs scalars near the alignment limit
depend on the nature of the extended Higgs sector. 
The SM Higgs boson is a {\it vacuum state} 
in that it carries the quantum numbers of the vacuum.  
Additional neutral 
Higgs bosons may include {\it pure vacuum states} with the quantum 
numbers of the vacuum, allowing mixing with the SM Higgs, but without 
intrinsic coupling to gauge bosons or fermions. 
Mixing with new pure vacuum states 
modify the SM-like Higgs boson 
couplings in a model-independent way by diluting all couplings 
uniformly. The real singlet extension of the Higgs sector provides a natural 
example of a pure vacuum state, with one new CP-even Higgs boson 
whose couplings to SM fermions and gauge bosons are uniformly suppressed 
in the alignment limit. In this case, proximity to the alignment limit implies 
suppression of all production modes for the additional boson at the LHC. 

Alternately, additional neutral Higgs bosons may simply include new {\it vacuum 
states} allowing both mixing with the SM-like Higgs {\it and} independent 
intrinsic couplings to massive gauge bosons and fermions. 
Such vacuum states modify the
SM-like 
Higgs boson couplings in a model-dependent way. Such vacuum states 
arise, for example, in CP-conserving Two-Higgs-Doublet Models (2HDM). 
The physical spectrum of 2HDM includes four additional Higgs bosons  -- 
a CP-even scalar $H$, CP-odd pseudoscalar $A$, and charged Higgs bosons 
$H^\pm$, of which the CP-even scalar $H$ without any 
quantum numbers can be identified as a vacuum state. 
The couplings of these additional Higgs bosons to SM bosons are suppressed 
in the alignment limit, while their couplings to SM fermions are generically 
unsuppressed but depend in detail on the coupling structure of the 2HDM. 
In this case, proximity to the alignment limit implies suppression of production 
and decay modes involving SM bosons, while production and decay via SM 
fermions may be appreciable. Consequently, typical searches such as 
$H\rightarrow ZZ$ become ineffective. Similarly, $H/A\rightarrow\tau\tau$ 
and $H^\pm\rightarrow\tau\nu$ searches may be effective in some scenarios, 
but are ineffective whenever the down-type fermionic couplings are not 
substantially enhanced over the SM Yukawas. 

In this paper we articulate a systematic strategy for searching individually 
for additional Higgs scalars in light of the properties of the 125 GeV Higgs. 
We focus on the phenomenology of CP-conserving scenarios with two 
Higgs doublets satisfying the Glashow-Weinberg (GW) condition
\cite{Glashow:1976nt}, 
as this describes the physics of many well-motivated extensions of the Higgs 
sector while still covering many of the key features of models with additional 
singlets or higher electroweak representations. We first summarize the state 
of limits on 2HDM at the LHC from direct searches for additional Higgs states 
and indirect constraints from Higgs coupling measurements (for recent related work, see \cite{Celis:2013rcs, Chiang:2013ixa, Grinstein:2013npa, Craig:2013hca, Chen:2013rba, Belanger:2013xza, Barger:2013ofa, Brownson:2013lka, Chen:2013qda, Chang:2013ona, Cheung:2013rva, Celis:2013ixa, Englert:2014uua, Dumont:2014wha, Kanemura:2014bqa, Carena:2014nza, Djouadi:2015jea}). We then identify 
and study three primary channels where, without being meaningfully constrained 
by existing searches, a second Higgs boson could exhibit $\mathcal{O}(1)$ 
signals:

\begin{enumerate}
\item The single production of a heavy scalar or pseudoscalar Higgs boson with 
decay to $t \bar t$.
\item The single production of a heavy scalar or pseudoscalar with decay to 
invisible final states.
\item The $t b$ associated production of a charged Higgs with decay to $t b$. 
\footnote{Although the $H^\pm tb$ process with dileptonic top decay is the 
subject of a recent CMS search at 8 TeV~\cite{CMS:2014pea}, there is room 
for improvement in the reach, particularly by inclusion of the semileptonic top 
decay channel studied here at 14 TeV. }
\end{enumerate}

In each case, ``single production'' includes both resonant production of a 
single heavy Higgs boson and potential associated production modes involving 
SM fermions in conjunction with a heavy Higgs boson. The combination of Higgs 
coupling measurements, ongoing searches for heavy Higgses, and the three 
search channels studied in this work should contribute
a rather comprehensive coverage 
of individual scalar states in extended electroweak symmetry breaking sectors.
 
The paper is organized as follows: In section~\ref{sec:limits} we first 
review the constraints from current Higgs coupling measurements on 
the parameter space of motivated 2HDM, then present the combined 
impact of existing direct searches for heavy Higgs states on the same 
parameter space. In section~\ref{sec:ttbar} we consider one of the most pressing 
signatures of additional Higgs scalars in light of current direct and indirect 
limits: the strong production of a heavy neutral Higgs boson followed by 
decay into $t \bar t$. 
This process has a distinctive interference 
between signal and SM $t \bar t$ background \cite{Dicus:1994bm}, 
but is hampered by the sizable SM background, 
and further complicated by the modest reconstruction resolution
available at the LHC  
of the invariant mass of the $t \bar{t}$ system.
Given the challenges of 
the search for resonant $t \bar t$ production, we also briefly consider 
$t \bar t$ and $b \bar b$ associated production of a heavy neutral Higgs 
followed by decay to $t \bar t$. In section~\ref{sec:invisible} we turn to 
invisible decays of heavy Higgses. We consider $t \bar t$ and $b \bar b$ 
associated production of an invisibly-decaying heavy Higgs boson, as proximity 
to the alignment limit renders ineffective traditional searches involving 
vector bosons. Finally, in section~\ref{sec:charged} we study the $t \bar b$ 
associated production and decay of charged Higgs bosons. We conclude 
in section~\ref{sec:conc} and reserve details of our Higgs fit, kinematics 
of the $H/A \to t \bar t$ process, and top quark reconstruction algorithm 
for a series of appendices.
 
\section{Direct and Indirect 2HDM Limits}
\label{sec:limits}
Direct searches at the LHC and indirect limits arising from Higgs coupling measurements impose constraints on the parameter space of 2HDM. As we will discuss in this section, the search for additional Higgses is guided by the complementarity of these direct and indirect constraints. 

In light of stringent flavor constraints, we focus on (CP-conserving) 2HDM satisfying the Glashow-Weinberg condition that all fermions of a given representation receive their masses via renormalizable couplings to a single Higgs doublet. There are four distinct possible configurations satisfying the GW condition; in this paper we will further focus on the two most common, known as Type 1 and Type 2 2HDM. In Type 1 2HDM all SM fermions couple to one doublet, while in Type 2 2HDM the up-type quarks and down-type quarks/leptons couple to separate doublets. These two types arise most frequently in motivated extensions of the SM, including composite Higgs models, little Higgs models, and supersymmetric models; the Higgs sector of the MSSM is an instance of the Type 2 2HDM.

In theories with two Higgs doublets $\Phi_1, \Phi_2$ and the most 
general renormalizable 
CP-conserving potential, there are nine free parameters that remain after minimizing the potential and fixing the symmetry breaking vev $v^2 = v_1^2 + v_2^2 = (246 \, {\rm GeV})^2$. There are various possible parameterizations. Here we use the conventions of \cite{Craig:2013hca}, taking for the free parameters the ratio $\tan \beta = |\langle \Phi_2^0 \rangle / \langle \Phi_1^0 \rangle|$, the mixing angle $\alpha$ that diagonalizes the neutral scalar mass matrix, the four physical masses $\{m_h, m_H, m_A, m_{H^\pm}\}$, and the dimensionless couplings $\lambda_{5,6,7}$.

The coupling of the physical states $h, H, A, H^\pm$ to SM fermions and gauge bosons are fully determined by the angles $\alpha$ and $\beta$, while the renormalizable couplings involving three or four physical Higgs bosons depend on the additional parameters of the potential.  The couplings of physical scalars to SM fermions and gauge bosons as a function of $\alpha$ and $\beta$ in Type 1 and Type 2 2HDM are summarized in table~\ref{tab:couplings}. In this work we will assume that the observed 125 GeV Higgs is the CP-even scalar $h$
with SM-like Higgs couplings, 
with the additional Higgs scalars $H, A, H^\pm$ parametrically heavier. The case of additional scalars lighter than the 125 GeV Higgs is also quite interesting but qualitatively distinct.

\begin{table}[t]
\begin{center}
\begin{tabular}{|c|c|c|} \hline
$y_{\rm 2HDM} / y_{\rm SM}$ & Type 1& Type 2  \\ \hline
$hVV$ & $s_{\beta - \alpha}$ &  $s_{\beta - \alpha}$  \\
$h Q u $ & $s_{\beta - \alpha} +  c_{\beta - \alpha} / t_\beta$ & $s_{\beta - \alpha} + c_{\beta - \alpha}/ t_\beta$   \\
$h Q d$ & $s_{\beta - \alpha} +  c_{\beta - \alpha}/ t_\beta$ & $s_{\beta - \alpha} - t_\beta c_{\beta - \alpha}$ \\
$h L e$ & $s_{\beta - \alpha} + c_{\beta - \alpha}/ t_\beta$ & $s_{\beta - \alpha} - t_\beta c_{\beta - \alpha}$  \\ \hline
$HVV$ & $c_{\beta - \alpha}$ & $c_{\beta - \alpha}$  \\
$H Q u$ & $c_{\beta - \alpha} -  s_{\beta - \alpha}/ t_\beta$ & $c_{\beta - \alpha} -  s_{\beta - \alpha}/ t_\beta$   \\
$H Q d$ & $c_{\beta - \alpha} -  s_{\beta - \alpha}/ t_\beta$ & $c_{\beta - \alpha} + t_\beta s_{\beta - \alpha}$   \\
$H L e$ & $c_{\beta - \alpha} - s_{\beta - \alpha}/ t_\beta$ & $c_{\beta - \alpha} + t_\beta s_{\beta - \alpha}$  \\\hline
$AVV$ & 0 & 0 \\
$AQu$ & $1/t_\beta$ & $1/t_\beta$  \\
$AQd$ & $-1/t_\beta$ & $t_\beta$   \\
$ALe$ & $- 1/t_\beta$ & $t_\beta$   \\ \hline
\end{tabular}
\caption{The coupling of Higgs bosons $h, H, A$ to SM bosons and fermions as a function of the angles $\alpha$ and $\beta$, expressed in terms of the alignment 
parameter 
$c_{\beta - \alpha} \equiv \cos(\beta - \alpha)$, and $t_\beta \equiv \tan \beta$. The coupling dependence of the charged scalars $H^{\pm}$ is
the same as the pseudo-scalar $A$. \label{tab:couplings}}
\end{center}
\end{table}%

It is apparent from table~\ref{tab:couplings} that couplings of the CP-even scalar $h$ become exactly SM-like in the limit $\cos(\beta - \alpha) \to 0$, which coincides with the alignment limit for 2HDM satisfying the Glashow-Weinberg condition. In the alignment limit the heavy CP-even neutral Higgs $H$ decouples from SM vector bosons, and its couplings become akin to those of the pseudoscalar Higgs $A$. Crucially, the Higgs bosons $H, A,$ and $H^\pm$ retain couplings to SM fermions in the alignment limit. These couplings ensure that the additional states have non-vanishing production channels and 
visible decay signatures involving SM fermions even in the limit where the 125 GeV Higgs is exactly SM-like.

At present, the SM-like nature of the 125 GeV Higgs boson implies proximity to the alignment limit commensurate with the precision of Higgs coupling measurements. 
In order to quantify the impact 
on the $(\alpha,\beta)$ parameter space of 2HDM, 
we perform a global fit to recent Higgs measurements reported by the ATLAS and CMS collaborations.\footnote{For this fit and for the interpretation of direct searches for heavy Higgs bosons, we use the programs \texttt{HIGLU/HDECAY} \cite{Spira:1996if} to determine the NLO dependence of the $h/H/A$ gluon fusion production cross section and partial widths $h/H \to gg, t \bar t, b \bar b, s \bar s, c \bar c, \mu \mu, \tau \tau, WW, ZZ,$ and $A \to gg$ on the parameters $\alpha$ and $\beta$. We use analytic NLO QCD expressions for the partial widths $h/H/A \to \gamma \gamma$ \cite{Harlander:2005rq} and $A \to t \bar t, b \bar b, c \bar c$ \cite{Chetyrkin:1995pd, Spira:1997dg}. We use the program \texttt{SusHi} \cite{Harlander:2012pb} to determine the NLO $b\bar b h/H/A$ production cross section and validate the \texttt{HIGLU} result for gluon fusion. We use \texttt{MadGraph 5} \cite{Alwall:2014hca} to determine the LO $t \bar t h/H/A$ production cross section with a $k$-factor of 1.18 \cite{Dittmaier:2011ti}. We use leading order results for the partial widths $H \to hh$ and $A \to Zh, \tau \tau, \mu \mu$~\cite{Djouadi:2005gj}.}
We provide details of our fit procedure in appendix \ref{app:fit}. In figures~\ref{fig:fits} we show the result of global fits for Type 1 and Type 2 2HDM as a function of $\tan \beta$ and $\cos(\beta - \alpha)$. We refer the reader to \cite{Craig:2013hca} for discussion of the physics underlying the shape of these fits.

\begin{figure}[t]
  \centering
 \includegraphics[height=0.4\textwidth]{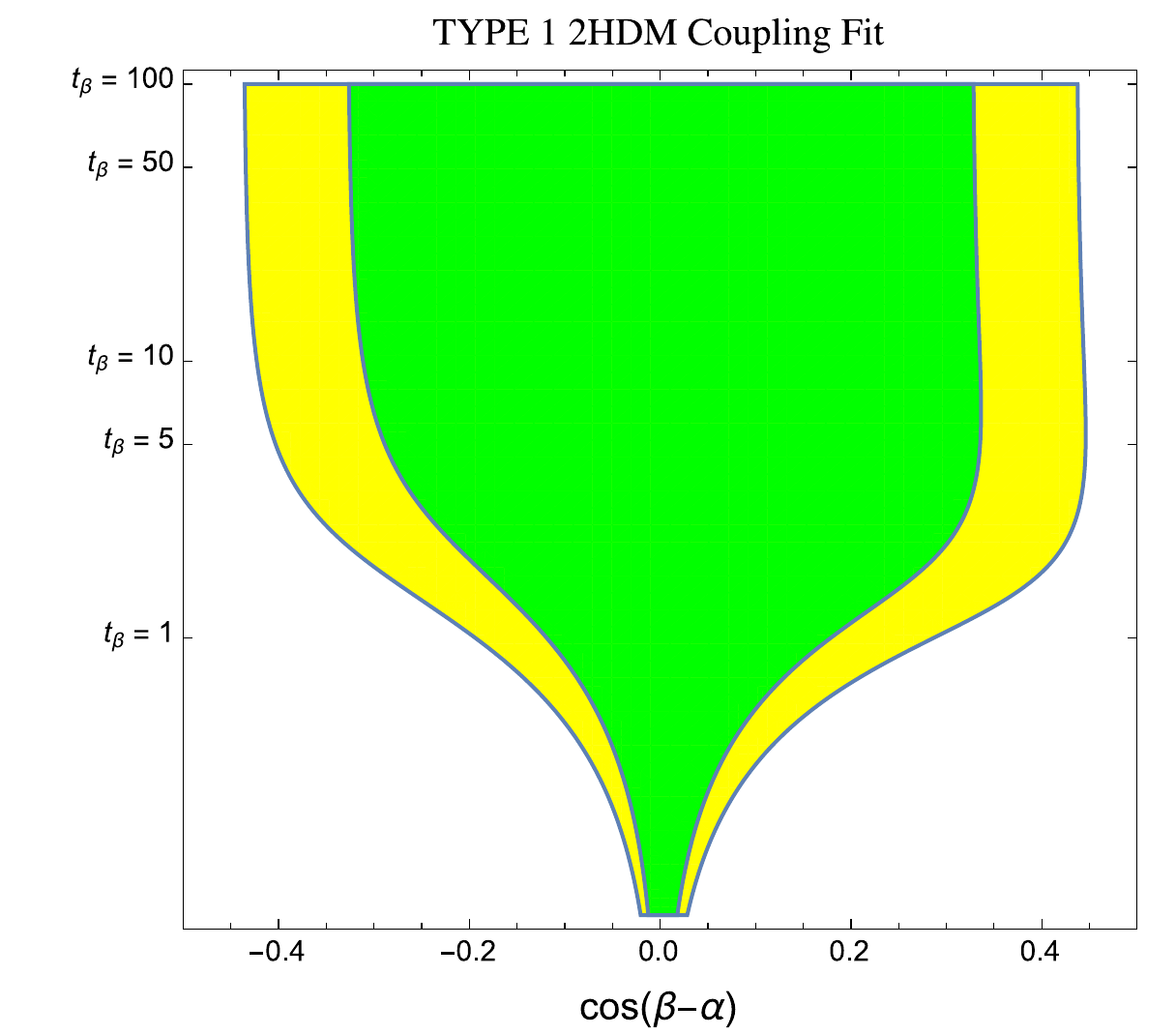}
  \includegraphics[height=0.4\textwidth]{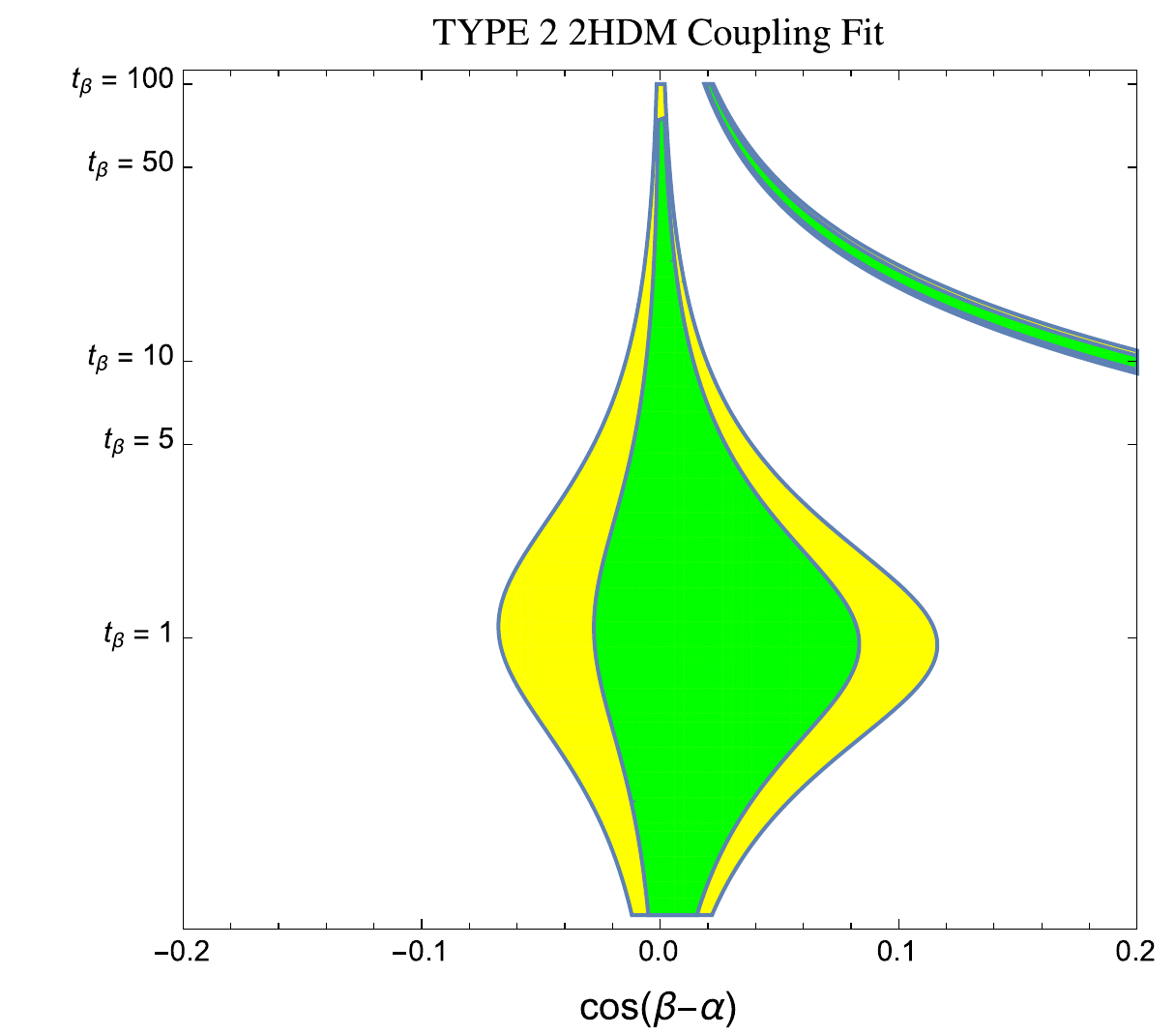}
  \caption{Coupling fits in the 2HDM 
  parameter space of $\tan \beta$ and $\cos(\beta - \alpha)$ in Type 1 (left) and Type 2 (right) 2HDM. Details of the fit procedure are discussed in appendix \ref{app:fit}.}
  \label{fig:fits}
\end{figure}

The proximity to the alignment limit implied by coupling measurements of the SM-like Higgs provides a natural organizing principle for the signatures of additional Higgs bosons. The implications for production modes 
are particularly transparent. In Type 1 2HDM, current fits require $\cos(\beta - \alpha) \lesssim 0.4$, suggesting that vector associated production modes of $H$ such as $ZH$ associated production or vector boson fusion (VBF) are suppressed by at least a factor $\sim 0.2$ relative to a SM Higgs of the same mass. In contrast, strong production modes may remain appreciable. Gluon fusion production of $H$ or $A$ proceeds through fermion loops as in the SM, uniformly proportional to $\cot^2 \beta$ in the alignment limit. The same is true of $t \bar t H/A$ and $b \bar b H/A$ associated production and $t \bar b H^\pm$ associated production, indicating that these channels remain promising in the alignment limit of Type 1 2HDM. 

In Type 2 2HDM the suppression implied by Higgs coupling fits is even more extreme, such that vector associated production modes of $H$ are at most $\sim 1\%$ of a SM Higgs of the same mass. As in the case of Type 1 2HDM, strong production modes are still appreciable. Gluon fusion production of $H$ and $A$ again proceeds through fermion loops, with the top loop contribution proportional to $\cot^2 \beta$ and the bottom loop contribution proportional to $\tan^2 \beta$ at leading order in the alignment limit. The $t \bar t H/A$ associated production mode again scales as $\cot^2 \beta$, while the $b \bar b H/A$ associated production mode scales as $\tan^2 \beta$.  Production of the charged Higgs is a function of both $\tan \beta$ and $\cot \beta$ in the alignment limit.

The impact on branching ratios of heavy Higgs bosons is somewhat more subtle. As discussed in detail in \cite{Craig:2013hca}, although proximity to the alignment limit implies suppression of couplings to SM bosons, these longitudinally-enhanced partial widths are competing only with relatively small fermionic partial widths. As such, decays into SM bosons may remain appreciable close to the alignment limit. In the exact alignment limit, tree-level decays into massive SM bosons (including the 125 GeV Higgs $h$) vanish in favor of decays into SM fermions and the massless gauge bosons.\footnote{We do not consider loop-level decays into massive vector bosons, which are nonzero in the exact alignment limit but sufficiently small to avoid influencing the tree-level result.}


\begin{table}[h]
\begin{center}
\begin{tabular}{ccccl} \hline \hline 
&&   \\
 Single Heavy Higgs && ${\cal O}(g_s^4 \lambda_f^2)$ && $gg \to ~ \!  H ~\!, ~ \! A$     \\
    Strong Production & \\
& \\ 
 Single Heavy Higgs && ${\cal O}( g_s^4 \lambda_f^2)$  && 
    $ gg \to ~ \! bbH ~ \! , ~ \!  bbA ~ \! , ~ \! tb H^{\pm} ~ \! , ~ \! ttH ~ \! , ~\! ttA$   \\
    Associated Strong Production & \\ 
& \\
  Single Heavy Higgs && ${\cal O}(g_s^2 g_w^4 \lambda_f^2)$ && 
     $  g q \to ~ \! b q' ~ \! b H^{\pm} ~ \! ,  ~\! bq ~ \! t H ~ \! , ~ \!  bq ~ \! t A$  \\
    Associated Weak Production &&   \\
 & \\
 Double Heavy Higgs && ${\cal O}(g_w^4)$ && 
     $q \bar{q} \to ~\! HA ~\! , ~ \! H H^{\pm}
          ~ \! , ~ \! A H^{\pm} ~\!  , ~ \!  H^+ \! H^-$ \\
 Weak Production && \\
& \\
 Light  +  Heavy Higgs  && ${\cal O}(g_s^4 \lambda_f^4)$ & & 
  $ gg \to ~ \! hH ~ \! , ~ \! hA$ \\
    Strong Production & \\
  & \\
 Double Heavy Higgs && ${\cal O}(g_s^4 \lambda_f^4)$ & & 
  $ gg \to ~ \! HH ~ \! , ~ \! HA ~ \! , ~ \! AA ~ \! , ~ \! H^+ \! H^-   $ \\  
  Strong Production & \\
& \\
\hline \hline
\end{tabular}
\caption{
Hierarchy of heavy Higgs leading LHC production channels 
that do not vanish in the 2HDM alignment limit. 
 \label{tab:prod-channels} }
\end{center}
\end{table}


\begin{table}[h]
\begin{center}
\begin{tabular}{cccccccc} \hline \hline 
&& &   & & & &  \\
&& & &  $H$  &  $A$  &  $  H^\pm$ & \\
&&   & & & & \\
\hline 
&&&   & & & & \\
& Standard Model && $WW,ZZ$ & $-$  & & &  \\
& Decay Channels && $tt, bb, \tau \tau, \mu\mu$ & \checkmark & \checkmark &  &   \\
&&& $\gamma \gamma $ & \checkmark & \checkmark & &    \\
& && & & & & \\
&&& $Zh$ &  & $-$ &  &   \\
&&& $ hh$ & $-$ & & &  \\
&&& $Wh $ &  & & $-$  &    \\
&&& $tb, \tau \nu$ &   &  & \checkmark &    \\
&&& & & & &\\
\hline \hline
\end{tabular}
\caption{
Standard Model
decay channels of 2HDM heavy Higgs bosons.
A checkmark indicates that the partial decay width approaches a constant 
in the alignment limit, while a dash indicates that the decay width vanishes in the alignment 
limit. 
 \label{tab:channels} }
\end{center}
\end{table}

In table \ref{tab:prod-channels} we summarize the the leading LHC production channels for heavy Higgs bosons in 2HDM that are non-vanishing in the alignment limit, ordered by their relative size at LHC energies. These include resonant production of heavy neutral Higgses by gluon fusion; single production of heavy neutral or charged Higgses in association with top and bottom quarks; heavy Higgs pair production via Drell-Yan processes; heavy-light Higgs boson production via gluon fusion; and heavy Higgs pair production via gluon fusion. Other production modes that vanish in the alignment limit are {\it significantly} suppressed near the alignment limit, rendering them unpromising in the parameter space currently allowed by Higgs coupling fits. We likewise summarize the Standard Model decay channels of heavy Higgs bosons in table \ref{tab:channels}. In contrast with production modes, decay modes that vanish near the alignment limit may still be appreciable near the alignment limit, given the relatively small partial widths of competing decays.

Given proximity to the alignment limit, there is a natural ordering 
of searches for additional Higgs bosons obtained by combining the dominant production and decay modes. Many of the single heavy Higgs boson 
production 
channels are covered by existing searches, including searches for gluon fusion production of $H/A$ with decay to 
$b \bar b, \tau \tau,\gamma \gamma, \mu \mu$
as well as $WW,ZZ,Zh, hh$; searches for $b\bar{b}H/A$ associated production with decay to $b \bar b, \tau \tau, \mu \mu$; and $t \bar b H^\pm$ associated production with decay to $\tau \nu$ and $\bar t b$. However, several key channels remain uncovered, particularly gluon fusion 
with decay to $t \bar{t}$; 
associated production of $b\bar b H/A$  
 followed by decay to 
$\gamma \gamma$ and $WW,ZZ,Zh,hh$ as well as 
$t \bar t$; 
and associated production of $t\bar t H/A$ with decay to 
$b \bar b, \tau \tau,\gamma \gamma, \mu \mu$ as well as 
$WW,ZZ,Zh,hh$ and $t \bar{t}$. 
 Once decay into $t \bar t$ becomes kinematically accessible, it becomes one of the primary decay modes of heavy neutral Higgs bosons near the alignment limit, and this decay channel
may entirely dominate the visible signatures of additional Higgses. 
Similarly, $t \bar b H^\pm$ associated production with decay to  $\bar t b$ is likely to be a dominant signature of charged Higgses at the LHC
when this decay channel is open. 
Although there is a search for this mode at $\sqrt{s} = 8$ TeV~\cite{CMS:2014pea}, there is room for improvement in this channel.

In addition to decays into SM final states, it is possible for new Higgs bosons to decay into non-SM final states. These processes include both invisible decays and potentially visible decays that do not fall into the acceptance of existing searches. Given the suppression of vector associated production modes in the alignment limit, the most promising potential channels are 
$t \bar t H/A$ or $b \bar b H/A$ associated production with decay to invisible final states.

\begin{table}[t]
\small
\centering
\renewcommand{\arraystretch}{1.1}
\begin{tabular}{| l | c | c |}
\hline
Channel & Collaboration & Reference \\
\hline\hline 
$gg \to \Phi \to \gamma \gamma$ & ATLAS, 20.3 fb$^{-1}$ & \cite{Aad:2014ioa} \\
$gg \to \Phi \to \gamma \gamma$ & CMS, 19.7 fb$^{-1}$ & \cite{CMS:2014onr} \\
$gg \to \Phi \to \tau \tau$ & ATLAS, 20.3 fb$^{-1}$ & \cite{Aad:2014vgg} \\
$b \bar b \to \Phi \to \tau \tau$ & ATLAS, 20.3 fb$^{-1}$ & \cite{Aad:2014vgg} \\
$gg \to \Phi \to \tau \tau$ & CMS, 19.7 fb$^{-1}$ & \cite{Khachatryan:2014wca} \\
$b \bar b \to \Phi \to \tau \tau$ & CMS, 19.7 fb$^{-1}$ & \cite{Khachatryan:2014wca} \\
$gg \to A \to Zh \to \ell \ell + (b \bar b, \tau \tau)$ & ATLAS, 20.3 fb$^{-1}$ & \cite{Aad:2015wra} \\
$gg \to A \to Zh \to \ell \ell + b \bar b$ & CMS, 19.7 fb$^{-1}$ & \cite{CMS:2014yra} \\
$gg \to H \to hh \to b\bar b + \gamma \gamma$ & ATLAS, 20 fb$^{-1}$ & \cite{Aad:2014yja} \\
$gg \to H \to hh \to b \bar b + b \bar b$ & CMS, 17.9 fb$^{-1}$ & \cite{Khachatryan:2015yea} \\
$gg \to H \to hh \to b \bar b + \gamma \gamma$ & CMS, 19.7 fb$^{-1}$ & \cite{CMS:2014ipa} \\
$gg \to H \to ZZ \to 4 \ell$ & ATLAS, 20.7 fb$^{-1}$ & \cite{ATLAS:2013nma} \\
$gg \to H \to ZZ$ & CMS, 19.7 fb$^{-1}$ & \cite{Khachatryan:2015cwa} \\
$gg \to H \to WW$ & CMS, 19.7 fb$^{-1}$ & \cite{Khachatryan:2015cwa}   \\
\hline
\end{tabular}
\caption{Relevant ATLAS and CMS searches for heavy Higgs bosons at the $\sqrt{s} = 8$ TeV LHC. Here $\Phi = H,A$.
 \label{tab:searches} }
\end{table}

To fully characterize the state of coverage by direct searches, we interpret searches by the ATLAS and CMS collaborations for heavy Higgs states in the parameter space of Type 1 and Type 2 2HDM. The relevant search channels are summarized in table \ref{tab:searches}. These searches present limits in terms of single-channel cross sections times branching ratios that are amenable to reinterpretation. Powerful limits on $gg \to H \to hh$ and 
$gg \to A \to Zh$ for moderately heavy $H,A$ have also been obtained 
using multi-lepton and di-photon 
final states \cite{Khachatryan:2014jya}, but these bounds combine many exclusive  
channels with non-uniform scaling and acceptance 
across the 2HDM parameter space and cannot be easily reinterpreted in our framework. 

For each search, we consider the contribution of $H$ or $A$ separately (in contrast to e.g. the MSSM interpretation of searches in the $\tau \tau$ final state, which includes the sum of contributions from $h, H,$ and $A$). To determine the theory prediction for relevant cross sections times branching ratios across the 2HDM parameter space, we obtain the relevant cross sections and partial widths as a function of $\alpha$ and $\beta$ as discussed above. Here we assume that the total widths of $H$ and $A$ are determined purely by their decays into SM final states.

\begin{figure}[t]
  \centering
 \includegraphics[height=0.45\textwidth]{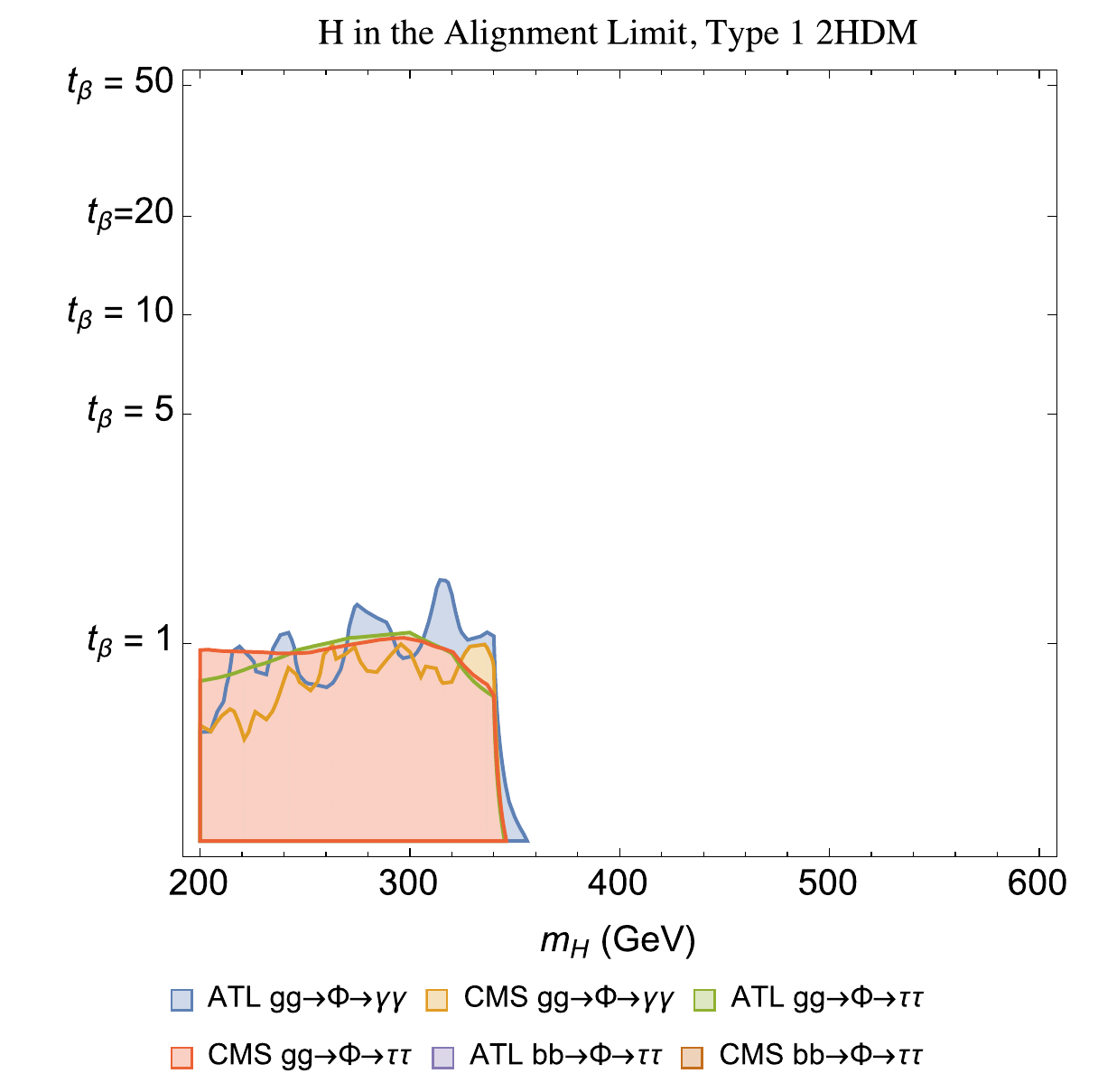}
  \includegraphics[height=0.45\textwidth]{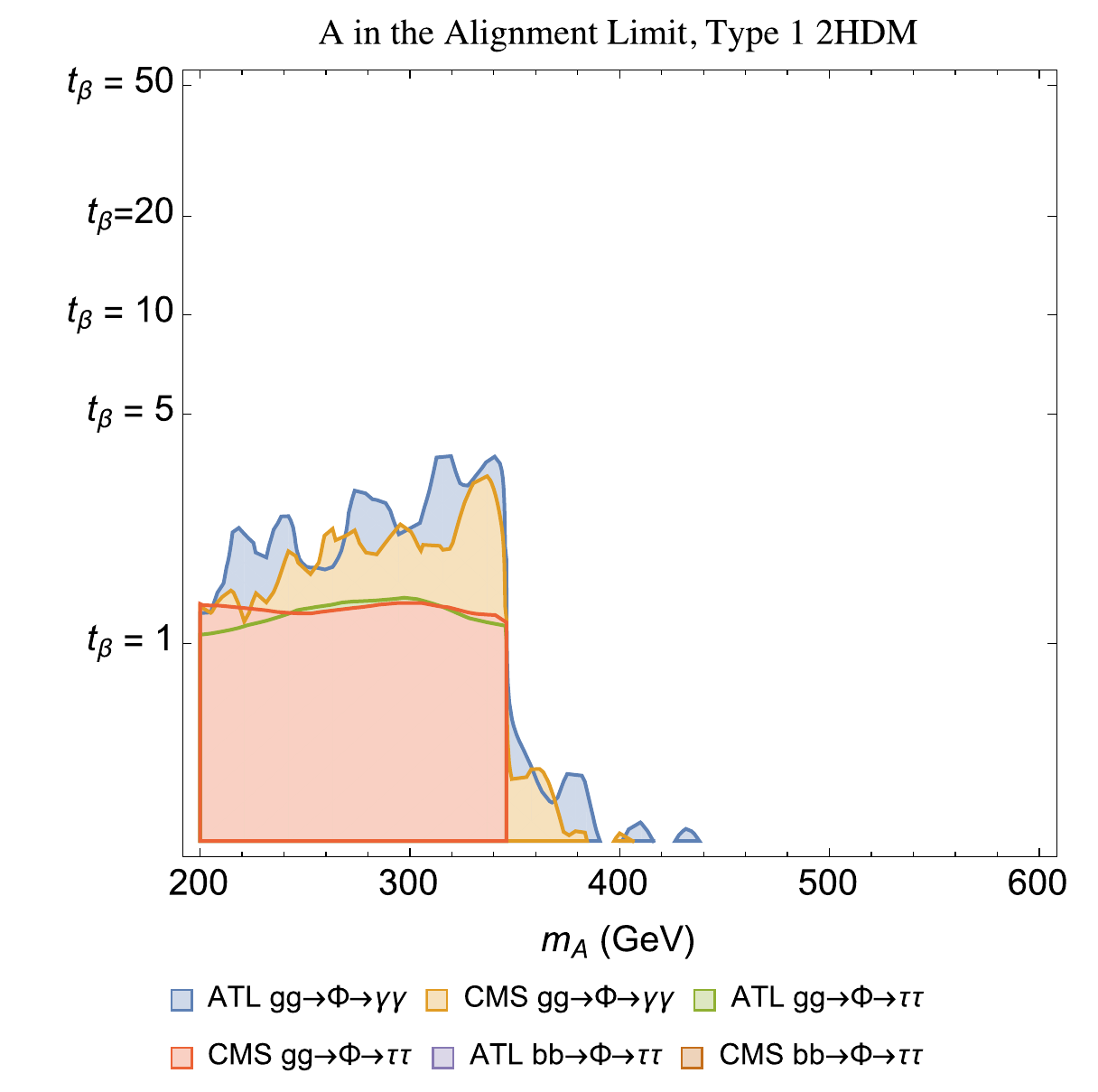}
   \includegraphics[height=0.45\textwidth]{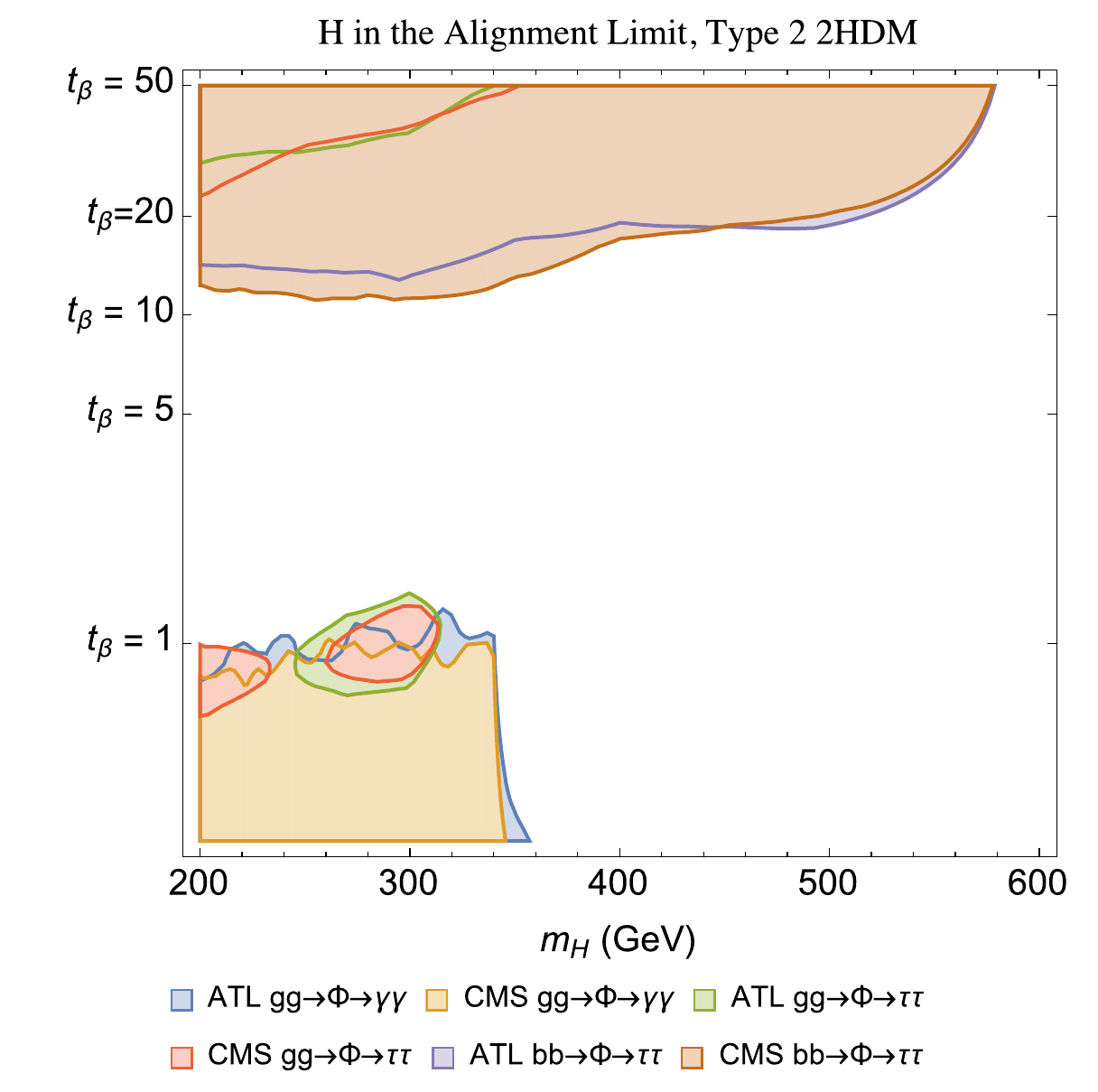}
  \includegraphics[height=0.45\textwidth]{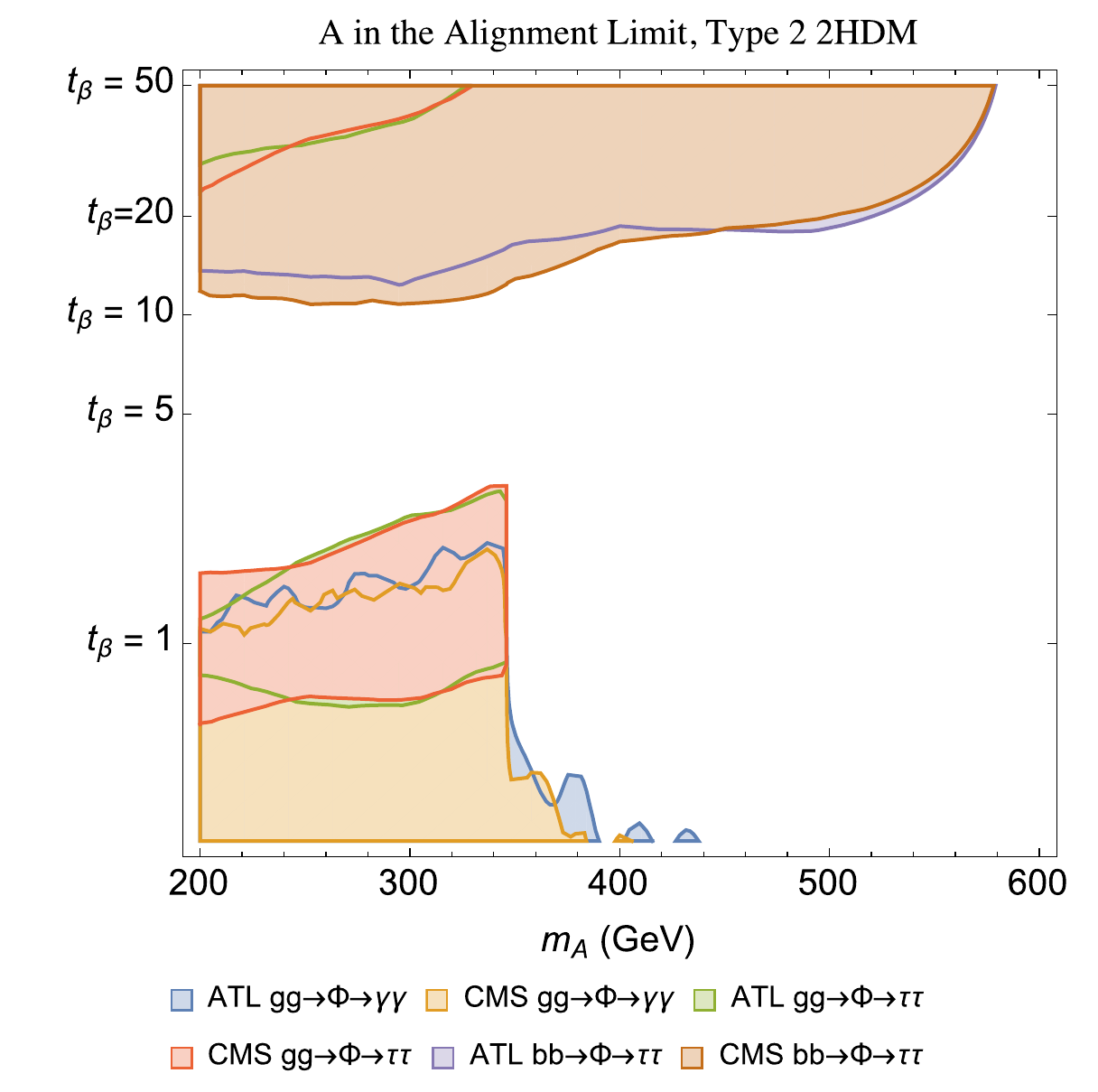}
  \caption{Top: Direct search limits on a heavy CP even neutral scalar $H$ (left) and CP odd neutral pseudo-scalar $A$ (right) as a function of mass in the alignment limit $\cos(\beta - \alpha) = 0$ in a Type 1 2HDM. Bottom: Same as above for a Type 2 2HDM.}
  \label{fig:t1ALlims}
\end{figure}

In figure~\ref{fig:t1ALlims} we present the state of current direct searches in the exact alignment limit $\cos(\beta - \alpha) = 0$ for heavy CP-even neutral scalar $H$ and CP-odd neutral pseudoscalar $A$ as a function of $\tan \beta$ and $m_{H/A}$ in Type 1 and Type 2 2HDM. In the exact alignment limit, only production and decay modes involving Higgs couplings to fermions (including gluon fusion production and decay into photons arising from top/bottom quark loops) contribute. In Type 1 2HDM all production modes involving fermions are suppressed at large $\tan \beta$, so that existing searches are only effective at low $\tan \beta$. The most sensitive search channels include inclusive production of $H/A$ followed by decay to $\gamma \gamma, \tau \tau$. These channels are modestly effective near $\tan \beta = 1$ for $m_{H/A} \lesssim 350$ GeV, but lose sensitivity for $m_{H/A} \gtrsim 2 m_t$ once decays into $t \bar t$ go on-shell.  In Type 2 2HDM both the gluon fusion and $b \bar b H/A$ associated production modes grow at large $\tan \beta$, providing additional sensitivity relative to the Type 1 scenario. Note that the exclusion due to our interpretation of searches in the $\tau \tau$ final state is somewhat weaker than the comparable MSSM exclusion plot. This is due to the fact that the MSSM interpretation combines contributions from $h, H,$ and $A$, whereas we consider only the contribution due to $H$ or $A$ individually. In both 2HDM types, the profound weakening of limits at low $\tan \beta$ in the alignment limit when the $H/A \to t \bar t$ channel becomes kinematically accessible highlights the need for effective searches in the $t \bar t$ final state.

\begin{figure}[t]
  \centering
 \includegraphics[height=0.5\textwidth]{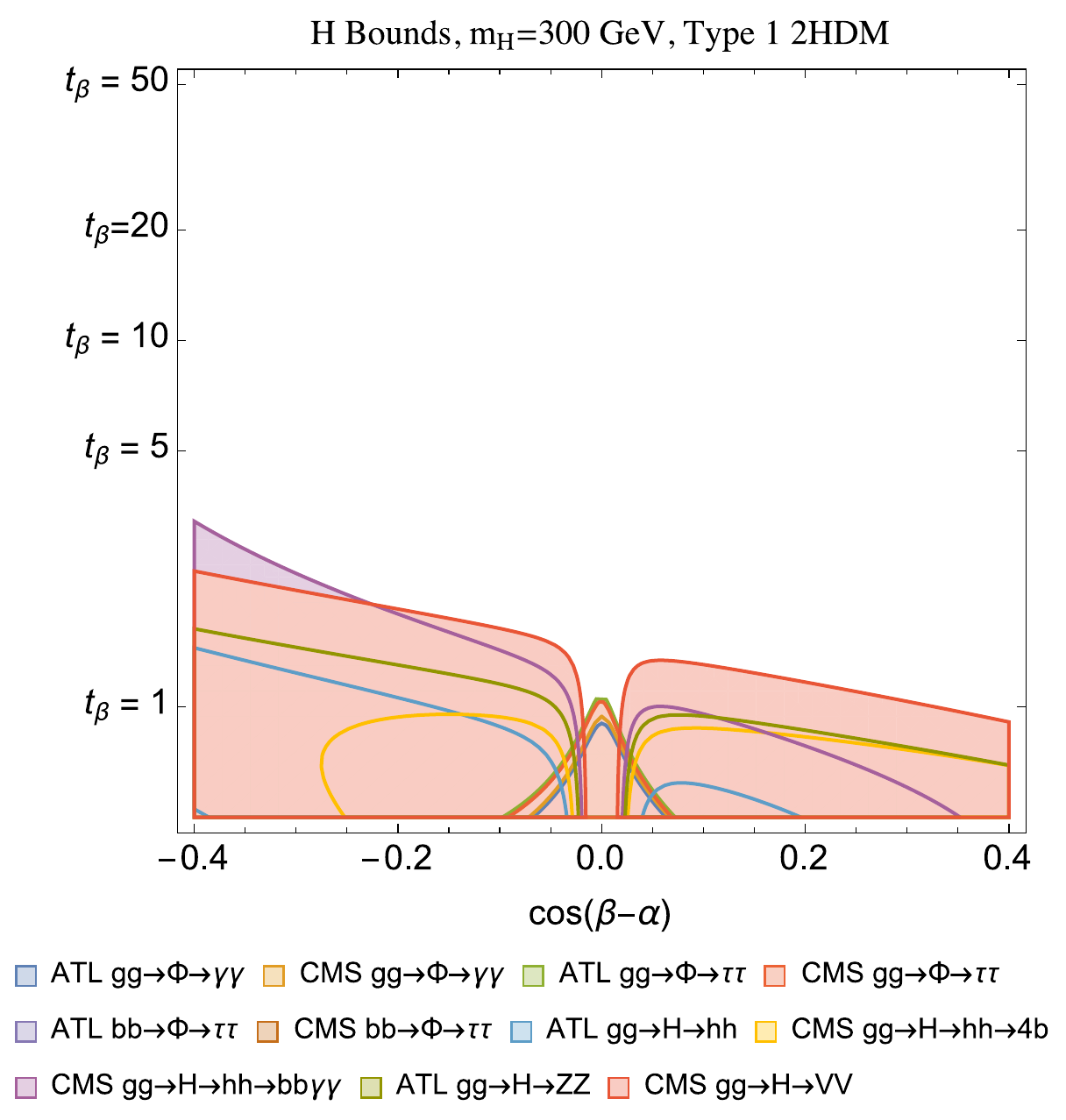}
  \includegraphics[height=0.5\textwidth]{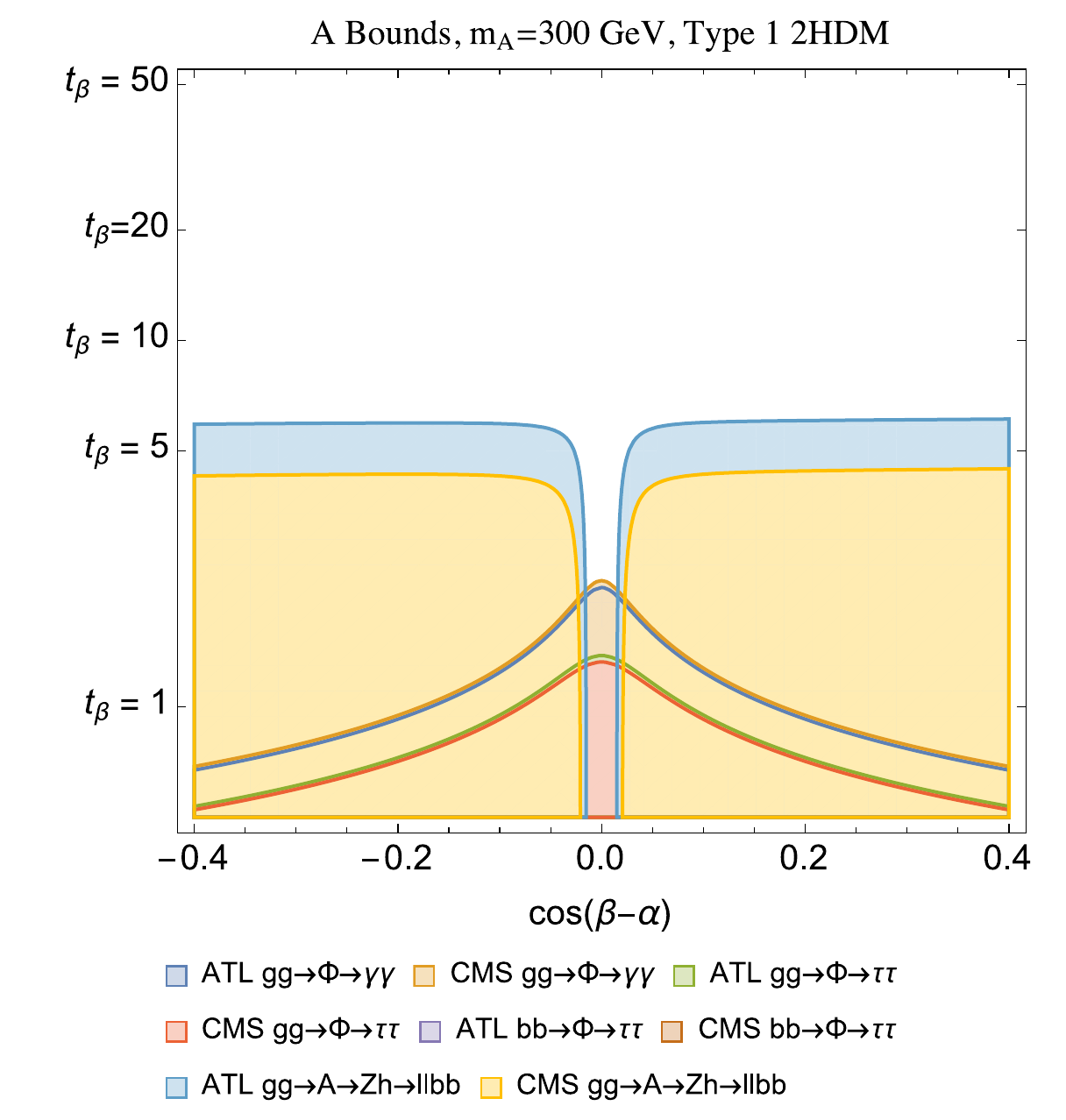}
   \includegraphics[height=0.5\textwidth]{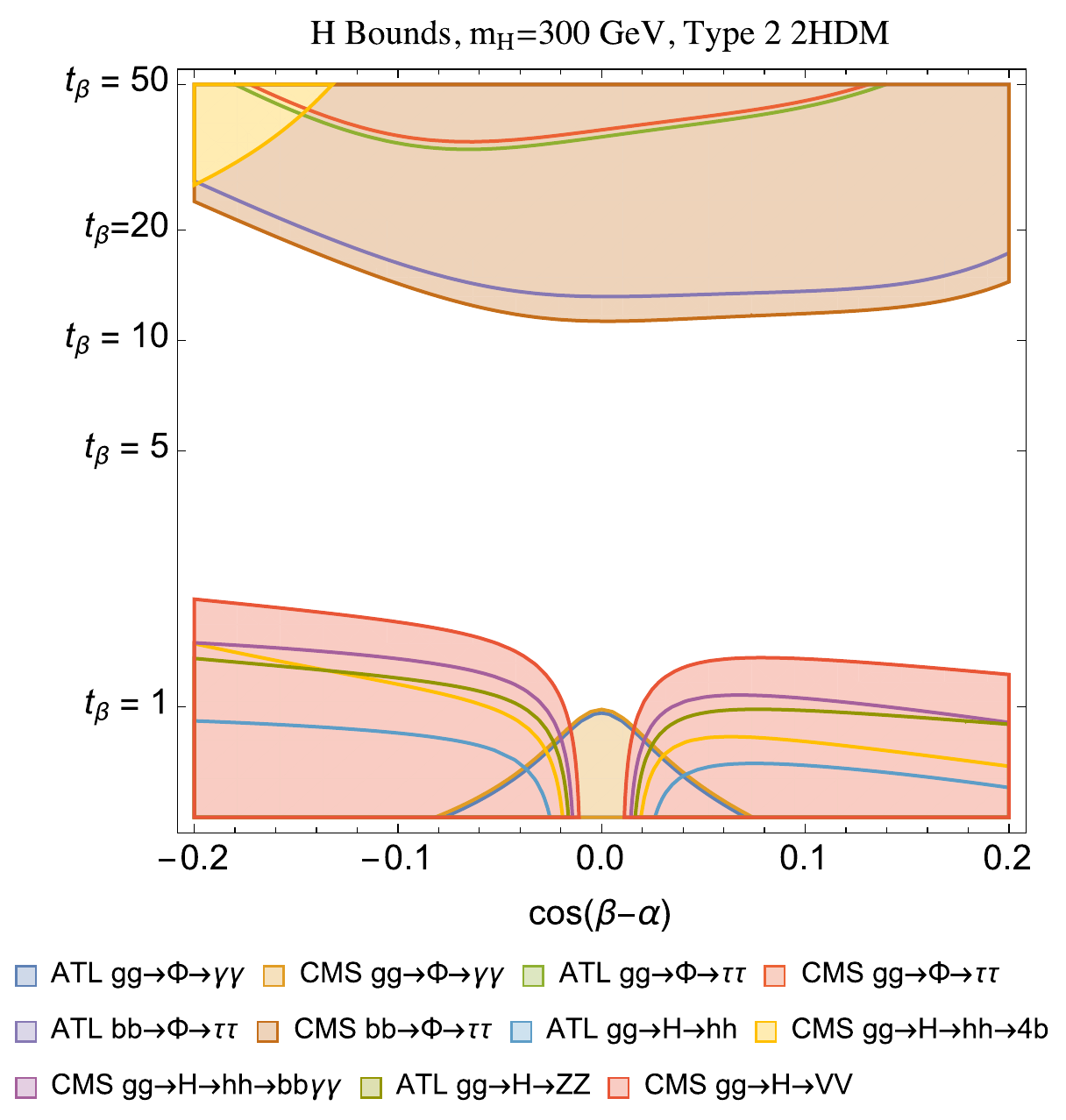}
  \includegraphics[height=0.5\textwidth]{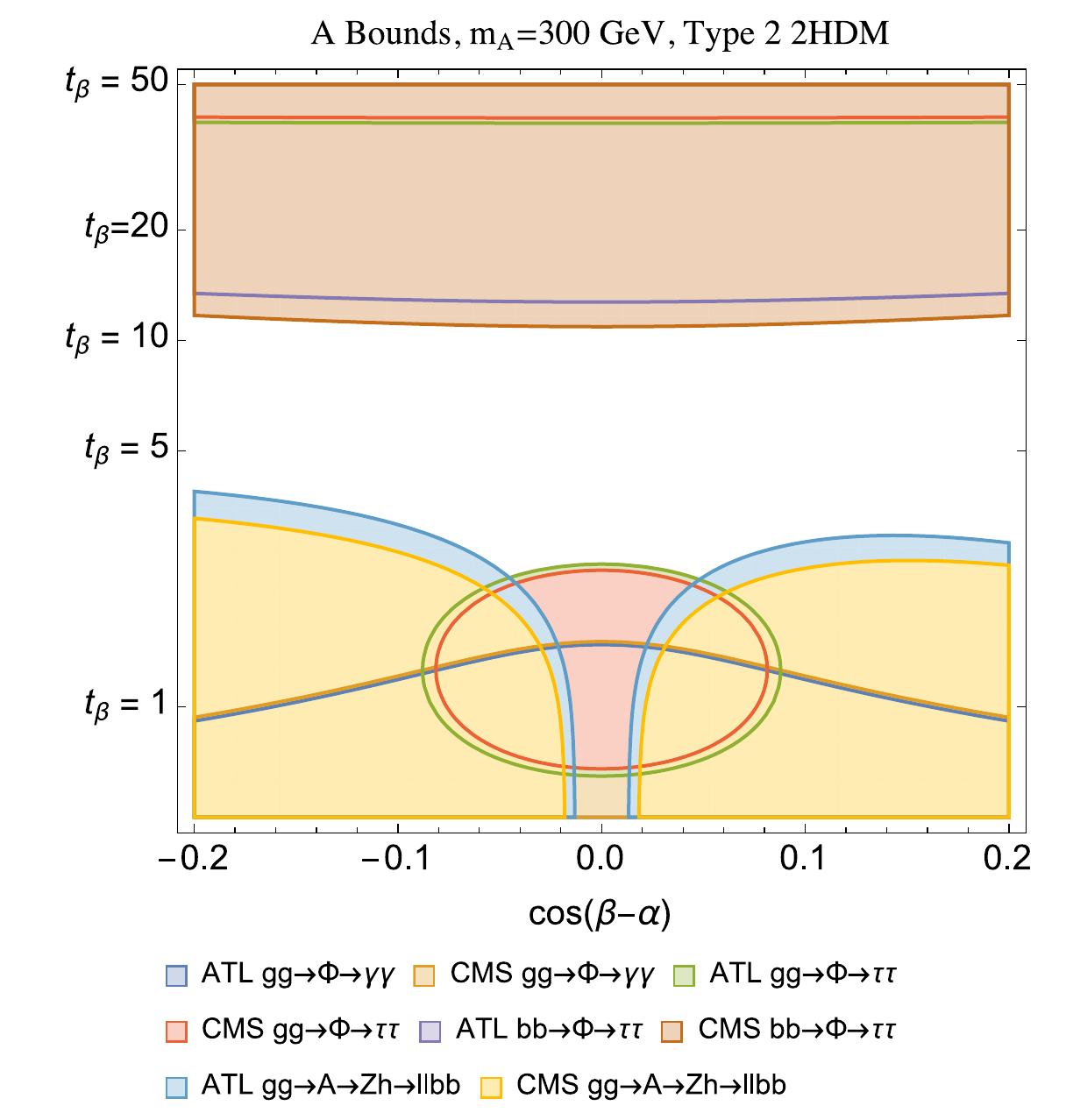}
  \caption{Top: Direct search limits on a 300 GeV CP even neutral scalar $H$ (left) and CP odd neutral pseudo-scalar $A$ (right) as a function of $\cos(\beta - \alpha)$ and $\tan \beta$ in a Type 1 2HDM. Bottom: Same as above for a Type 2 2HDM. Here we have taken $\lambda_{5,6,7} = 0$ in all plots. Note the different range of $\cos(\beta - \alpha)$ for Type 1 and Type 2 2HDM, motivated by the parameter space allowed by Higgs coupling fits.}
  \label{fig:masslims300}
\end{figure}

As we move away from the exact alignment limit, vector boson associated production modes remain unimportant, but decays into vectors can become appreciable. Given the sensitivity of searches for heavy scalars decaying into SM bosons, searches in these final states become significant relatively close to the alignment limit.  In figure~\ref{fig:masslims300} we present the state of direct searches for $H/A$ with $m_{H/A} = 300$ GeV as a function of $\tan \beta$ and $\cos(\beta - \alpha)$ in Type 1 and Type 2 2HDM. As in the case of the exact alignment limit, for Type 1 2HDM sensitivity falls off with increasing $\tan \beta$ due to the falling production cross section. The strongest limits on $H$ are provided by searches for gluon fusion production of $H$ followed by decays into $ZZ \to 4 \ell$, although these limits fall off near the alignment limit, where they are supplanted by searches for $H \to \gamma \gamma, \tau \tau$. The strongest limits on $A$ come from searches for gluon fusion production of $A$ followed by decay into $Zh \to \ell \ell b \bar b$. These limits likewise fall off near the alignment limit, where $A \to \gamma \gamma, \tau \tau$ provides complementary sensitivity. The situation for Type 2 2HDM is comparable to the Type 1 2HDM, save that searches in the $\tau \tau$ final state (either in gluon fusion or $b \bar b H/A$ associated production) become appreciable at large $\tan \beta$.

\begin{figure}[t]
  \centering
 \includegraphics[height=0.5\textwidth]{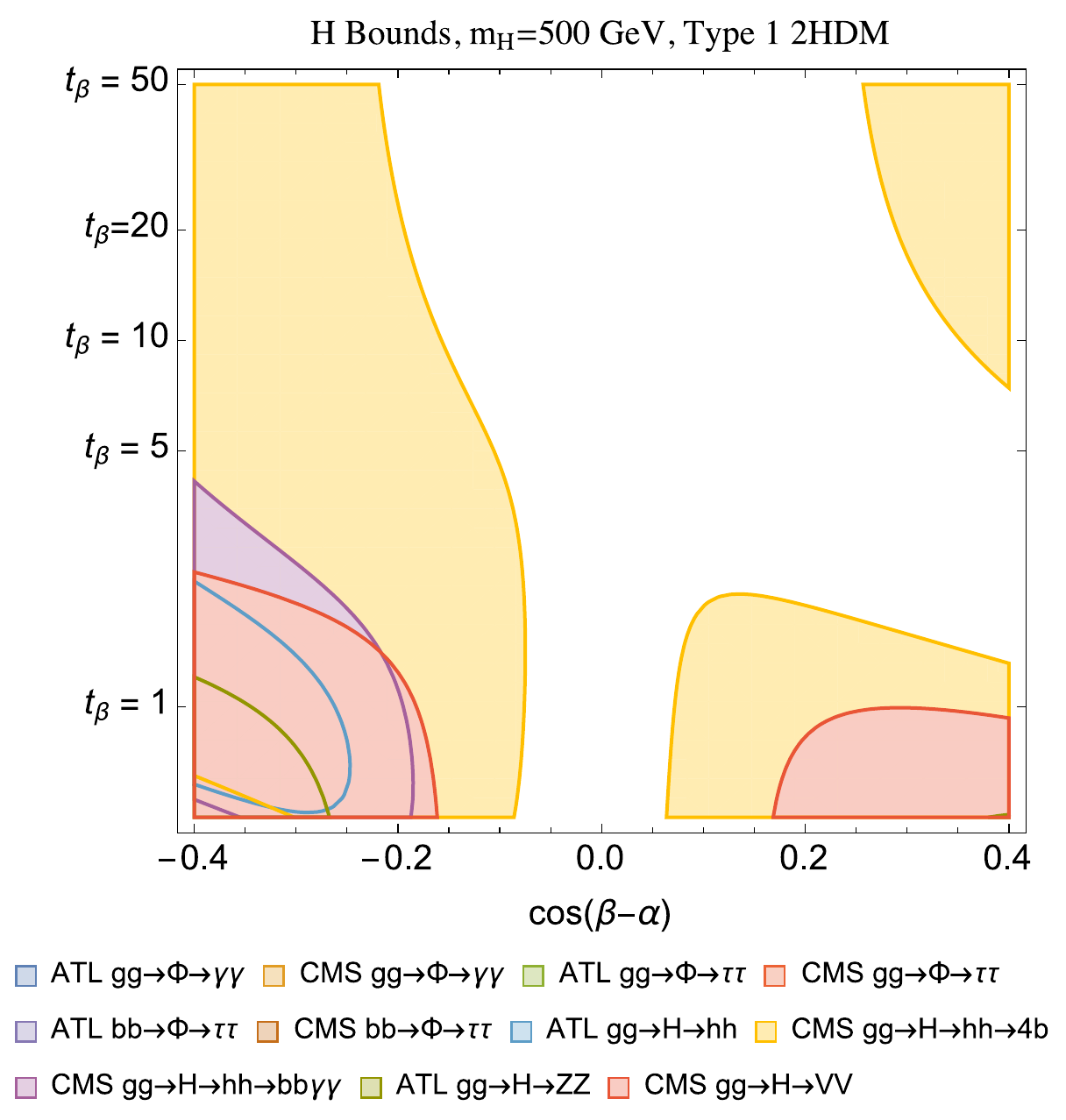}
  \includegraphics[height=0.5\textwidth]{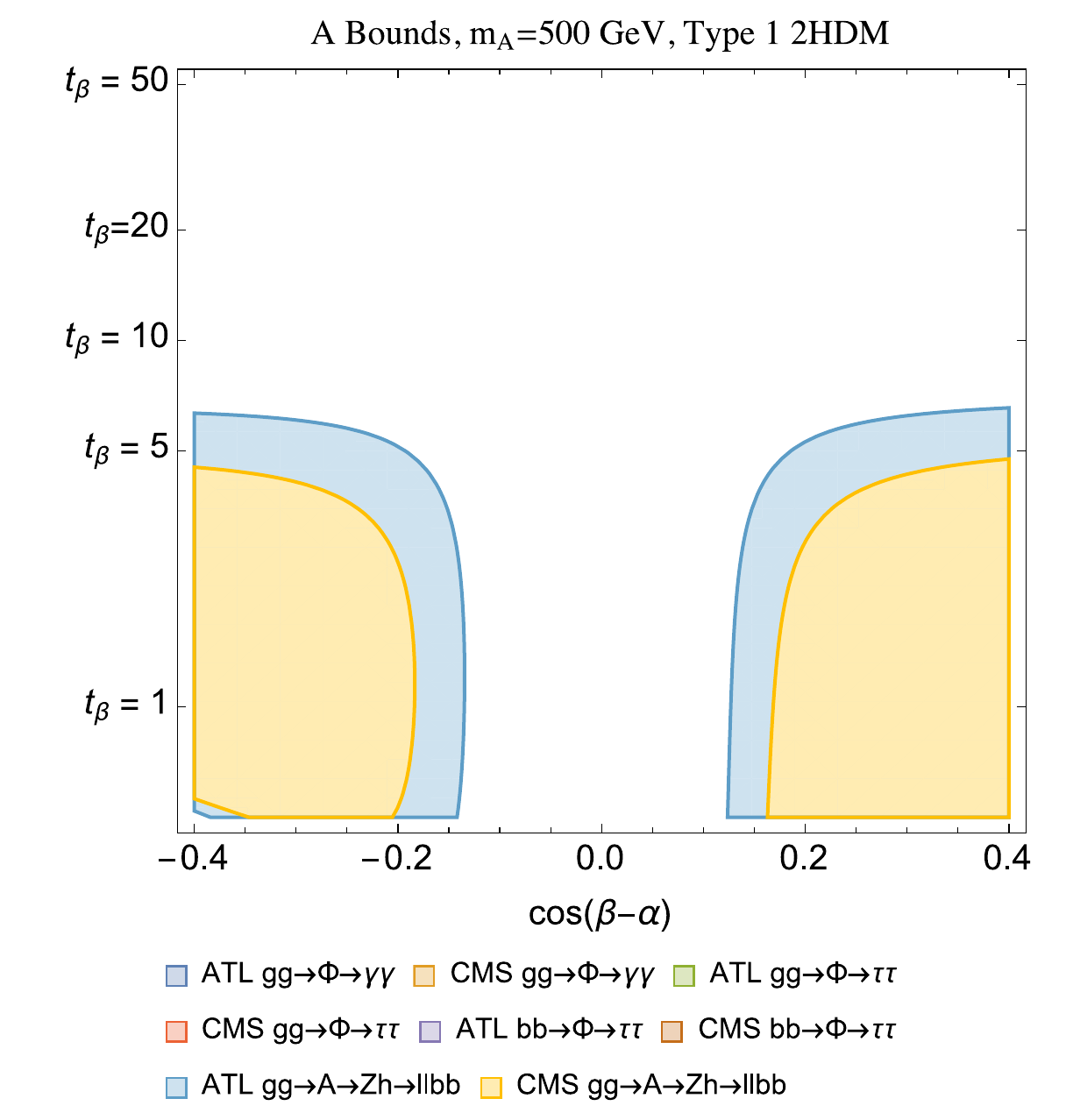}
   \includegraphics[height=0.5\textwidth]{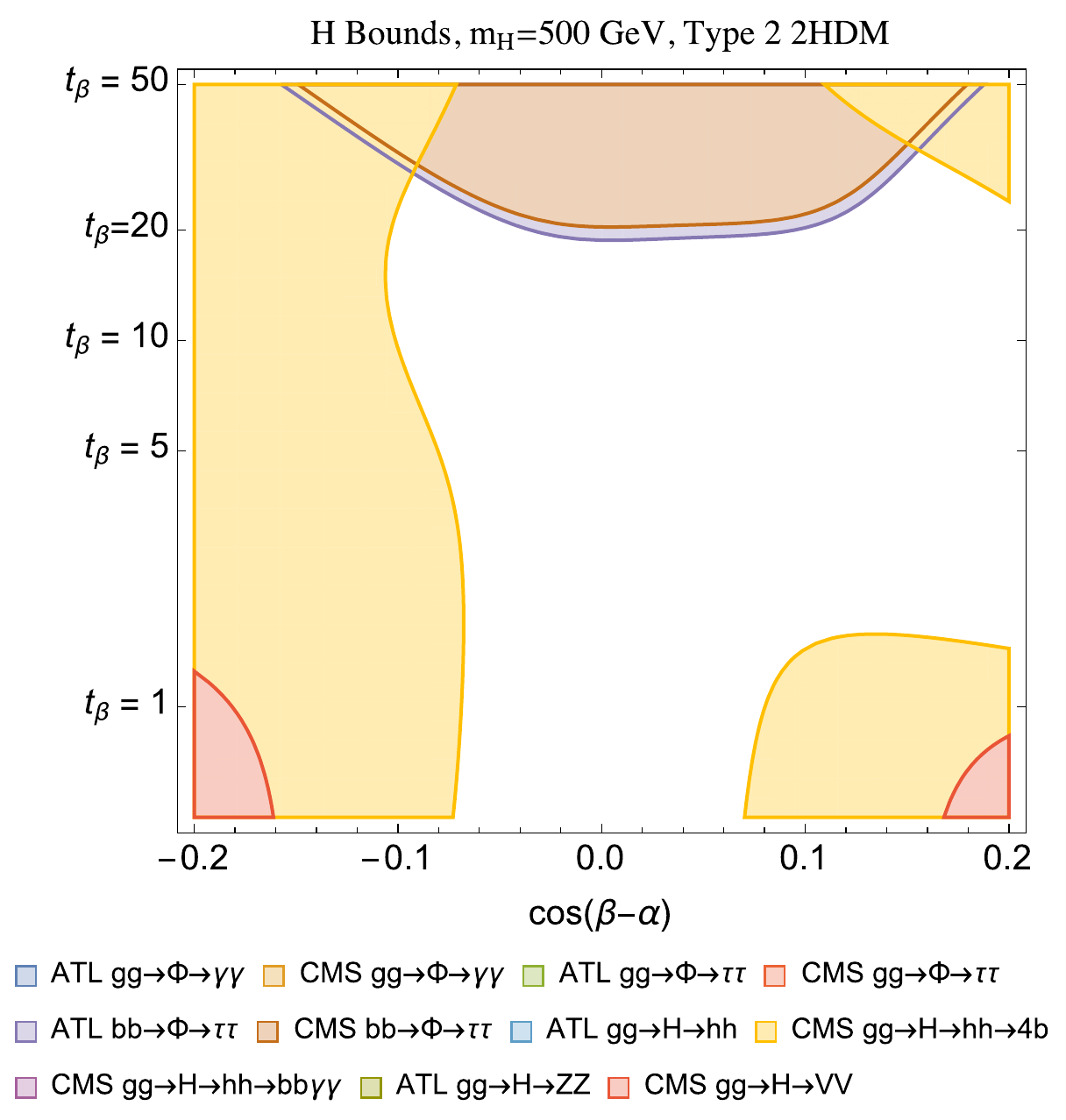}
  \includegraphics[height=0.5\textwidth]{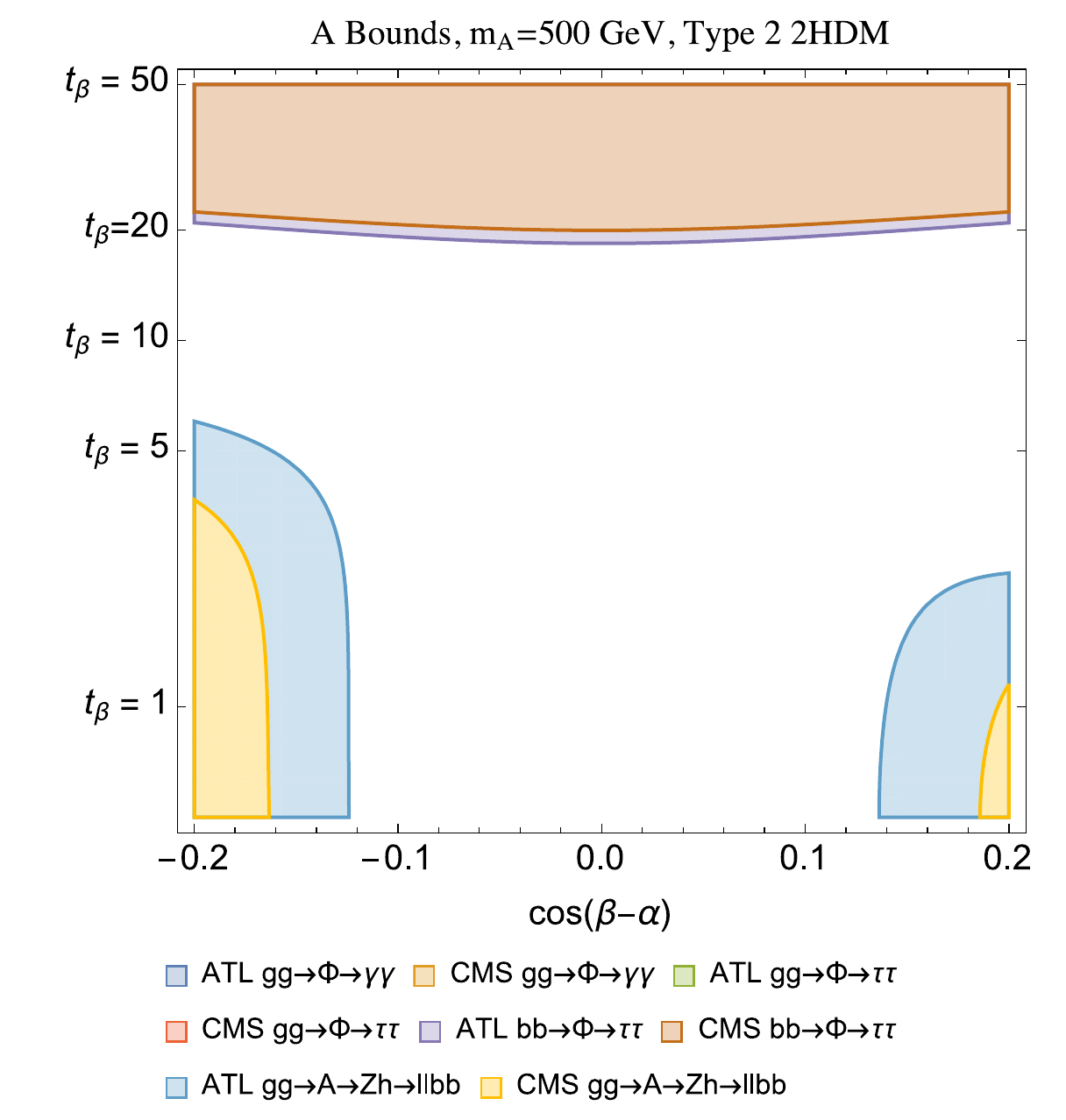}
  \caption{Top: Direct search limits on a 500 GeV CP even neutral scalar $H$ (left) and CP odd neutral pseudo-scalar $A$ (right) as a function of $\cos(\beta - \alpha)$ and $\tan \beta$ in a Type 1 2HDM. Bottom: Same as above for a Type 2 2HDM. Here we have taken $\lambda_{5,6,7} = 0$ in all plots. Note the different range of $\cos(\beta - \alpha)$ for Type 1 and Type 2 2HDM, motivated by the parameter space allowed by Higgs coupling fits.}
  \label{fig:masslims500}
\end{figure}

We repeat the process for heavier $H/A$ in Type 1 and Type 2 2HDM with $m_{H/A} = 500$ GeV as a function of $\tan \beta$ and $\cos(\beta - \alpha)$ in figure~\ref{fig:masslims500}. The limits are generally weaker compared to $m_{H/A} = 300$ GeV, due both to falling signal cross sections and additional contributions to the total width from $H/A \to t \bar t$. However, notable exceptions are the bounds on $H \to hh$ from the CMS $H \to hh \to 4b$ search and the bounds on $A \to Zh$ from the ATLAS $A \to Zh \to \ell \ell b \bar b$ search. In particular, the considerable improvement in $H \to hh \to 4b$ sensitivity at high mass is due to the boosted kinematics of the $4b$ final state \cite{Khachatryan:2015yea}.

The combination of diverse searches for heavy Higgs bosons demonstrates considerable and complementary coverage across a wide range of 2HDM parameter space, but also highlights the substantial holes in existing coverage. In particular, in Type 1 2HDM, searches lose effectiveness at large $\tan \beta$ due to falling signal cross sections, and more generally lose sensitivity near the alignment limit when $H/A$ can decay into $t\bar t$ pairs. In Type 2 2HDM there is additional sensitivity at large $\tan \beta$ due to enhanced gluon fusion and $b \bar b$ associated production, but sensitivity is poor at moderate $\tan \beta$. This is due to a combination of low production cross sections and, where kinematically available, missed decays into the $t \bar t$ final state. Among other things, these holes  demonstrate the need for an effective $H/A\rightarrow t\bar t$ search.

\section{Searching for a Neutral Higgs in $t \bar t$}
\label{sec:ttbar}

As we have discussed, the natural place to look for new Higgs states 
heavier than about 350 GeV and with SM-like coupling strength to the 
top quark is in $gg\rightarrow H/A\rightarrow t\bar t$. It has been known 
since the seminal work of~\cite{Dicus:1994bm} (and recently emphasized 
in \cite{Djouadi:2015jea}) that this channel provides an interesting and 
challenging opportunity for hadron colliders. In contrast to searches for 
spin-1 or spin-2 $t \bar t$ resonances, the spin-0 signal amplitude interferes 
with the QCD background, producing a characteristic peak-dip structure. 
As such, existing searches for $t \bar t$ resonances cannot be meaningfully 
reinterpreted to place a constraint on additional Higgs bosons in the 
$t \bar t$ final state.

In this section we begin by revisiting the analysis of~\cite{Dicus:1994bm} 
with an eye towards the impact of detector effects in a realistic collider 
environment. 
Given the size of the SM $t \bar t$ background, we introduce 
a novel technique to efficiently model both 
detector and event reconstruction effects with adequate statistics.
We find that because smearing of the reconstructed invariant mass 
of the $t \bar{t}$ system is typically the same 
order as (or larger than) the widths and range of interference 
effects of the new Higgs states, the peak-dip 
structure is largely washed out. While the statistical significance of the 
residual excess can become large at high luminosity, systematic effects 
are likely to render this channel unviable using 
standard reconstruction techniques.  

We are thus motivated to consider ancillary probes of the $t \bar t$ 
final state that are not subject to the same interference effects, namely 
the associated production channels $t\bar t H/A \rightarrow t\bar t t \bar t$ 
and $b\bar b H/A\rightarrow b\bar b t \bar t$. We do not perform complete 
14 TeV studies of these channels here, but we argue that -- based on 
current 8 TeV trilepton limits -- the four top quark 
channel in particular is likely 
to have sensitivity to moderate-mass scalars. 

\subsection{$pp\rightarrow H/A\rightarrow t\bar t$}

\begin{figure}
\begin{center}
\includegraphics[width=2.935in]{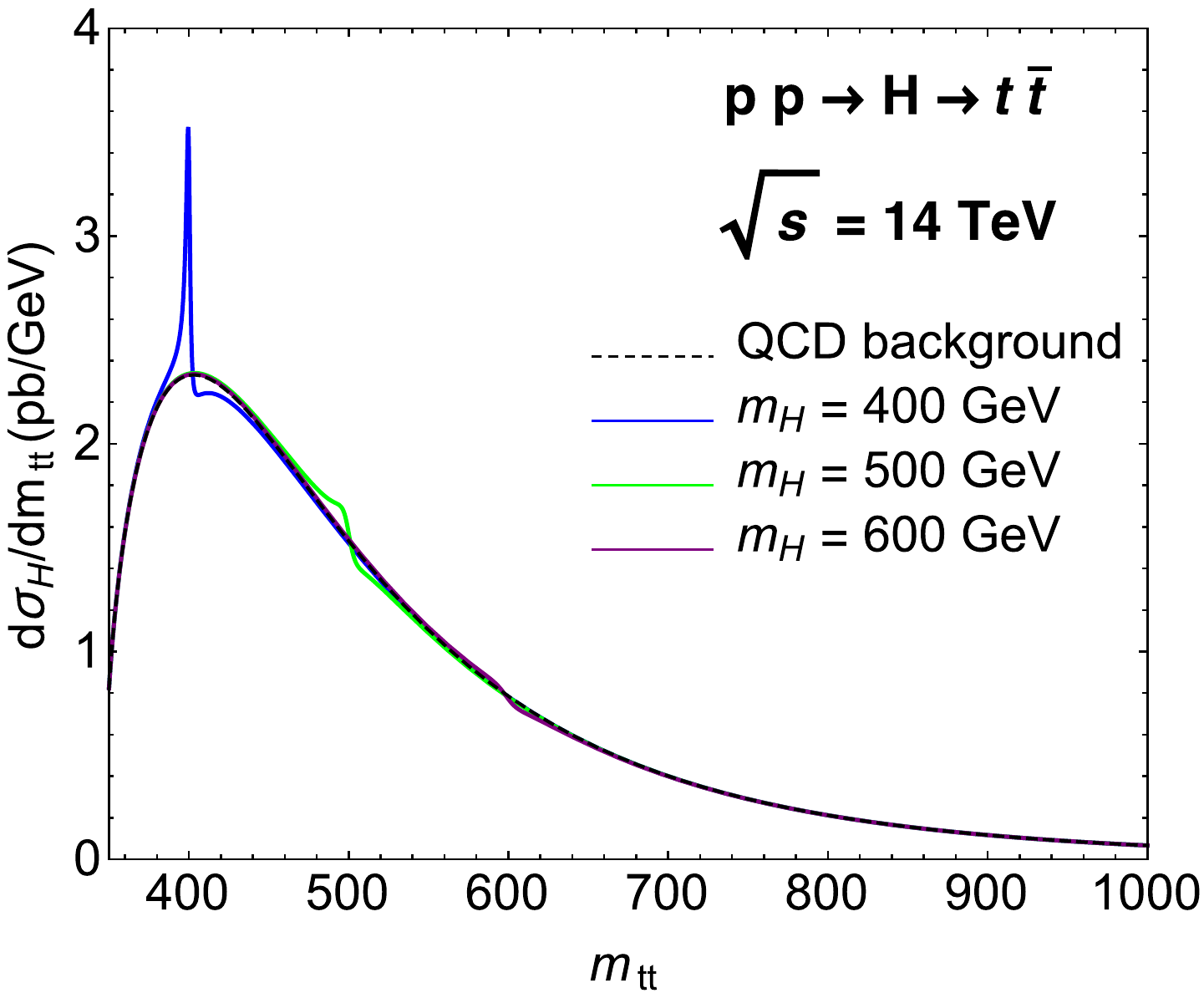} $\quad$ 
\includegraphics[width=2.935in]{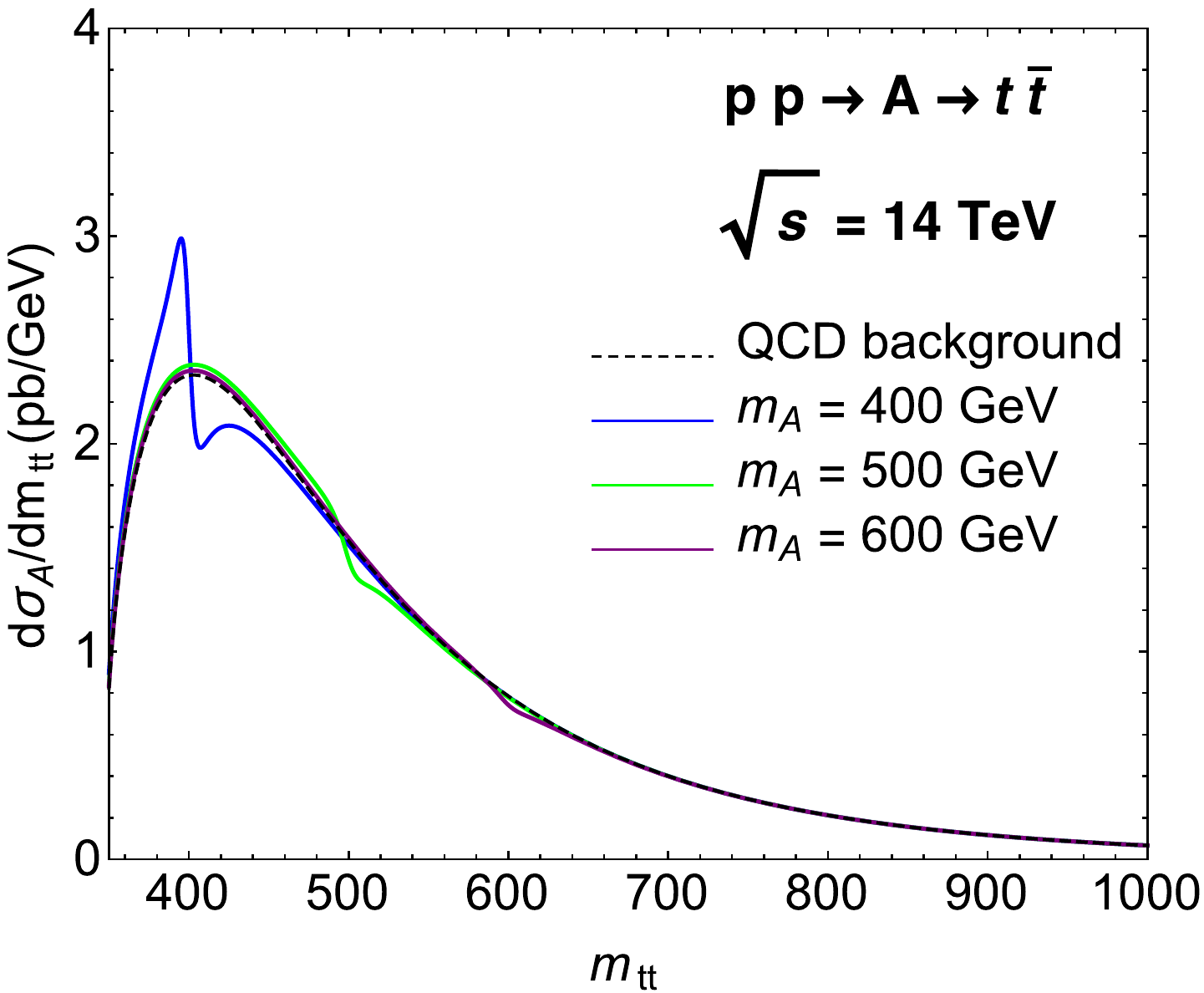}
\end{center}
\caption{Cross sections vs $t\bar t$ invariant mass for $p p \rightarrow 
\Phi \rightarrow t \bar{t}$, where $\Phi = H$ ($A$) in the left (right) panel.  
The dashed black line shows the QCD background, and the different 
solid lines are associated to different values of the $\Phi$ mass.}
\label{fig:ppHAttsigma}
\end{figure}

We begin by considering the leading-order interference effects between the 
$pp \to H/A \to t \bar t$ signal and the SM continuum $t \bar t$ 
background. In figure~\ref{fig:ppHAttsigma} we reproduce the differential rates for $p p \rightarrow H/A \rightarrow t \bar{t}$, combining the parton-level cross sections computed in~\cite{Dicus:1994bm} with the parton distribution functions (PDFs) evaluated in~\cite{Martin:2009iq}. The coupling strengths are set by the SM top Yukawa and $m_t=173$ GeV
(the full $gg \to t \bar{t}$ 
differential cross section including all interference effects 
for general 2HDM couplings
is given in appendix~\ref{appx:angular}).
 The characteristic 
peak-dip interference structure is apparent, particularly for heavier (pseudo)scalars; 
the signal-background interference term dominates the pure signal term for 
all heavy Higgs boson masses. This highlights the challenge facing searches 
for $H/A \to t \bar t$ at hadron colliders even before finite detector resolution 
is taken into account.

\begin{figure}
\begin{center}
\includegraphics[scale=0.6]{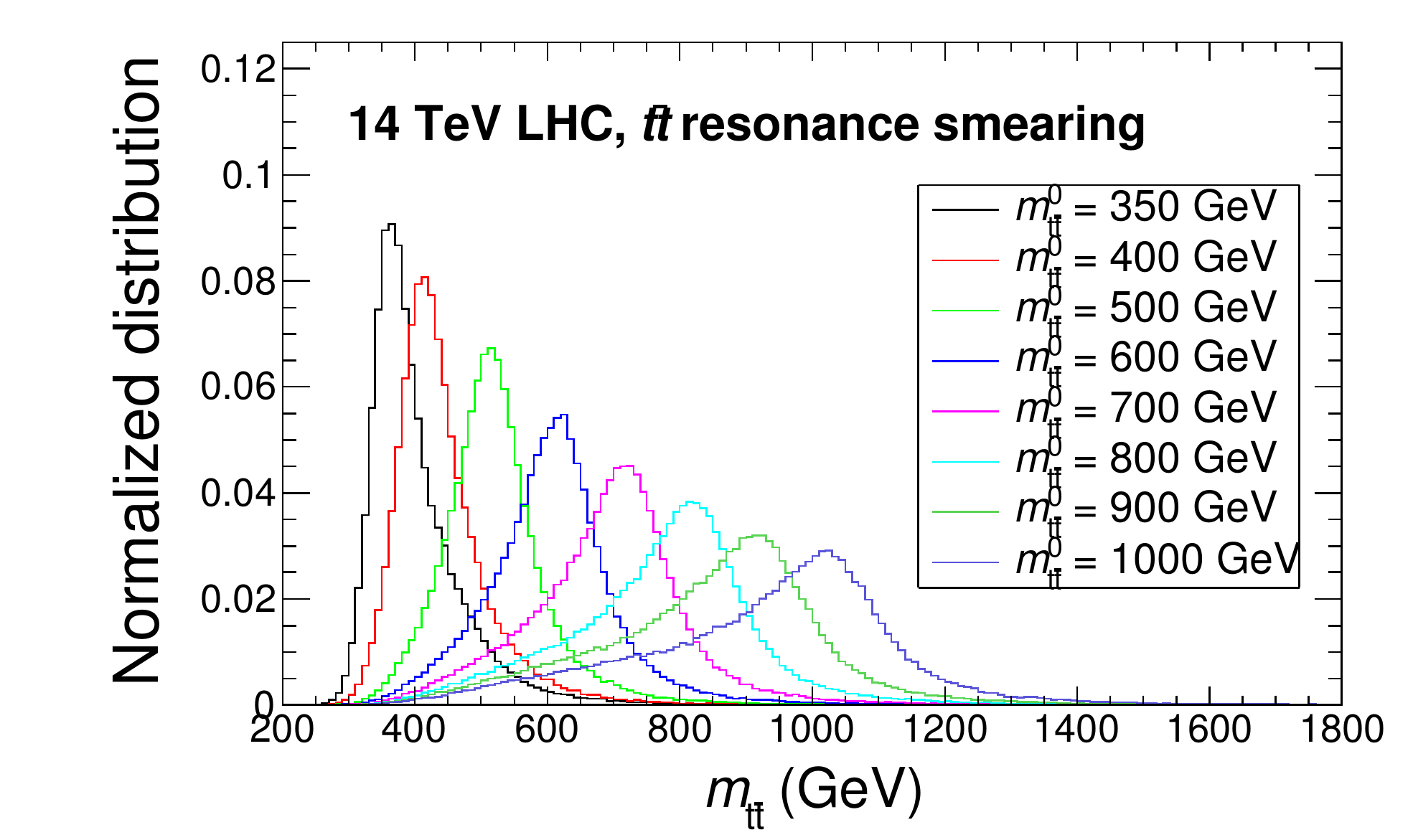} 
\end{center}
\caption{Distribution of $m_{t\bar t}$ after detector effects and $t \bar t$ 
reconstruction for different values of the produced top quark pair mass 
$m_{t\bar t}^0$.}
\label{fig:mttinstogram}
\end{figure}

\begin{figure}
\begin{center}
\includegraphics[width=2.935in]{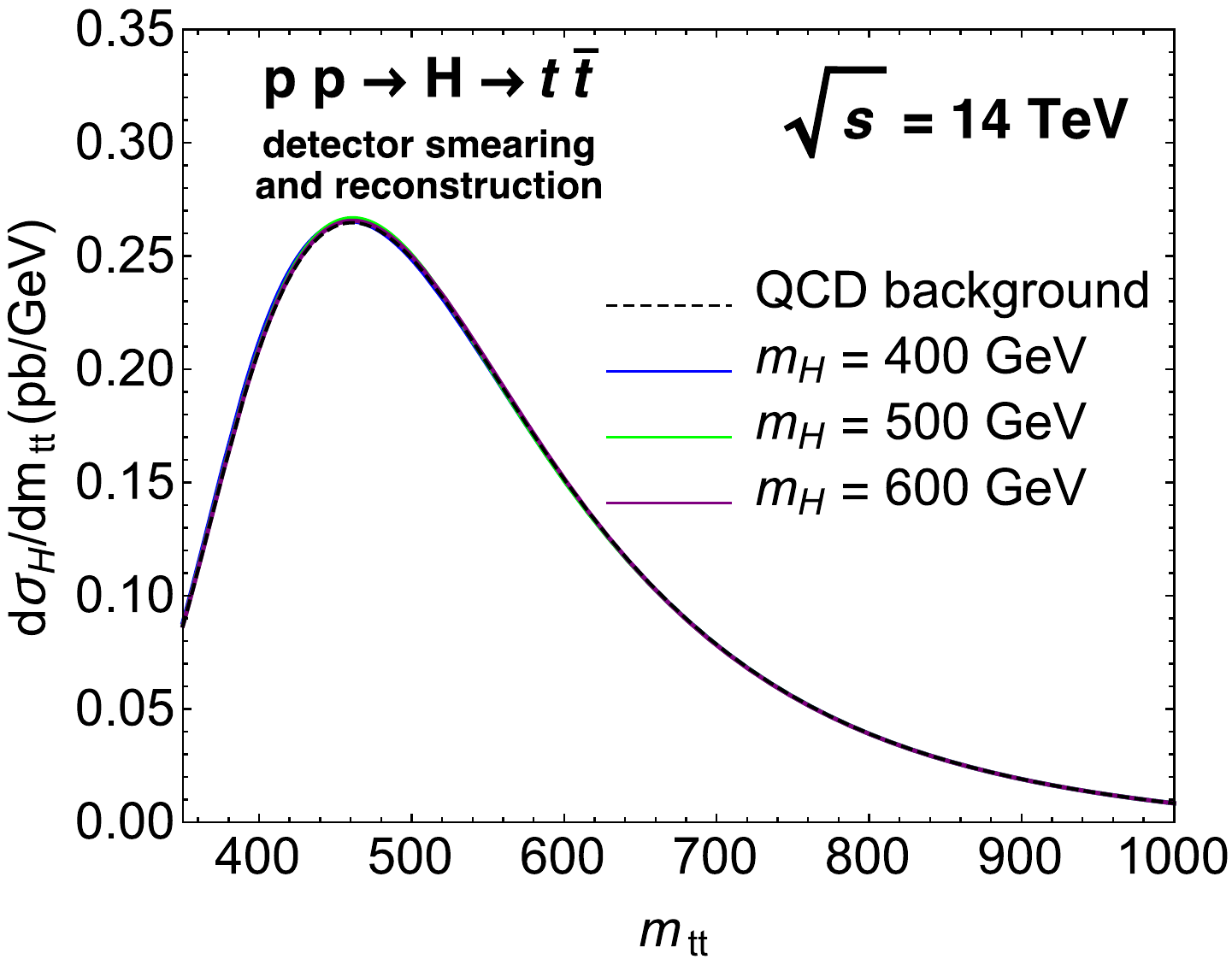} $\quad$ 
\includegraphics[width=2.935in]{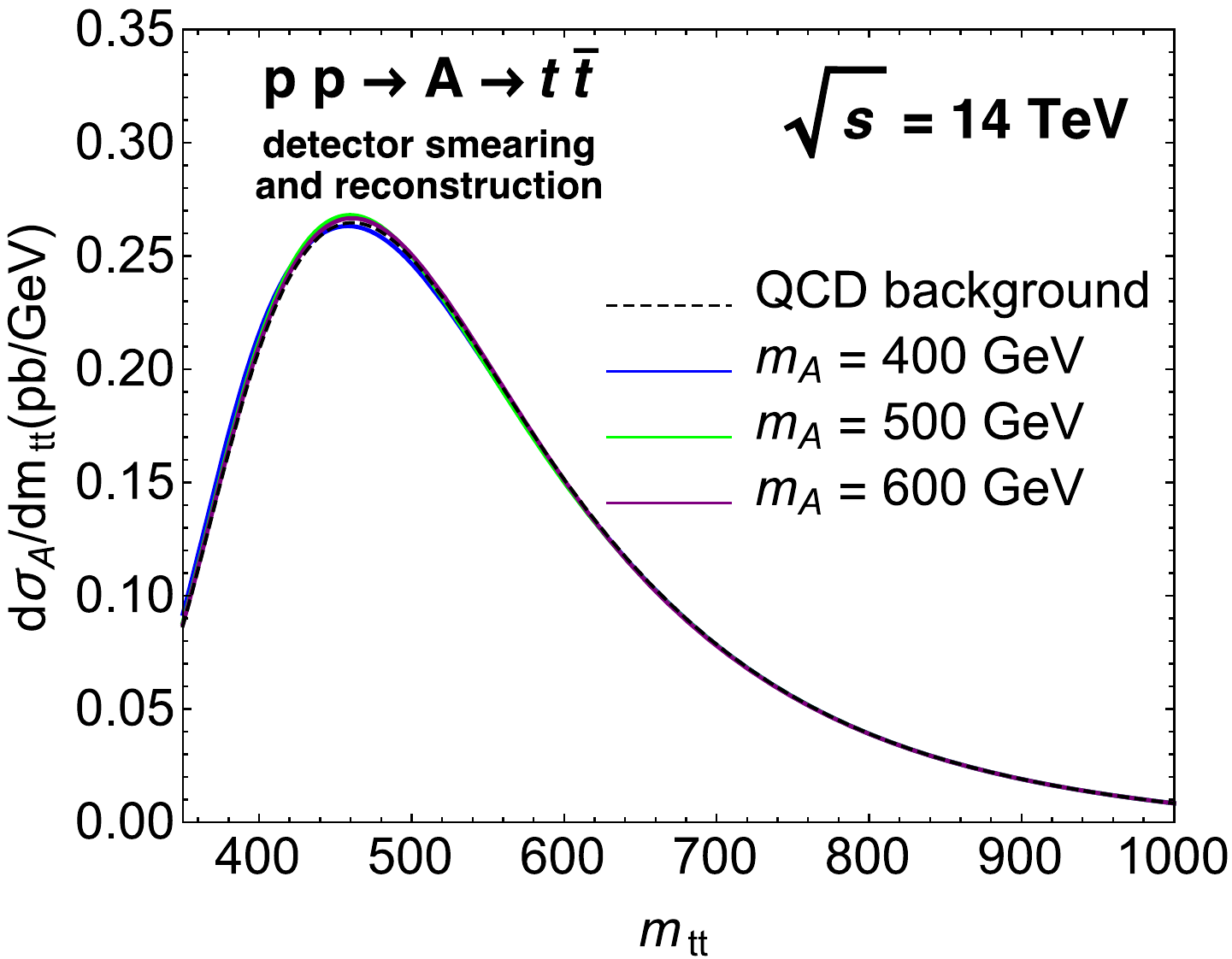}
\end{center}
\caption{Cross sections vs $t\bar t$ invariant mass for $p p \rightarrow \Phi 
\rightarrow t \bar{t}$, where $\Phi = H$ ($A$) in the left (right) panel.  
Relative to figure~\ref{fig:mttinstogram}, we now include detector
and reconstruction effects. 
In figure~\ref{fig:ppHAttsigmasmearedDIF} we plot the difference 
between the background and signal+background curves.}
\label{fig:ppHAttsigmasmeared}
\end{figure}

Given the size of the SM $t \bar t$ background and delicacy of the 
signal-background interference, it is crucial to incorporate detector effects 
with adequate Monte Carlo statistics. To efficiently simulate detector effects, 
we derive composite 
smearing functions for $t \bar t$ events as follows: We consider 
seven different reference values for the top quark pair invariant mass $m_{t\bar 
t}^0$, and for each we generate $10^6$ QCD $t\bar t$ events in Madgraph
\cite{Alwall:2014hca}, requiring $|m_{t\bar t}-m^0_{t\bar t}|<0.5$ GeV. We then 
shower with PYTHIA6.4~\cite{Sjostrand:2006za} and process the events through 
Delphes3~\cite{deFavereau:2013fsa,Cacciari:2011ma}. We then reconstruct the semi-leptonic $t \bar t$ system using mass-shell constraints as detailed in appendix~\ref{appx:reconstr}, thereby obtaining a response 
function mapping $m^0_{t\bar t}$ to an $m_{t\bar t}$ distribution. 
In \Fig{fig:mttinstogram} we plot histograms of these $m_{t\bar t}$ distributions. 
Interpolating numerically in $m_{t\bar t}^0$ and $m_{t\bar t}$, we obtain a 
kernel $P(m_{t\bar t}^0, m_{t\bar t})$ against which we can convolve the 
PDF-smeared parton-level differential cross section. This allows us to model 
the effects of detector resolution and $t \bar t$ reconstruction on the peak-dip 
structure without being limited by Monte Carlo statistics. We plot the results 
in the two panels of \Fig{fig:ppHAttsigmasmeared} for the scalar and the 
pseudoscalar. 

\begin{figure}
\begin{center}
\includegraphics[width=2.935in]{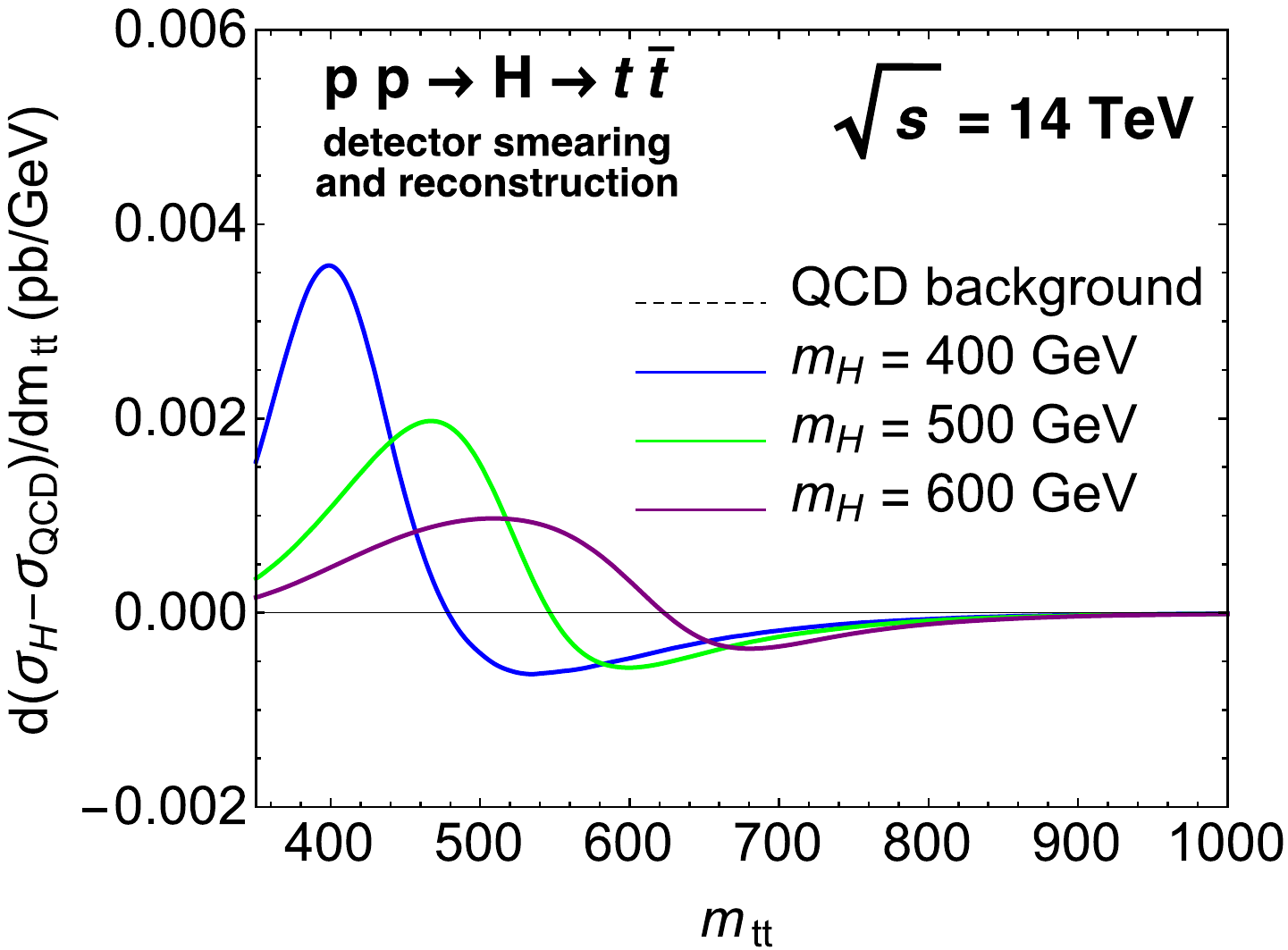} $\quad$ 
\includegraphics[width=2.935in]{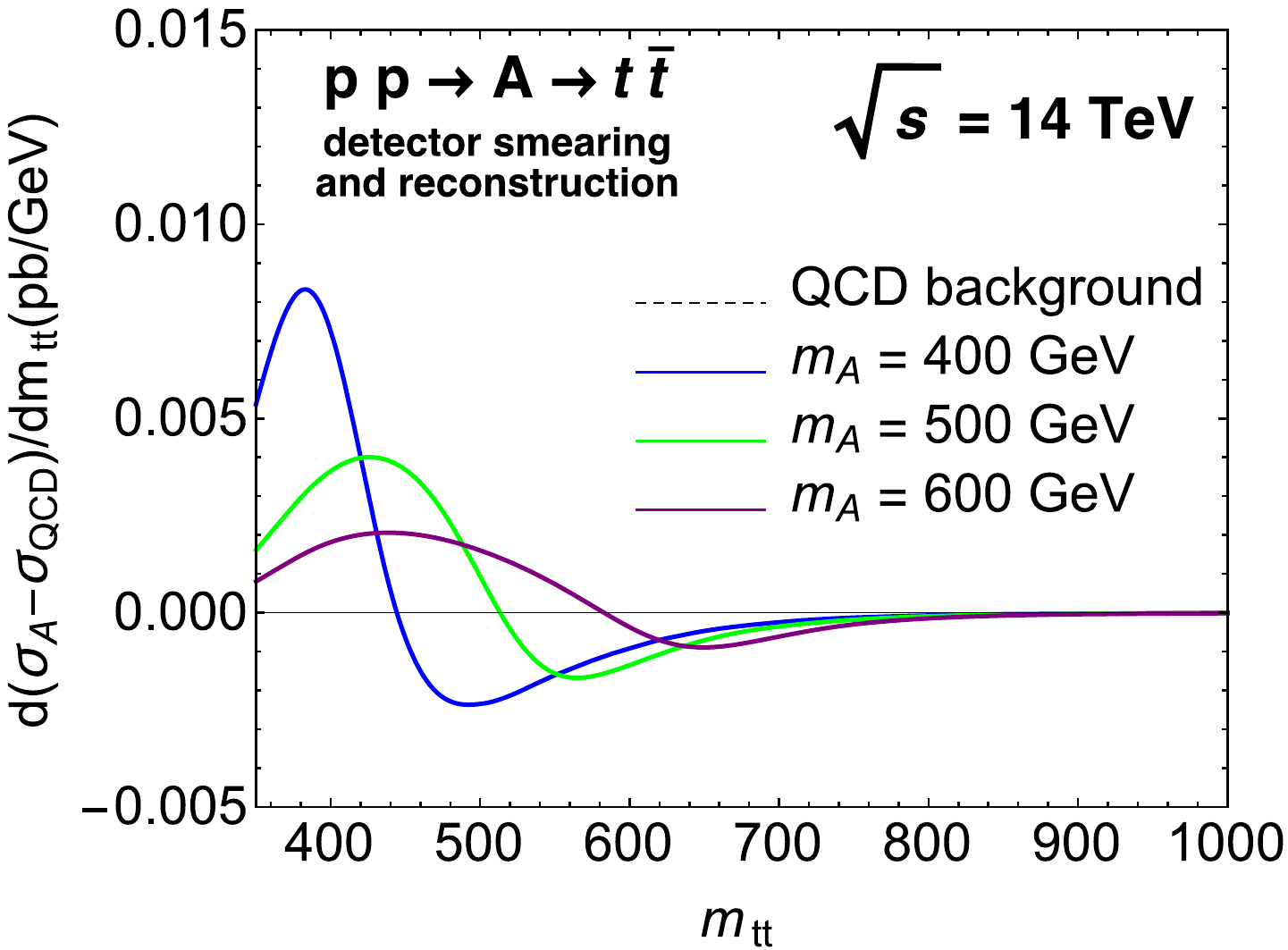}
\end{center}
\caption{Difference between the background-only and signal+background 
cross section curves shown in figure~\ref{fig:ppHAttsigmasmeared} for 
$p p \rightarrow \Phi \rightarrow t \bar{t}$, where $\Phi = H$ ($A$) in the 
left (right) panel.}
\label{fig:ppHAttsigmasmearedDIF}
\end{figure}

\begin{figure}
\begin{center}
\includegraphics[width=2.935in]{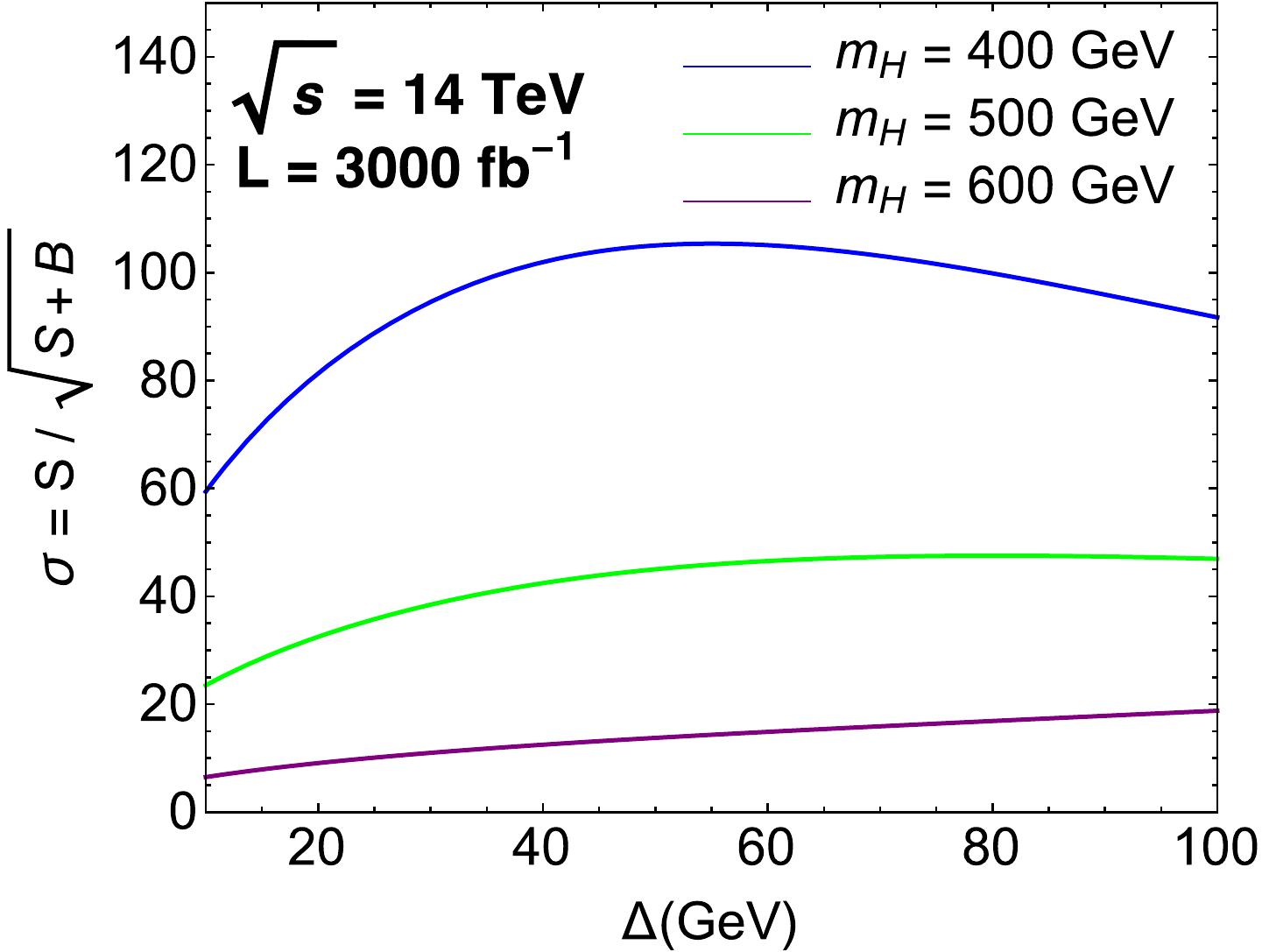} $\quad$ 
\includegraphics[width=2.935in]{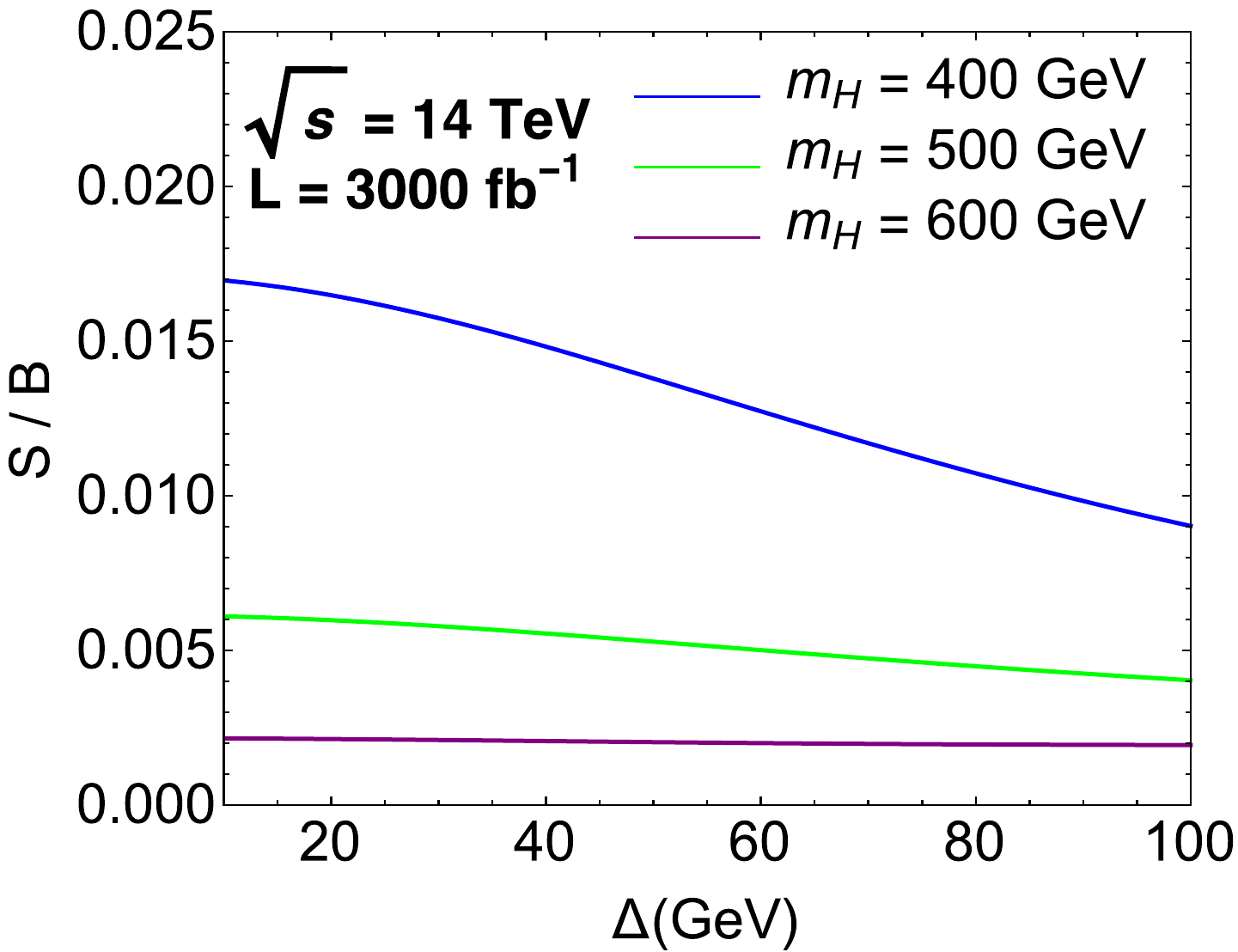}
\end{center}
\caption{Left: Best-$m_{t\bar t}$-bin statistical significance expected at 
3000 fb$^{-1}$ for the scalar case as a function of bin width $\Delta$. Right: the corresponding $S/B$. 
Qualitatively similar results hold for the pseudoscalar resonance.}
\label{fig:sigttbar}
\end{figure}

Detector resolution and $t \bar t$ reconstruction completely erode the 
peak-dip structure in the presence of a heavy Higgs, leaving behind 
only modest shifts in the $t \bar t$ invariant mass distribution relative to 
the QCD prediction. In figure~\ref{fig:ppHAttsigmasmearedDIF} we 
plot the difference between the smeared invariant mass spectra predicted 
by QCD with a heavy Higgs boson and pure QCD. The best-$m_{t\bar t}$-bin 
statistical significances $\sqrt{\Delta\chi^2}$ at 3000 fb$^{-1}$ and the 
corresponding $S/B$ are shown for the scalar resonance in figure~\ref{fig:sigttbar} 
as a function of bin size; qualitatively similar results hold for the pseudoscalar. 
From these figures, we conclude that although the high-luminosity LHC 
will have sufficient statistical power to observe $H/A\rightarrow t\bar t$ in 
principle, systematic uncertainties (even at the percent level) will almost 
certainly prevent any significant detection. Although we have only considered 
signal and background and leading order (as full next-to-leading-order (NLO)
expressions for signal+background do not yet exist), it is unlikely that 
the inclusion of NLO effects will significantly alter these conclusions.

Of course, there is more information in the $t\bar t$ final state than just 
the invariant mass; angular distributions and spin correlations may 
provide additional handles. In appendix~\ref{appx:angular} we present 
a parametrization of the $t\bar t$ differential cross section in terms 
of a well behaved scattering variable 
that affords some additional discrimination between signal 
and background.  
A full multivariate analysis employing all this ancillary information 
would increase the sensitivity incrementally,  but 
we do not expect this would substantially alter our conclusions.

\subsection{$pp \to b\bar b H/A \to b \bar b t \bar t$ and $pp \to 
t \bar t H/A \to t \bar t t\bar t$ }

Given the considerable challenges facing a search in the $t \bar t$ 
final state, it is useful to consider associated production modes in 
conjunction with $H/A\rightarrow t\bar t$. In the alignment limit, 
vector associated production modes for $H$ such as Higgs strahlung 
or VBF are strongly suppressed, while such modes 
are entirely nonexistent for $A$. This suggests focusing on fermionic 
associated production modes such as $t \bar t H/A$ or $b\bar b H/A$. 
The former is appreciable at low $\tan \beta$ in both 2HDM types, 
while the latter is appreciable at moderate to large $\tan \beta$ in 
Type 2 2HDM such as the Minimal Supersymmetric Standard Model 
(MSSM). 

Let us first consider $t \bar t$ associated production of $H$ or $A$ 
followed by decay to $t \bar t$. On one hand, the resulting 4-top 
final state provides an abundance of promising signal channels. 
On the other hand, the massive three-body kinematics of $t \bar t$ 
associated production with $m_{H/A} \gtrsim 2 m_t$ lead to at 
most a $\mathcal{O}$(fb) rate at 8 TeV. Prospects improve significantly 
at $\sqrt{s} = 14$ TeV, but even here the production cross section 
is at most in the tens of femtobarn. 

There are a variety of searches at $\sqrt{s} = 8$ TeV that are sensitive 
to the 4-top final state, particularly those involving same-sign dileptons 
(SSDL) or multileptons with additional $b$-tagged jets. To estimate the 
reach of the 4 top channel at $\sqrt{s} = 8$ TeV, we reinterpret an 8 TeV 
CMS multilepton search~\cite{CMS-PAS-SUS-13-008}. We use SRs 8, 
18, and 28 of~\cite{CMS-PAS-SUS-13-008}. These signal regions are 
all characterized by trilepton events with a $Z$-veto, at least four jets 
with two $b$-tags, and $H_T>200$ GeV, and are distinguished by $\met$ 
in the range 50-100 GeV, 100-200 GeV, and $\geq 200$ GeV, respectively.
We compute the acceptance times efficiency for $t \bar t H/A \to t \bar t 
t \bar t$ signals in these signal regions using Madgraph/Pythia/Delphes 
as above. Signal regions 8 and 18 are the most sensitive, with SR 28 
contributing additional sensitivity for larger values of $m_{H/A}$. To set 
limits using the observed event counts in~\cite{CMS-PAS-SUS-13-008}, 
we treat each bin as an independent Poisson variable and combine 
limits from individual bins using a Bayesian algorithm with a flat prior 
on signal strength, marginalizing over a normally-distributed background 
uncertainty. We neglect potential uncertainties on the signal cross section.

The 8 TeV data is insufficient to set a limit on a SM-like $Ht\bar t$ coupling. 
For example, we find for $m_H = 350$ GeV that the combination of SRs 8, 
18, 28 exclude $\sigma \cdot {\rm Br} \gtrsim 160$ fb. By contrast, for 
$m_H = 350$ GeV and $\tan \beta = 1$ we have $\sigma(p p \to t \bar t H) 
\simeq 5$ fb at $\sqrt{s} = 8$ TeV, placing the signal cross section more 
than an order of magnitude below the current exclusion in this channel. 
While the inclusion of additional channels such as SSDL would improve 
sensitivity, the disparity between signal cross section and exclusion too 
great to hope for meaningful sensitivity in this channel at $\sqrt{s} = 8$ TeV. Although we have explicitly considered the case of $pp \to t \bar t H$, the rate for $pp \to t \bar t A$ is comparable, and degenerate $H/A$ in the alignment limit would lead to a doubling of the signal.

\begin{figure}
\begin{center}
\includegraphics[height=0.45\textwidth]{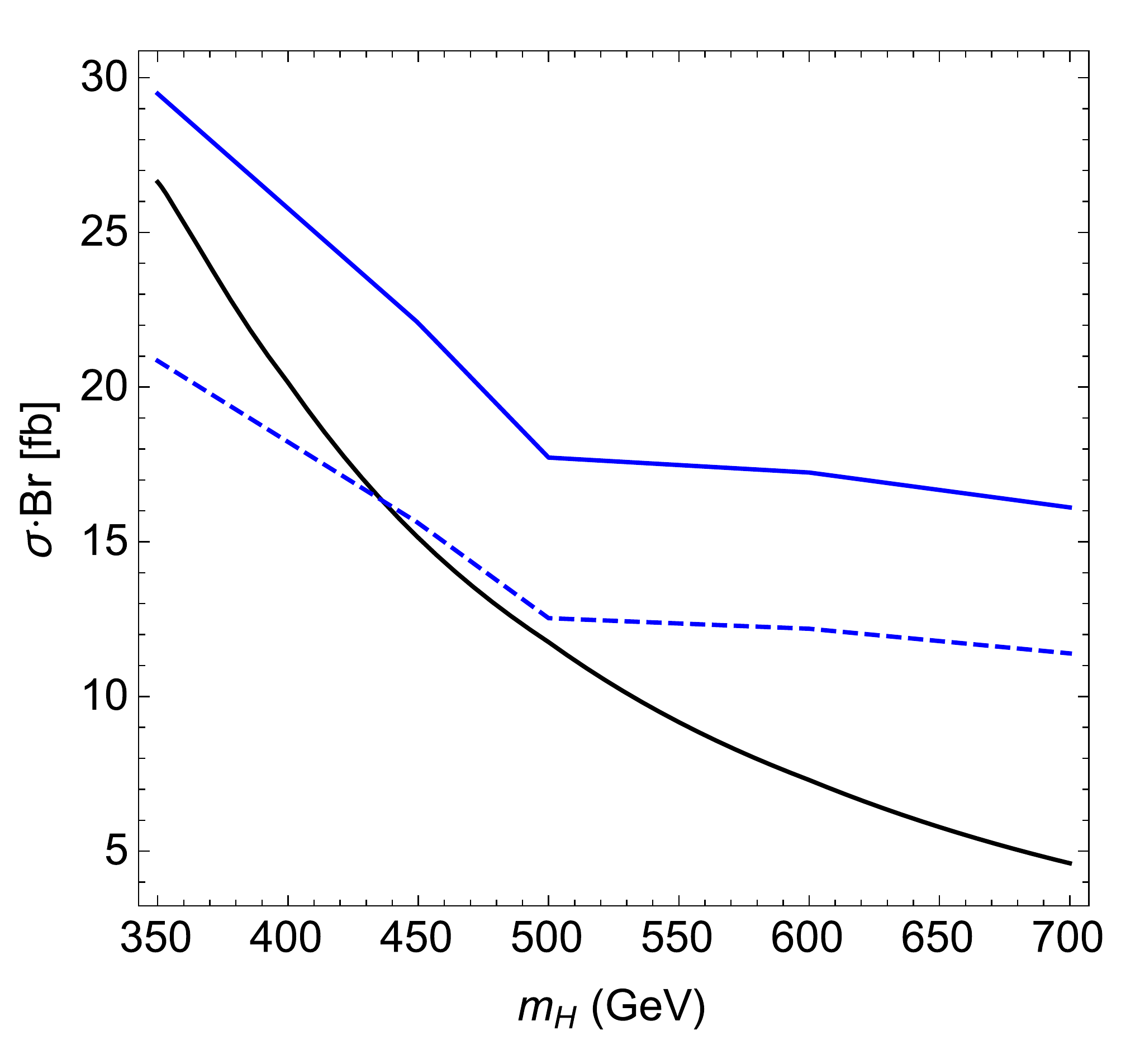} $\quad$ 
\includegraphics[height=0.45\textwidth]{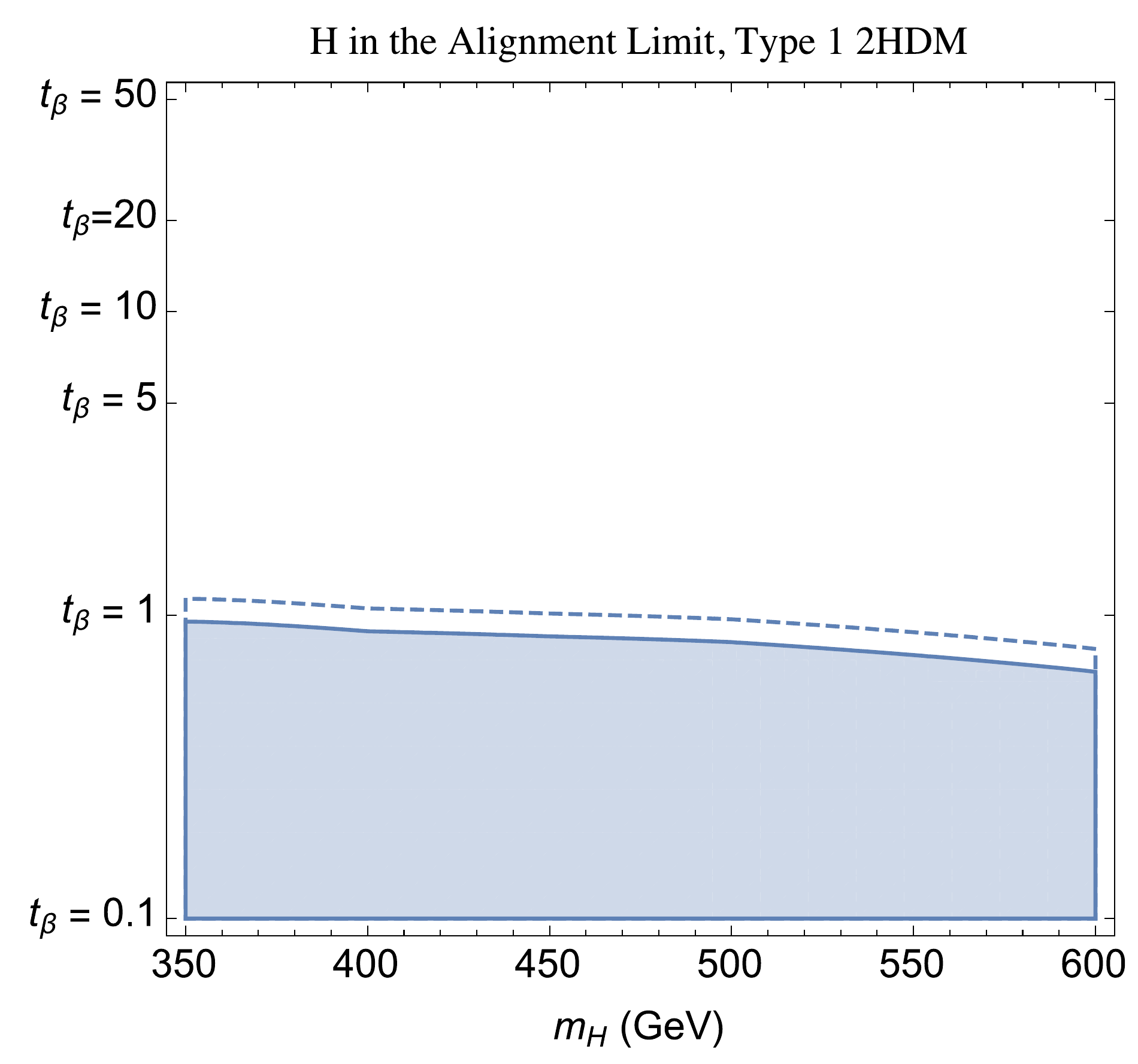}
\end{center}
\caption{Left: Projection for the excluded cross section times branching 
ratio for $pp \to t \bar t H \to t \bar t t \bar t$ at $\sqrt{s} = 14$ TeV, 3000 
fb$^{-1}$ as a function of $m_H$. The projection is obtained from the naive scaling of the combination 
of $3 \ell + 2 b$ channels (solid blue line) and with a further $\sqrt{2}$ 
improvement from the addition of SSDL channels (dashed blue line), and is 
compared to the signal cross section $\sigma(pp \to t \bar t H)$ at 
$\tan \beta = 1$ (solid black line). Right: Projected exclusion for a CP-even 
neutral scalar $H$ in the alignment limit of a Type 1 2HDM from the 
naive scaling of the combination of $3 \ell + 2 b$ channels (solid blue) 
and with a further $\sqrt{2}$ improvement from the addition of SSDL 
channels (dashed blue).}
\label{fig:ttHprojection}
\end{figure}

At $\sqrt{s} = 14$ TeV the prospects for a search in the 4-top channel 
improve considerably, as the $t \bar t H/A$ associated production cross 
section increases by an order of magnitude for $m_{H/A} \gtrsim 2 m_t$ 
relative to $\sqrt{s} = 8$ TeV. However, reliably estimating sensitivity 
at the level of a theory study is challenging since the largest backgrounds
to SSDL and multi-lepton searches in this final state typically originate 
from $t \bar t$ or $W/Z$ + jets events with an additional fake lepton. 
As such, we estimate the 14 TeV $\sigma\times$Br exclusion reach with 
luminosity and background cross section rescaling of~\cite{CMS-PAS-SUS-13-008}, 
assuming efficiencies remain comparable to 8 TeV. In particular, 
we rescale background cross sections by the ratio of $t \bar t$ cross 
sections at 14 TeV and 8 TeV, since $t \bar t$ + lepton fakes comprise 
the dominant background in signal regions 8, 18, and 28 
of~\cite{CMS-PAS-SUS-13-008}.

The impact of this projected sensitivity at $\sqrt{s} = 14$ TeV is shown 
in figure~\ref{fig:ttHprojection}, both for the naive scaling of the sensitivity 
from the combination of $3 \ell + 2 b$ channels, and with an additional 
factor-of-$\sqrt{2}$ improvement in the $\sigma \cdot Br$ reach to emulate 
the potential improvement from including SSDL channels. Even at 
$\sqrt{s} = 14$ TeV this remains a challenging channel, but offers hope 
for meaningful sensitivity at low $\tan \beta$ for $2 m_t \lesssim m_H 
\lesssim 500$ GeV.

We turn next to $b \bar b$ associated production of $H$ or $A$ followed 
by decay to $t \bar t$. This process may be significant in Type 2 2HDM 
where the $b \bar b H/A$ associated productiion grows with $\tan \beta$. 
However, in this case the partial width $H/A \to t \bar t$ also falls as with 
$\tan \beta$, suggesting the rate for $b \bar b H/A \to b \bar b t \bar t$ will 
peak at moderate values of $\tan \beta$. To estimate the LHC sensitivity 
to $pp \to b \bar b H/A \to b \bar b t \bar t$, we design and simulate a search 
in the semi-leptonic final state at $\sqrt{s} = 14$ TeV. In contrast to the 
multi-lepton search for the four-top final state, the dominant backgrounds 
for a $t \bar t b \bar b$ final state can be reliably simulated with available 
Monte Carlo techniques. 

We generate parton level signal and backgrounds events using 
MadGraph5 \cite{Alwall:2014hca} to leading order with 
CTEQ6L1 PDFs \cite{Pumplin:2002vw}. 
The events are showered with PYTHIA6.4 \cite{Sjostrand:2006za} 
and Delphes3 \cite{deFavereau:2013fsa,Cacciari:2011ma} 
is used to simulate detector effects. Jets are reconstructed using the
anti-$k_T$ algorithm with $R=0.5$ and are required to satisfy
$p_T^j>20{\text{GeV}},~|\eta^j|<4.5$.
Charged leptons (electrons and muons) are required to have
$p_T^\ell>15{\text{GeV}},~|\eta^\ell|<2.5,~I_{iso,\mu}\left(\Delta R=0.3\right)<0.1$.
The $b$-tagging efficiency is chosen to be 70\% 
with a 25\% (2\%) mistagging rate for charm (light) jets \cite{CMS:2013xfa}. 
$b$-tagged jets satisfy $p_T^b>40{\text{GeV}},~|\eta^j|<2.5$.

\begin{figure}
\begin{center}
\includegraphics[height=0.43\textwidth]{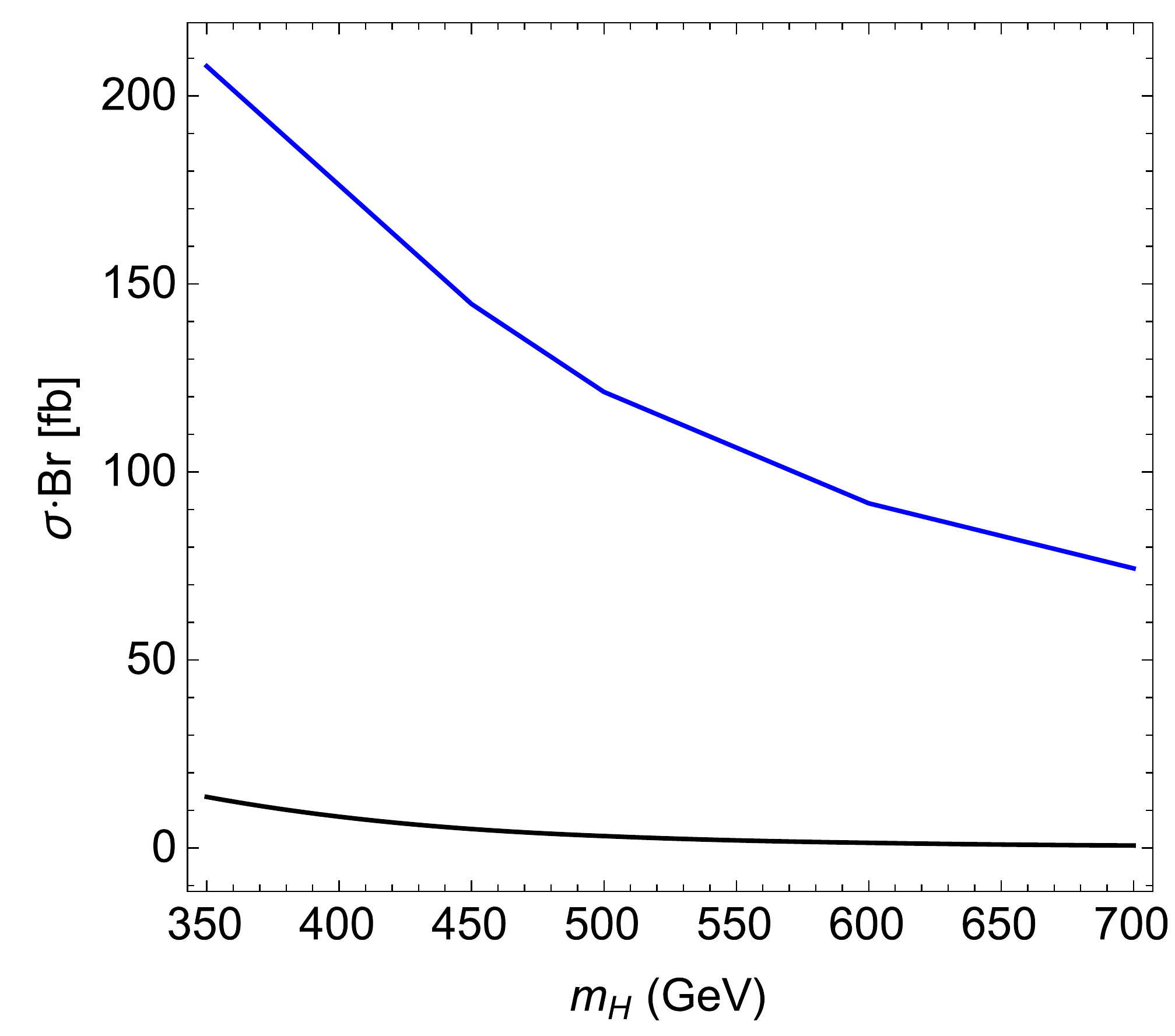} $\quad$ 
\includegraphics[height=0.43\textwidth]{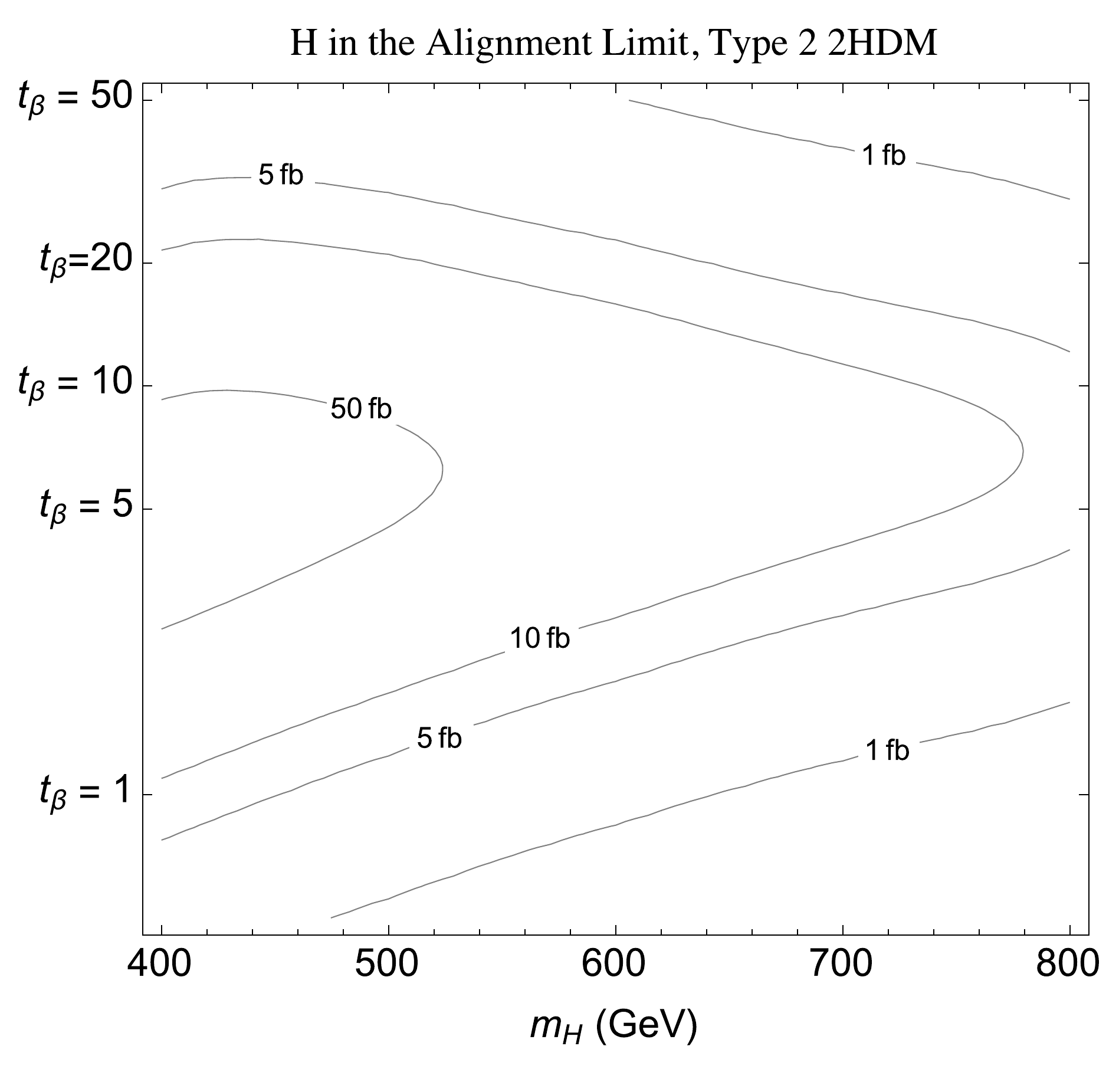}
\end{center}
\caption{Left: Projection for the excluded cross section times branching 
ratio in femtobarns for $pp \to b \bar b H \to b \bar b t \bar t$ at $\sqrt{s} = 14$ TeV, 
3000 fb$^{-1}$ as a function of $m_H$ (solid blue line). For comparison we show the signal 
cross section $\sigma(pp \to b \bar b H)$ at $\tan \beta = 1$ (solid black line). 
Right: Contours of $\sigma \cdot {\rm Br}(pp\to b \bar b H \to b \bar b t \bar t)$ 
in femtobarns for a  CP-even neutral scalar $H$ in the alignment limit of a
Type 2 2HDM.}
\label{fig:bbHprojection}
\end{figure}

In addition to the single lepton requirement, we require at least 6 jets with at least 
4 $b$-tags in the final state to suppress SM backgrounds. We also apply a missing 
transverse energy cut of $\met > 30$ GeV and veto events with more than one 
charged lepton. Top quarks and $W$ bosons are reconstructed from the mass-shell 
constraints, with small corrections for detector effects (for details, see appendix~
\ref{appx:reconstr}). After top quark reconstruction, we require that the signal 
events satisfy $\chi^2<5.0$, where $\chi$ is a variable characterizing the quality 
of the reconstruction (see appendix~\ref{appx:reconstr}). The irreducible background 
is $pp \to t\bar t bb,$ while the dominant reducible backgrounds for this analysis are
$pp \to t\bar t bj$ and $pp\to t\bar t jj$, with light jets faking bottom quarks. The 
backgrounds from $t\bar th$, $t\bar tZ$, single top production and vector boson 
plus multijets are subdominant \cite{CMS:2013vui}.

To set an exclusion limit, we use the likelihood function 
\be \label{eq:likelihood}
L\left(x|n\right)=\prod_{j=1}^{N}\frac{x_j^{n_j}}{n_j!}e^{-x_j},
\ee
where $x_j$ is the binned $m_{t\bar t}$ distribution predicted by the 
model (with or without signal) and $n_j$ is the observed 
distribution. The $2\sigma$ exclusion bound is obtained~\cite{Cowan:2010js} from
\be \label{eq:twosig}
\sqrt{-2\ln\left(\frac{L\left(\mu s+b|b\right)}{L\left(b|b\right)}\right)}=2.
\ee 
The results are shown in figure~\ref{fig:bbHprojection}. The 
excluded cross section ranges from 200-70 fb for $350 \, {\rm GeV} 
\lesssim m_H \lesssim 700$ GeV, while $\sigma(pp \to b \bar b H) \lesssim 
10$ fb for $\tan \beta = 1$ across the same range. Although the production 
cross section grows with $\tan \beta$, the branching ratio to $t \bar t$ falls, 
so that the peak rate $\sigma(pp \to b \bar b H \to b \bar b t \bar t) \sim 50$ 
fb is obtained around $\tan \beta \sim 5$.  Based on the results of our 
preliminary simulation, a meaningful limit cannot be set in Type 2 2HDM. 
However, given that sensitivity is of the same order as the peak rate for 
$350 \, {\rm GeV} \lesssim m_H \lesssim 500$ GeV, this channel deserves 
further experimental study at 14 TeV.

Finally, we note a third associated production channel that may prove 
useful in the hunt for $H/A \to t \bar t$, although we do not study it in 
detail here. The rate for electroweak production of the single-top $t(q)H/A$ 
final state via a $t$-channel $W$ boson exceeds that of $t \bar t H/A$ 
production around $m_{H/A} \simeq 340$ GeV \cite{Maltoni:2001hu} 
and has the same parametric scaling as a function of $\alpha$ and $\beta$. 
Consequently, this suggests the rate for $pp \to t (q) H/A \to t(q)t \bar t$ 
exceeds that of $pp \to t \bar t H/A \to t\bar t t \bar t$ in the entire region of 
interest for $H/A \to t \bar t$. The resulting three-top final state is particularly 
amenable to a search for same-sign dileptons with two or more $b$-tagged 
jets, and may provide a complementary probe of $H/A \to t \bar t$ at 
$\sqrt{s} = 14$ TeV.\footnote{For a recent study of single top production 
in association with the Higgs state at 125 GeV, 
see~\cite{Demartin:2015uha}.}

\section{Searching for an Invisible Neutral Higgs}
\label{sec:invisible}

In addition to decays of new scalars to SM fermions, it is also interesting to consider invisible decays, which have been actively studied for the SM Higgs boson following Run 1 (with current upper limits around 30\% coming from the VBF channel~\cite{ATLAS-CONF-2015-004}). As with the SM Higgs, any observation of an invisible width for a second Higgs state would provide a window into new physics, possibly signaling the first laboratory production of dark matter. Invisible decays could also provide a discovery mode for new Higgses, since SM backgrounds can be strongly suppressed with a large missing energy cut. In this section, we continue the study of new scalars in channels involving tops and bottoms, adding the ingredient of large missing energy. 

In section~\ref{mettheory} we briefly discuss some of the theoretical motivation for searching for invisibly decaying new scalars in association with tops and bottoms. There are many possible UV completions. For simplicity we focus on one amusing example, the MSSM Higgs sector benchmark points often used in reporting limits on $H/A\rightarrow\tau\tau$. Due to choice of neutralino mass parameters in these benchmark points, the new neutral scalars associated with the second Higgs doublet possess branching fractions into pairs of the lightest $R$-odd particle.  With some variation of the parameters, the invisible branchings can become substantial.

In section~\ref{bbmet}, we place a limit on $b\bar b H\rightarrow b\bar b+\met$ by reinterpreting an 8 TeV sbottom search~\cite{Aad:2013ija}. We give the corresponding limit in the parameter space of Type 2 2HDM, where the $b\bar b H$ coupling is $\tan\beta$-enhanced, and argue that it is likely to be a stronger limit on this parameter space than one derived from monojet searches. Previously, in a study focused on dark matter simplified models, Ref.~\cite{Buckley:2014fba} obtained a limit on the $b\bar b$-associated invisible scalar channel by reinterpreting an ATLAS effective operator study at 8 TeV~\cite{Aad:2014vea}. We have checked that the sbottom and effective operator cut flows provide very similar reach, and present our results on natural parameter spaces for new scalar searches.

In section~\ref{ttmet} we study the reach of semileptonic $t\bar t H\rightarrow t\bar t+\met$ at 14 TeV.
Previously, Ref.~\cite{Zhou:2014dba} obtained a limit on this channel at 8 TeV by reinterpreting a CMS stop search~\cite{Chatrchyan:2013xna}. Ref.~\cite{Haisch:2015ioa} also performed an 8 TeV analysis, and furthermore estimated a 14 TeV limit by a parton-level reinterpretation of an ATLAS stop study~\cite{ATL-PHYS-PUB-2013-011}.  We complete the phenomenological analysis of semileptonic $t\bar t H\rightarrow t\bar t+\met$, performing a full 14 TeV analysis with optimized cuts and detector simulation. We also argue that this channel is likely to be competitive with monojet searches for new invisibly-decaying scalars with masses below $2m_t$.

\subsection{Models with $H/A\rightarrow \met$}
\label{mettheory}

In the most model-independent spirit of simplified models, any search for an invisibly decaying new scalar is of interest: the topologies are simple and capitalize on the small SM backgrounds. The production channels studied here are motivated by the alignment limits of new weakly-coupled scalar models, which preserve the SM-like properties of the light observed Higgs boson at the expense of suppressing traditional invisible scalar searches involving gauge couplings such as $ZH\rightarrow \ell\ell+\met$ and $qqH\rightarrow qq+\met$. It is not difficult to add additional theoretical structure, such as dark matter candidates or particles that are long-lived on detector timescales, into which new scalar states may decay invisibly with substantial branching fraction.

Searches in the $t\bar t +\met$ channel are most effective in models where new scalars couple to $t \bar t$ with coupling $y'_t\sim y_t^{\rm SM}$. In the alignment limits of Type 1 and Type 2 2HDM, $y'_t/y_t^{\rm SM}=\cos\beta$ and $y'_t/y_t^{\rm SM}=\cot\beta$, respectively, so the LHC reach is strongest when $\tan\beta\sim\mathcal{O}(1)$. A simple model of invisible decays can be obtained by coupling the second doublet $\Phi_2$ to a new massive singlet scalar through a portal-type coupling:
\begin{align}
V\supset \kappa|\Phi_2|^2S^2+\frac{1}{2}\mu_s^2 S^2+\dots
\end{align}
In the alignment limit, the dominant decays of the neutral components $H,A$ of $\Phi_2$ will be into Standard Model fermions and $S$ pairs. Let us assume that the singlet mass is small and that $m_H\lesssim 2m_t$, so that the dominant SM decays are into $b\bar b$. Then the ratio of invisible to visible partial widths is approximately:
\begin{align}
\frac{\Gamma_{SS}}{\Gamma_{b\bar b}}\simeq \frac{\kappa^2 v^2\cos^2\beta}{3m_H^2 {y'_b}^2}\;.
\end{align}
In Type 1 2HDM, $y'_b=y_b^{\rm SM}\cos\beta $ in the alignment limit, so this ratio is $\mathcal{O}(1)$ already for $\kappa\sim 0.1$ in the range of $m_H$ considered. In Type 2 2HDM, $y'_b$ is $\tan\beta-$enhanced in the alignment limit, but for $\tan\beta\sim\mathcal{O}(1)$ the invisible decay is still substantial for small $\kappa$. Simple models of this type are most effectively probed by the $t\bar t +\met$ channel.

Since $y_b^{\rm SM}$ is small, searches for $b\bar b +\met$ are most effective in cases where the new scalar has enhanced coupling $y'_b$ to $b\bar b$.  The most well-known example is the $\tan\beta$ enhancement of the Type 2 2HDM. In the toy model above, the ratio of partial widths is suppressed by $(\tan\beta)^{-4}$, so $\kappa$ must be large in order to obtain a substantial invisible width. Another Type 2 2HDM with the potential for invisible decays is the MSSM. In fact, traditional benchmark scenarios used to study $H/A\rightarrow\tau\tau$ actually have regions of parameter space where $H/A\rightarrow{\rm inv}$ occurs at non-negligible rates~\cite{Carena:2013qia}.

Supersymmetric Higgs bosons couple to neutralinos in the form Higgs - Higgsino - Electroweakino. Therefore, for $\mu\sim M_i\ll m_A$ ($i=1$ or $2$), the heavy neutral Higgs states can decay invisibly into the lightest neutralino through its gaugino/higgsino mixings. The invisible branching ratios for $H$ are maximized for $\tan\beta\sim 5$; at higher values $H\rightarrow b\bar b$ dominates, while at lower values we deviate substantially from the alignment limit and $H\rightarrow VV$ becomes important. In figure~\ref{fig:ttHMT2Whist} we show the $H/A\rightarrow\rm{ inv}$ branching ratios in an MSSM benchmark point with $\tan\beta=5$, $M_2=300$ GeV, $M_1=({5 s_w^2}/{3c_w^2})M_2=143$ GeV, and decoupled gluino and scalars. Sizable invisible branching fractions are possible for both states when the LSP is well-mixed and the $t\bar t$ channel is kinematically forbidden.

We do not pursue model building further in this work, but for completeness we note that LUX limits~\cite{Akerib:2013tjd} rule out most of the interesting parameter space if the neutralino is dark matter and $\mu>0$. If it is only a subcomponent of dark matter, or if it is stable on detector timescales but decays outside of the detector (say, through RPV couplings, or to a gravitino), then the direct detection limits do not apply. Another intriguing possibility is that $\mu<0$. Ref.~\cite{Huang:2014xua} observed that for $|\mu|\sim m_{LSP}$, and $\tan\beta\sim$few, there is a blind spot on the $m_H$ axis where the tree-level direct detection amplitudes from SM Higgs and MSSM Higgs exchange cancel with each other.

\begin{figure}[t]
\begin{center}
\includegraphics[width=2.935in]{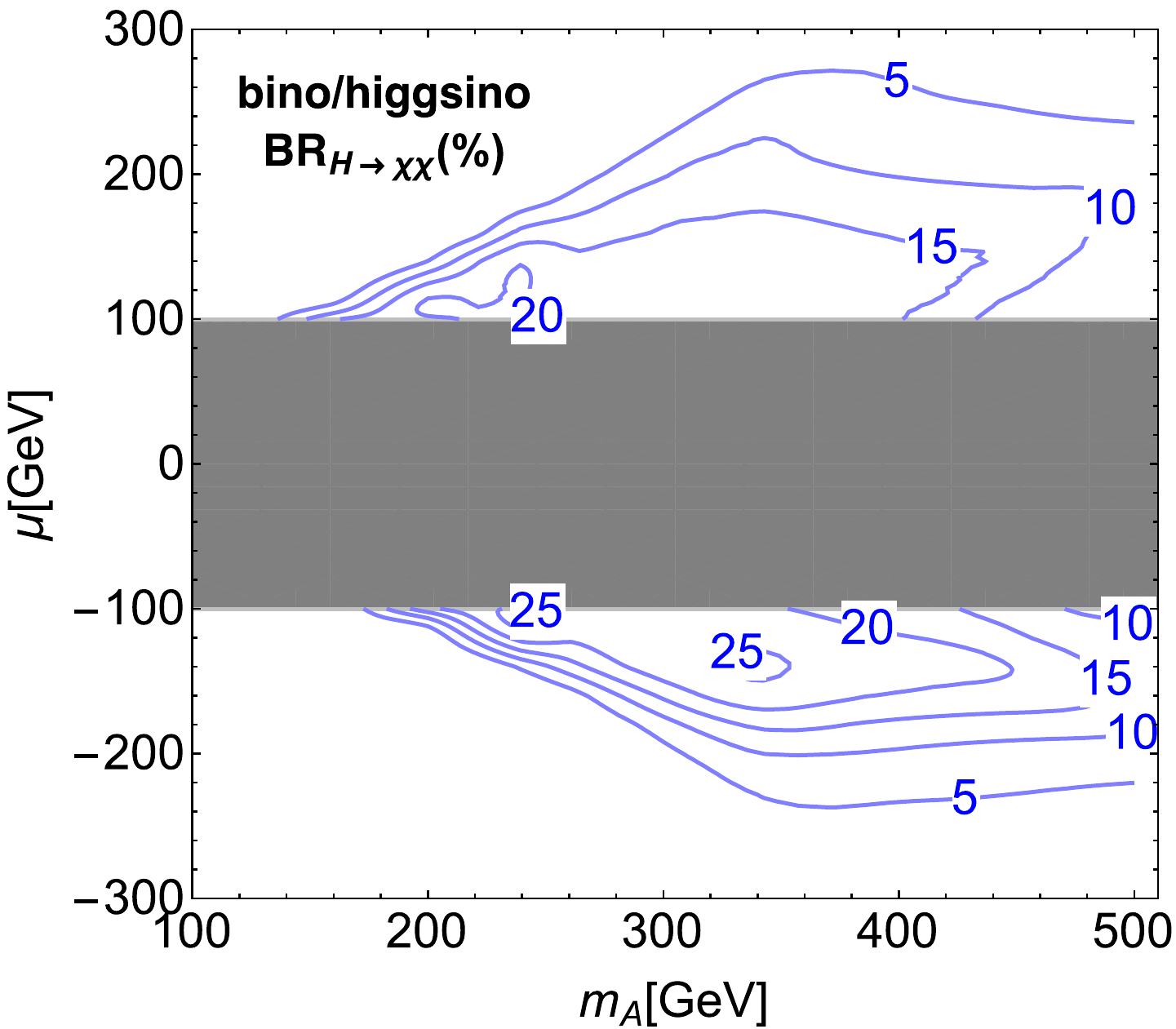}$\quad$
\includegraphics[width=2.935in]{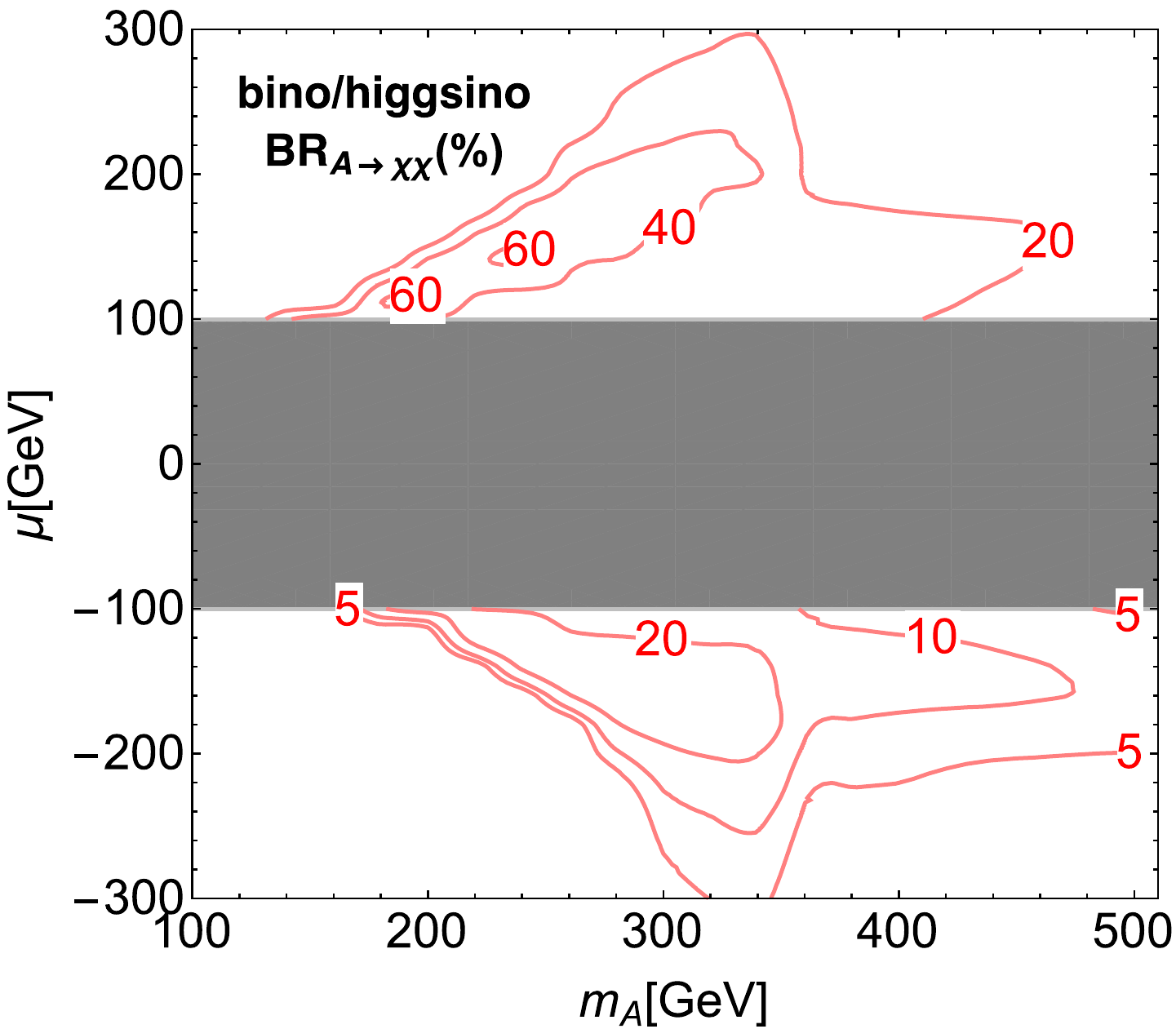}
\end{center}
\caption{Contours of the $H/A\rightarrow\rm{ inv}$ branching ratios in an MSSM benchmark point with $\tan\beta=5$, $M_2=300$ GeV, $M_1=({5 s_w^2}/{3c_w^2})M_2=143$ GeV, and decoupled gluino and scalars. The gray region denotes the LEP exclusion on light charginos.}
\label{fig:ttHMT2Whist}
\end{figure}

\subsection{$b \bar b+ \met$ at 8 TeV}
\label{bbmet}

\begin{figure}[t]
\begin{center}
\includegraphics[scale=0.5]{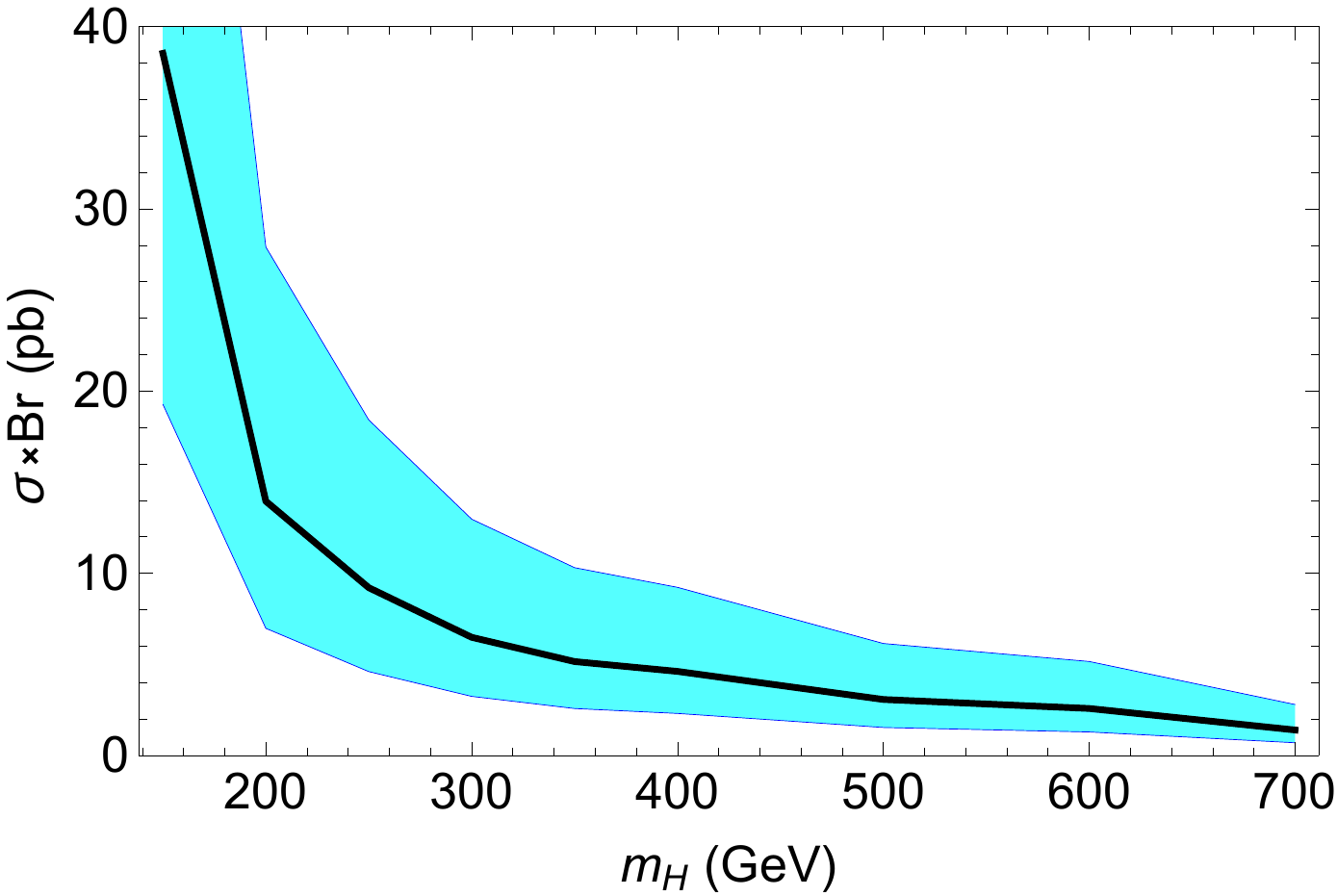} $\quad$ 
\includegraphics[scale=0.5]{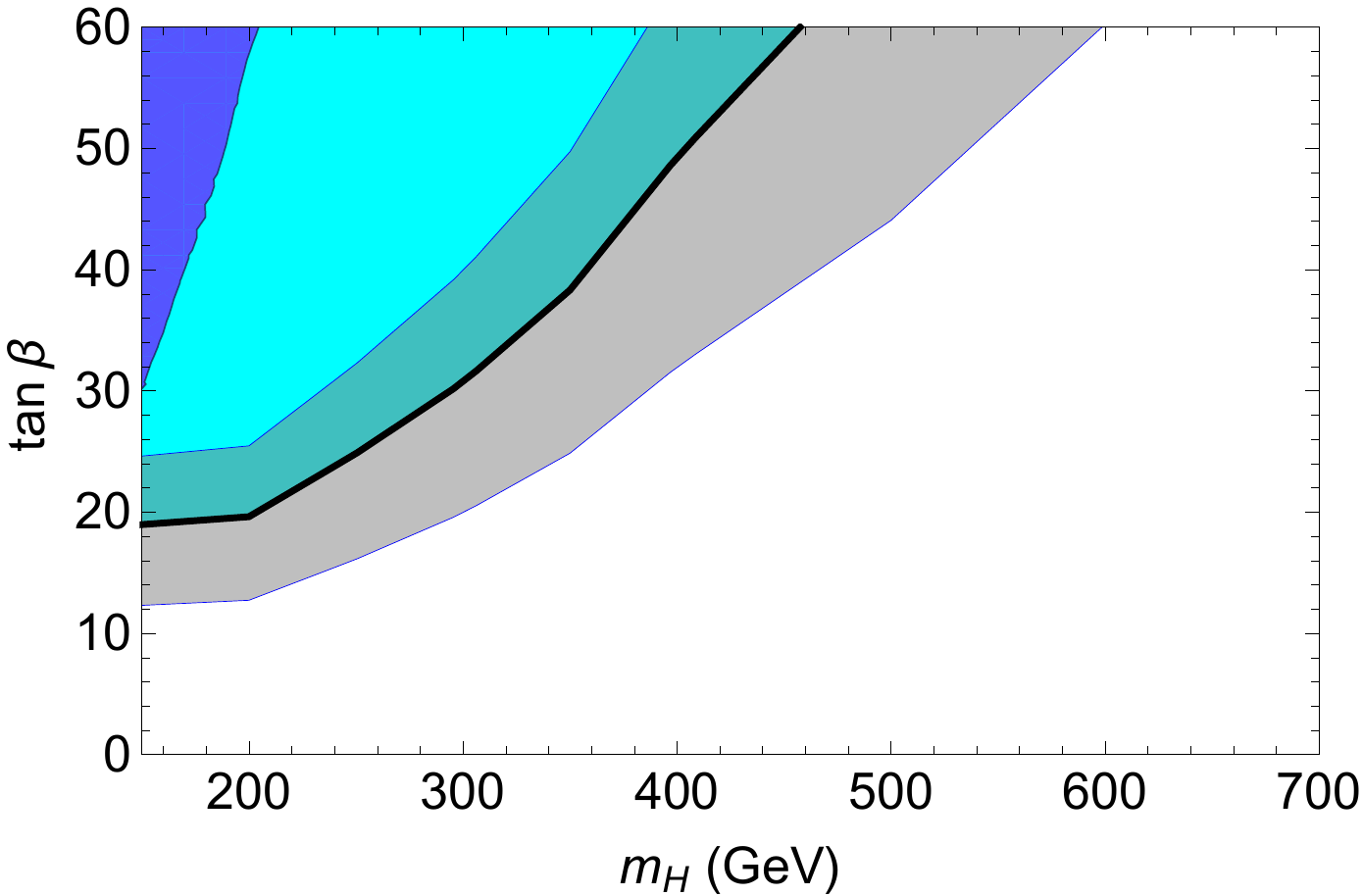}
\end{center}
\caption{Left: the 8 TeV $\sigma\times$Br limit on the production of invisibly decaying new scalars in association with bottom quarks. Right: the corresponding limit on the Type 2 2HDM plane. The lighter blue region in the upper left is the $b\bar b +\met$ exclusion and the darker blue region is the estimated monojet exclusion, rescaling the results of ~\cite{Harris:2014hga}.}
\label{fig:bbHMET}
\end{figure}
In this section, we reinterpret the ATLAS sbottom search ($pp\rightarrow b\bar b +\met$) performed in Ref.~\cite{Aad:2013ija} into a limit on $pp\rightarrow  H b\bar b$, where $H$ is a new heavy neutral CP-even scalar that decays invisibly. We unfold the signal efficiency factors from the cross section limits $\sigma_{\rm vis}$ given in Table 7 of~\cite{Aad:2013ija}. We focus on Signal Region A (SRA), defined by the cut flow of Table 1 of~\cite{Aad:2013ija}, because in that region the leading two jets are b-tagged. SRA is divided into five subregions based on the contransverse mass cut~\cite{Tovey:2008ui} (with ISR correction given in~\cite{Polesello:2009rn}), and the ATLAS observed limits on the cross sections in SRA range from $0.26-1.9$~fb depending on the $m_{CT}$ cut.

For signal generation and detector simulation we use Madgraph/Pythia/Delphes matched to one jet. 
We optimize the unfolded $\sigma_{pp\rightarrow b\bar b H}\times {\rm Br}(H\rightarrow{\rm inv})$ over the $m_{CT}$ cut, finding that the softest cut used in the sbottom search ($m_{CT}>150$ GeV) gives the best limit on $bbH$ until $m_H\gtrsim600$ GeV. We attempt to validate our analysis by reproducing the sbottom signal efficiencies given in the ancillary data of~\cite{Aad:2013ija}. We find that near the exclusion limit our analysis yields signal efficiencies approximately a factor of 2 better than those found by ATLAS. Therefore, to be conservative we assign a factor of $(1/2,2)$ uncertainty band to our $bbH$ limit.

Our results for the $\sigma\times{\rm Br}$ limit are given in the left-hand panel of figure~\ref{fig:bbHMET}.  The cross sections are relatively large, suggesting that this search mode is most effective in constraining new scalar models where the scalar coupling to bottom quarks is enhanced over that of the SM Higgs.  Such enhancements occur, for example, in the alignment limit of Type 2 2HDM, where at large $\tan\beta$ the $Hbb$ and $Abb$ couplings are a factor of $\tan\beta$ larger than the SM Higgs $hbb$ coupling. In the right-hand panel of figure~\ref{fig:bbHMET}, we reinterpret the $\sigma\times{\rm Br}$ limit on the parameter space of Type 2 models with Br$(H\rightarrow{\rm inv})=1$. 

We note that a monojet signal arises in this scenario by closing the bottom loop and radiating an additional jet. In the case of $t\bar t+\met$, the monojet signature obtained in this way is competitive and can outperform reinterpreted stop searches~\cite{Haisch:2015ioa}. However, in the $b\bar b+\met$ case, we expect that monojet is less powerful for two reasons. First, unlike the $t\bar t+\met$ case, the bottom quarks in the final state do not suffer a large phase-space suppression. Second, closing the loop costs a factor of $m_f$ in the amplitude, which is a large suppression in the $b\bar b$ case, even if the coupling to bottom quarks is $\tan\beta$-enhanced. (Moreover, in some models, like Type 2 2HDM, the new scalar coupling to top quarks is $\tan\beta$-suppressed.) In the dark blue region of the right-hand panel of figure~\ref{fig:bbHMET} we estimate the monojet exclusion by rescaling the results of~\cite{Harris:2014hga} into the Type 2 plane, and we see that it is much weaker than $b\bar b + \met$. Our rescaling uses the ratio of the LO $\Gamma(H\rightarrow gg)$ loop functions in the 2HDM relative to the SM, and is therefore rather crude. However, it is likely to be conservative, since the $p_T$ spectrum of the extra jet in the bottom-dominated process is harder than that of the jet in the top-dominated case~\cite{Field:2003yy}, suggesting that the actual monojet limits will be weaker.

\subsection{$t \bar t+ \met$ at 14 TeV}
\label{ttmet}

We now turn to invisible Higgs states produced in association with top quarks, where limits at 8 TeV have been obtained~\cite{Zhou:2014dba,Haisch:2015ioa} by reinterpretation of a CMS stop search~\cite{Chatrchyan:2013xna}, and perform an optimized projection for the reach of the 14 TeV LHC. We simulate signal and background at leading order in Madgraph/Pythia/Delphes, and apply cuts requiring large missing energy, four jets including two $b$-tags, and a veto on $n_{lep}\neq 1$. The dominant SM background processes are semileptonic and dileptonic $t \bar t$, $Zt\bar t\rightarrow\nu\nu\ell\nu jj b\bar b$, and $Wjjb\bar b$. 

\begin{figure}[t]
\begin{center}
\includegraphics[scale=0.68]{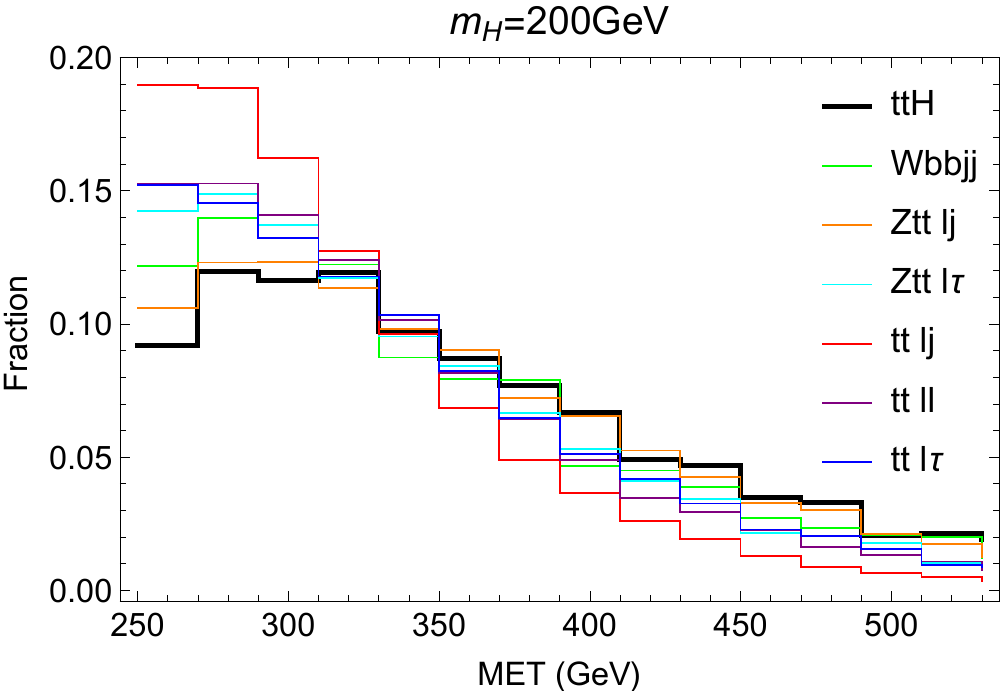}
\includegraphics[scale=0.68]{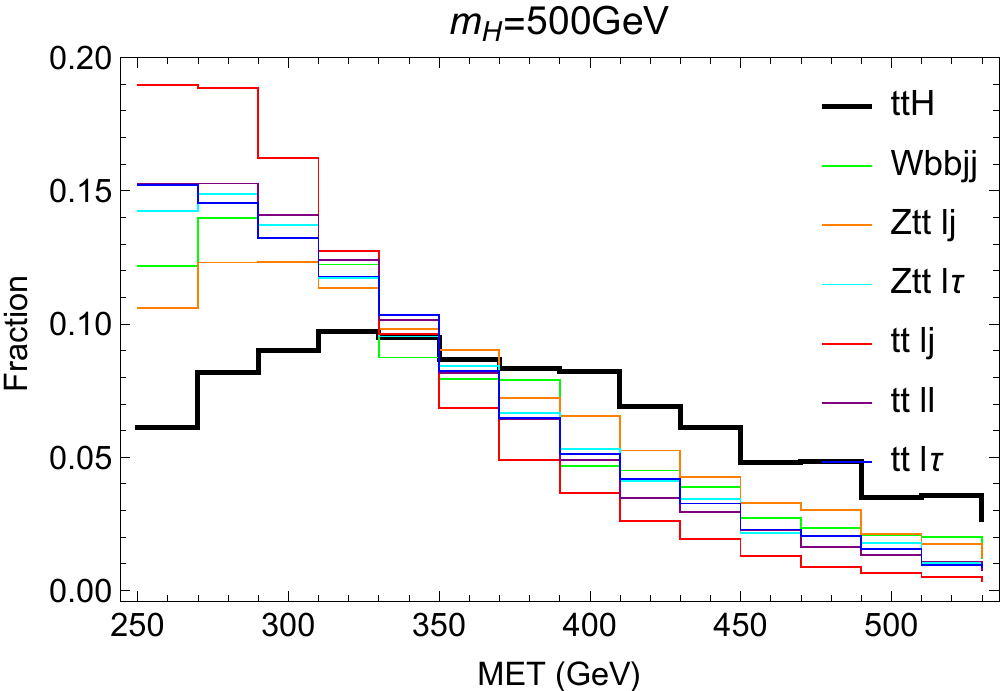}
\end{center}
\caption{The $\met$ distributions of the signal and backgrounds after the jet selection cuts described in the text. Left: $m_H=200$ GeV. Right: $m_H=500$ GeV.}
\label{fig:ttHMEThist}
\end{figure}

\begin{figure}[t]
\begin{center}
\includegraphics[scale=0.68]{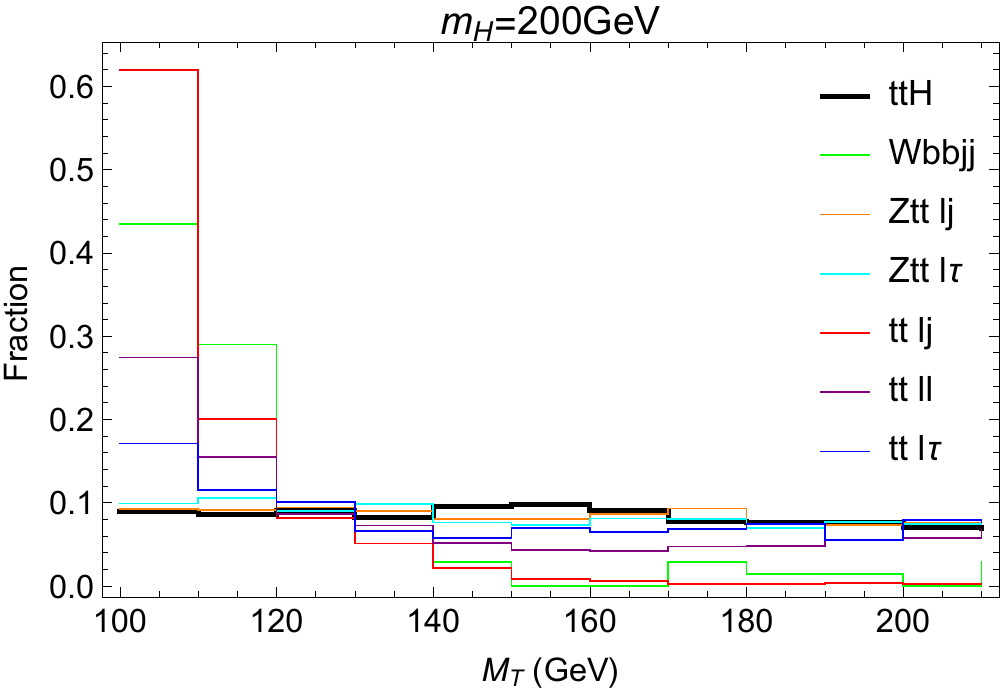}
\includegraphics[scale=0.68]{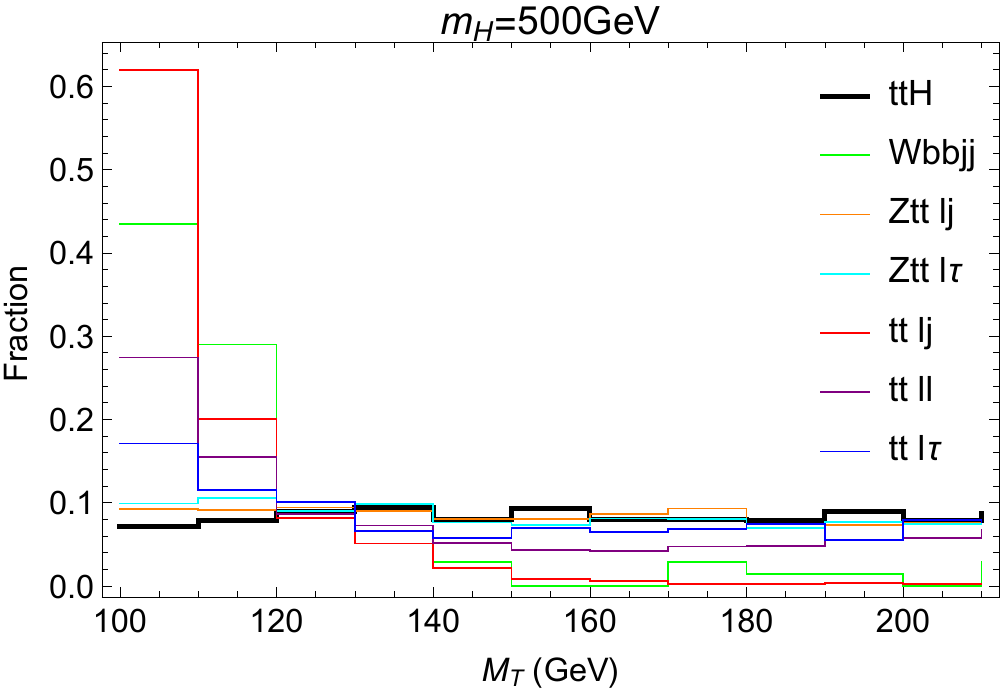}
\end{center}
\caption{The $m_T$ distributions of the signal and backgrounds after the jet selection cuts described in the text. Left: $m_H=200$ GeV. Right: $m_H=500$ GeV.}
\label{fig:ttHMThist}
\end{figure}

For our preselection and jet selection cuts, we require $\met>250$ GeV, 1 lepton, at least four central jets with $p_T>\{130,50,50,30\}$, respectively, two $b$-tags, and $\Delta\phi(j,\met)>0.8$ for the two hardest jets. In stop searches, hard cuts on the transverse mass $m_T$ and the variable $m^W_{T_2}$~\cite{Bai:2012gs} may be used to suppress semileptonic and dileptonic $t\bar t$, respectively~\cite{Chatrchyan:2013xna}. The distributions of $\met$, $m_T$, and $m^W_{T_2}$ after the jet selection cuts are given in Figs.~\ref{fig:ttHMEThist},~\ref{fig:ttHMThist}, and~\ref{fig:ttHMT2Whist} for backgrounds and signal for two values of the scalar mass. Subsequently we optimize cuts on $\met$, $m_T$,  and $m^W_{T_2}$, choosing $\met>300$ GeV, $m_T>140$ GeV, and $m^W_{T_2}>200$ GeV. After all cuts, $Zt\bar t$ is the dominant background. Subsequently we validate the $Zt\bar t$ backgrounds against the stop search study in Ref.~\cite{Chatrchyan:2013xna}.

\begin{figure}
\begin{center}
\includegraphics[scale=0.5]{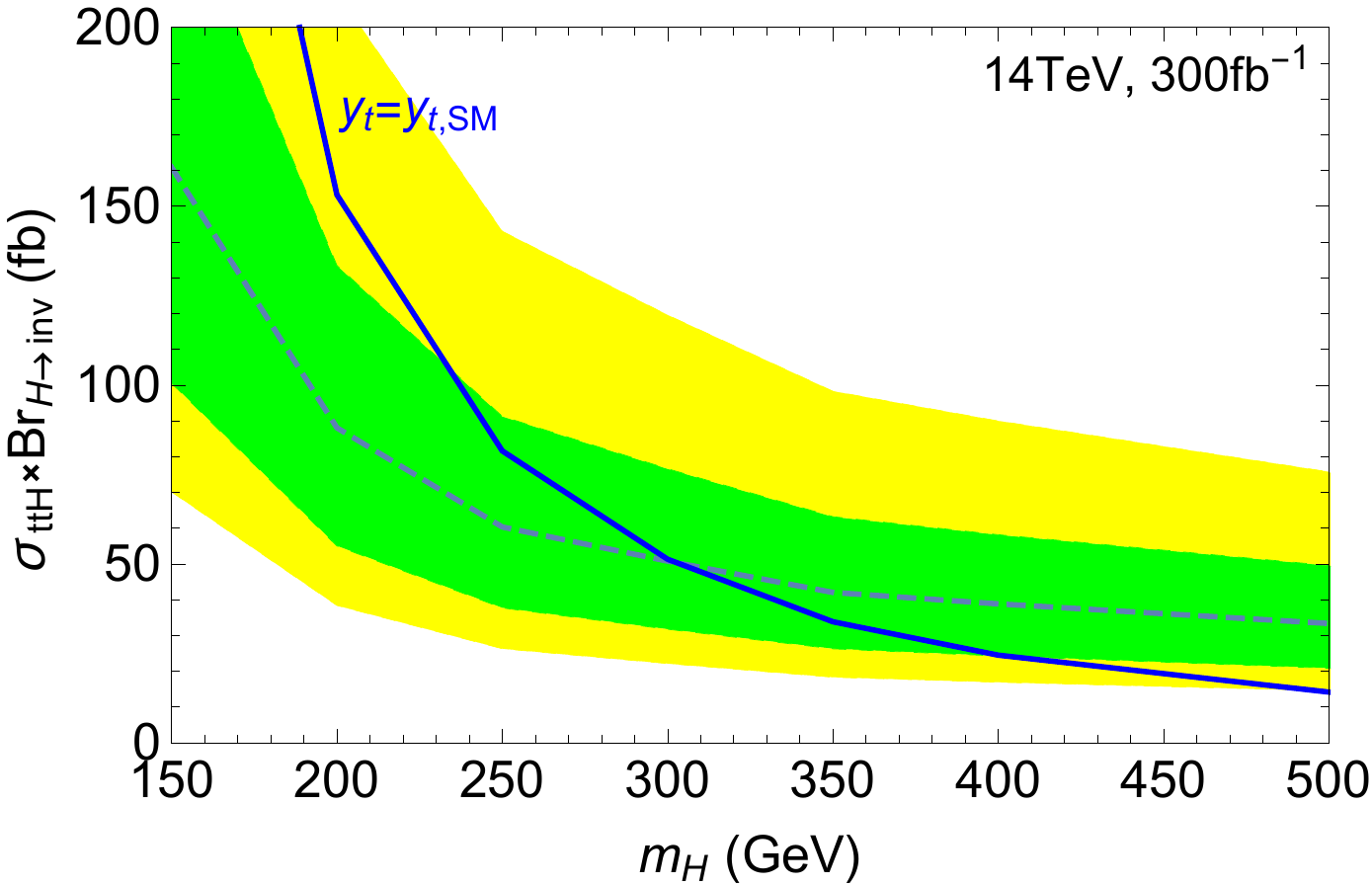}$\quad$ 
\includegraphics[scale=0.68]{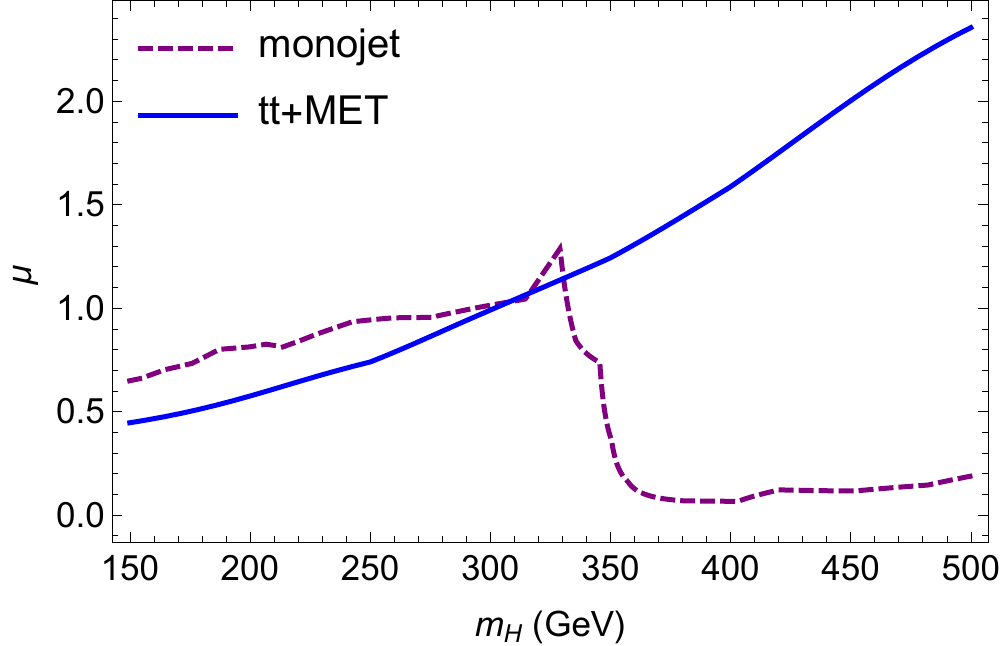}
\end{center}
\caption{Left: the 14 TeV, 300 fb$^{-1}$ expected limit on the $\sigma\times$Br of invisibly decaying new scalars in association with top quarks. Right: comparison with the expected limit from monojet exclusion~\cite{Harris:2014hga}, both normalized to the respective SM cross sections.}
\label{fig:ttHMET}
\end{figure}

Our projected limit ($CL_s=0.05$) with 300fb$^{-1}$ is given in the left-hand panel of figure~\ref{fig:ttHMET}. It is also interesting to compare with the monojet reach, which we take from the scalar analysis performed in~\cite{Harris:2014hga}. Unlike in the $b\bar b+\met$ case, the $t\bar t+\met$ search is most effective when the new states have SM-like couplings to top quarks. Therefore, rather than comparing the reaches in the parameter space of a Type 2 2HDM, we compare the reaches relative to a SM-like fiducial model with a new scalar decaying invisibly with unit branching ratio. (Of course, above the $t\bar t$ threshold, such a large invisible branching ratio would require a very large coupling to the invisible states; the fiducial model is meant only for the comparison of the two search modes.) The right-hand panel of figure~\ref{fig:ttHMET} shows that the associated production mode is expected to be competitive with monojet limit until around the $t\bar t$ threshold, where the threshold generates a bump in the SM monojet cross section. 

\begin{figure}[t]
\begin{center}
\includegraphics[scale=0.68]{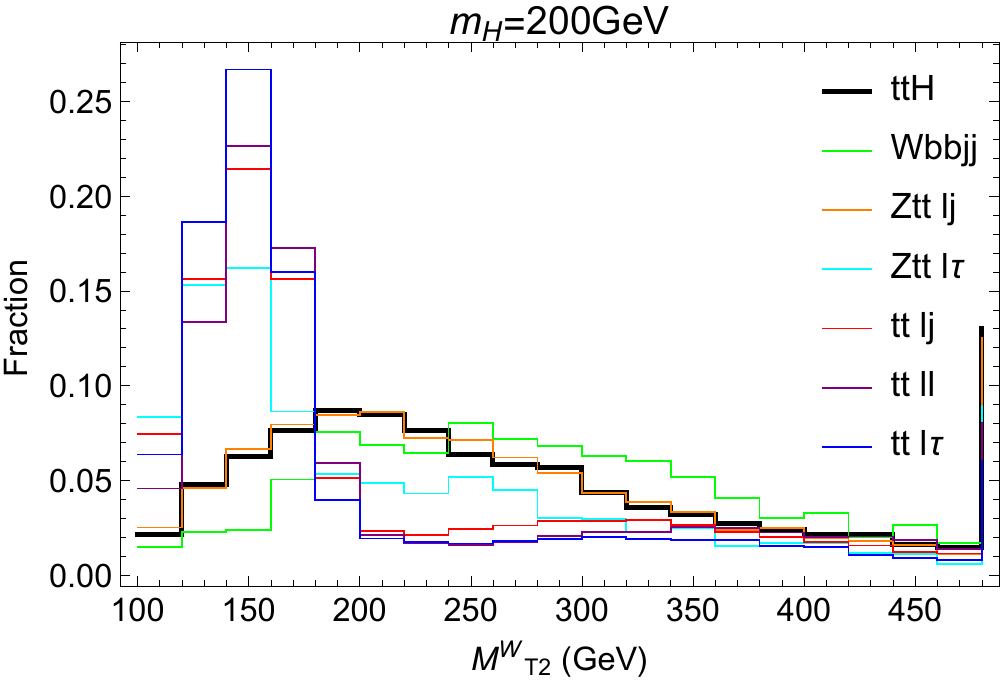}
\includegraphics[scale=0.68]{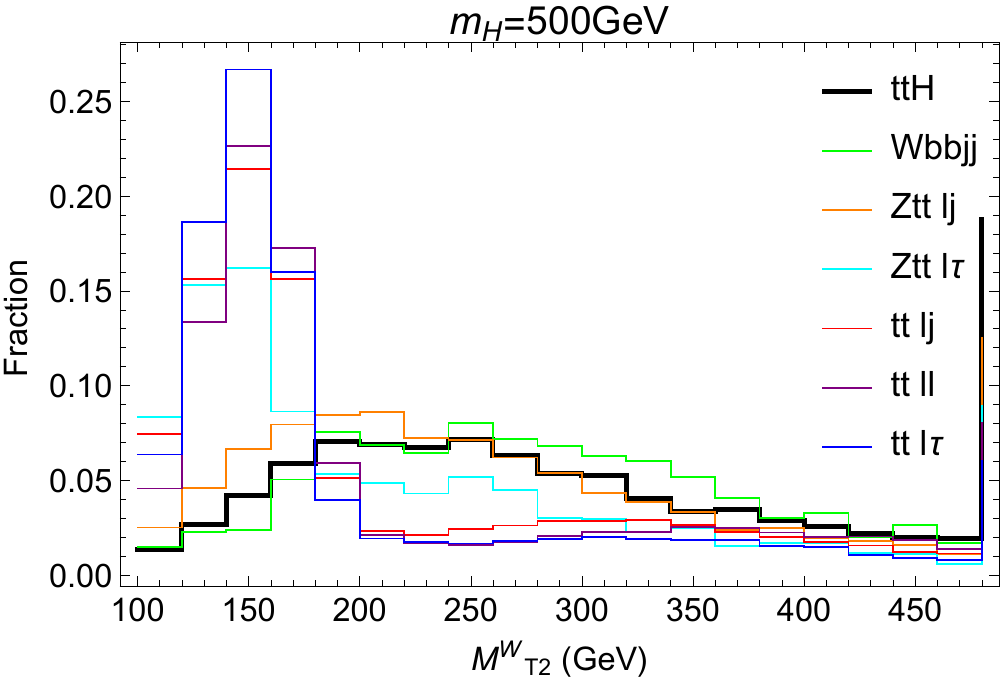}
\end{center}
\caption{The $m^W_{T_2}$ distributions of the signal and backgrounds after the jet selection cuts described in the text. Left: $m_H=200$ GeV. Right: $m_H=500$ GeV.}
\label{fig:ttHMT2Whist}
\end{figure}

\section{Searching for a Charged Higgs in $t \bar b$}
\label{sec:charged}

Thus far we have focused largely on the signatures of the vacuum states $H$ and $A$ near the alignment limit. In this section, we turn to the signatures of charged Higgs bosons $H^\pm$ in the alignment limit, which may provide an alternative handle on 2HDM in regions of parameter space where $H$ and $A$ are hard to find. In particular, we analyze the LHC reach for a charged scalar $H^\pm$ that couples to the SM top and bottom quark through a Yukawa 
interaction of the form
\be
\mathcal{L}_{\text{eff}}=y_{tb} H^+\bar t\left(P_L\sin\theta+P_R\cos\theta\right)b
+{\text{h.c.}}
\label{eq:interaction}
\ee
Near the alignment limit, $\bar{t} b$ associated production with decay to $t \bar{b}$ is the dominant channel for single production of a charged Higgs boson, as the $Wh$ mode vanishes in the alignment limit and decays into $t \bar{b}$ swamp those into $\tau \nu$. As such, we focus on the process $pp\to H^{+}\bar{t}b (H^{-}t\bar{b})+X$ with $H^+\rightarrow t\bar{b}(H^-\to\bar tb)$, and we employ the semi-leptonic decay of the $t\bar{t}$ pair.\footnote{
Early investigation of this channel at the LHC 
can be found in Ref. \cite{Assamagan:2004tt}.}
The CMS collaboration recently published a search for the charged 
Higgs at 8 TeV via the same production channel,  
using the dileptonic decay mode of the top pair \cite{CMS:2014pea}; our aim is to forecast sensitivity at $\sqrt{s} = 14$ TeV and demonstrate the added sensitivity available in a semi-leptonic search. Our results are insensitive to the value of $\theta$ in Eq.~(\ref{eq:interaction}), so for the results shown here we set $\theta=0$ \footnote{
The possibility of using this channel to investigate the chirality structure of the 
$H^+\bar tb$ vertex has been studied in \cite{Gong:2012ru,Cao:2013ud,Gong:2014tqa,
Xiao-ping:2014npa}.
}.

To suppress 
SM backgrounds, we require at least 4 $b$-tagged jets in the final state. The 
the irreducible background is
\bea
pp&\to& t\bar t bb,
\eea
while the dominant reducible backgrounds are
\bea
pp&\to&t\bar t bj ,\\
pp&\to&t\bar t jj ,
\eea
with light jets faking bottom quarks. The $t\bar th$, $t\bar tZ$, single top 
production and vector boson with multijets 
backgrounds are comparatively
negligible \cite{CMS:2013vui}.

We generate parton level signal and backgrounds events using 
MadGraph5 \cite{Alwall:2014hca} to leading order with 
CTEQ6L1 pdfs \cite{Pumplin:2002vw}. 
The events are showered with PYTHIA6.4 \cite{Sjostrand:2006za} 
and Delphes3 \cite{deFavereau:2013fsa,Cacciari:2011ma} 
is used to simulate 
the detector.\footnote{Jets are reconstructed using the
anti-$k_T$ algorithm with $R=0.5$ and are required to satisfy
$p_T^j>20{\text{GeV}},~|\eta^j|<4.5$.
Charged leptons (electrons and muons) are required to have
$p_T^\ell>15{\text{GeV}},~|\eta^\ell|<2.5,~I_{iso,\mu}\left(\Delta R=0.3\right)<0.1$.
The $b$-tagging efficiency is chosen to be 70\% 
with a 25\% (2\%) mistagging rate for charm (light) jets\cite{CMS:2013xfa}. 
$b$-tagged jets satisfy $p_T^b>40{\text{GeV}},~|\eta^j|<2.5$.} 
We require at least 6 jets with at least 4 $b$-tags, and require the leading $b$-jet to have $p_T>150$ GeV. We also apply a missing transverse 
energy cut of $\met > 30$ GeV and veto events with 
more than one charged lepton.

Top quarks and $W$ bosons are reconstructed from the mass-shell constraints, 
with small corrections for detector effects (for details, see appendix~\ref{appx:reconstr}).
After top quark reconstruction, we require that the signal events satisfy
$\chi^2<5.0$ and $\Delta R_{b_1b_2}>0.9$,
where $\chi$ is a variable characterizing the quality of the reconstruction (see appendix~\ref{appx:reconstr}), and $b_1$ ($b_2$) is the leading (sub-leading) $b$-jet
which is not recognized as emerging from a top quark decay.

\begin{figure}[!htb]
\includegraphics[scale=0.35,clip]{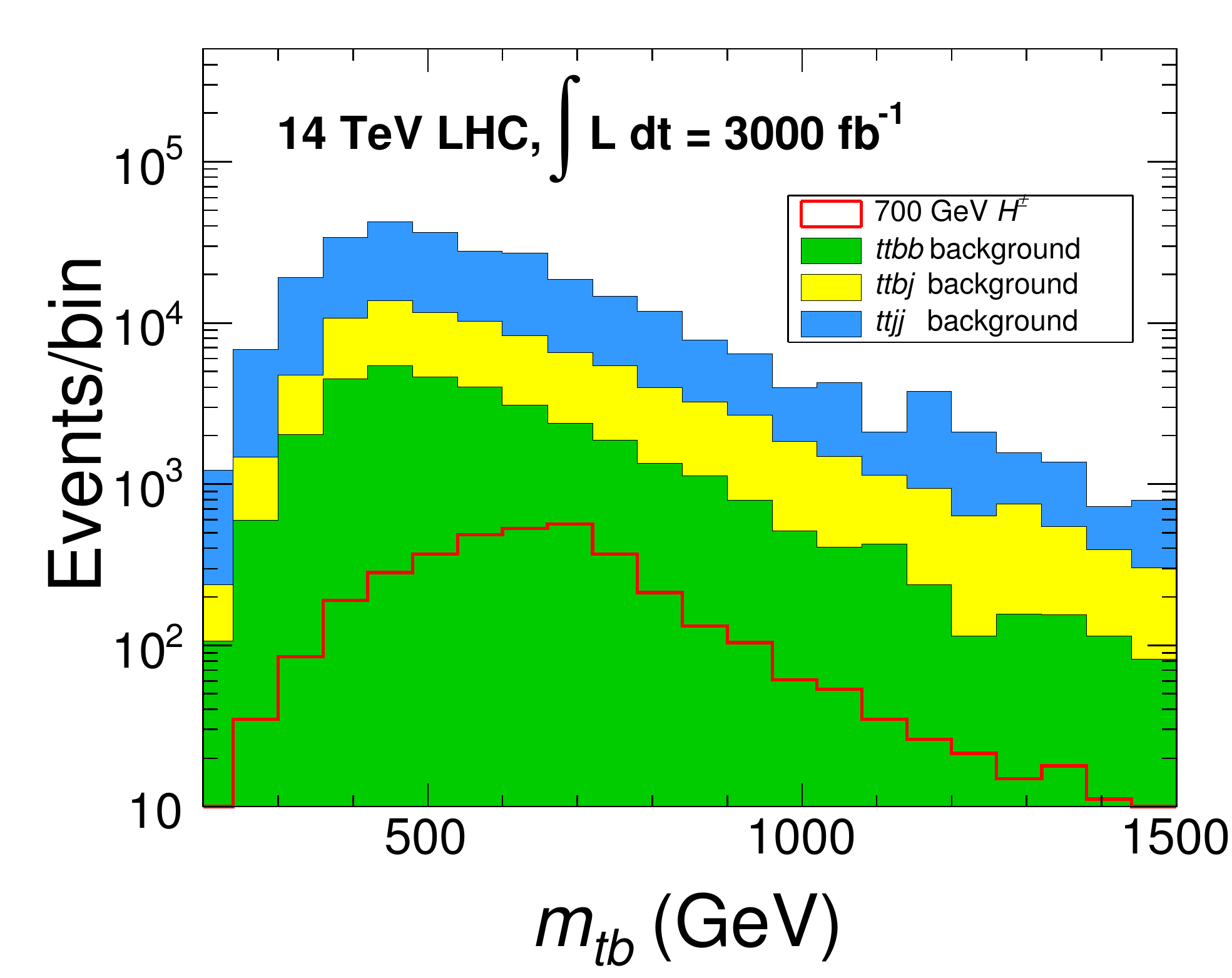}
\caption{The $tb$ invariant mass distribution of the backgrounds
and the signal from a 700 GeV $H^\pm$ with $y_{tb}=1$ at 14 TeV
LHC with 3000 fb$^{-1}$ integrated luminosity. 
\label{fig:mtb700} }
\end{figure}

The charged scalar invariant mass is reconstructed from the 
leading $b$-jet and the leading reconstructed top quark. In figure \ref{fig:mtb700},
we show the $tb$ invariant mass distribution of the backgrounds
and the signal from a 700 GeV $H^\pm$ with $y_{tb}=1$ at 14 TeV
LHC with 3000 fb$^{-1}$ integrated luminosity. 
\begin{figure}[!htb]
\includegraphics[scale=0.35,clip]{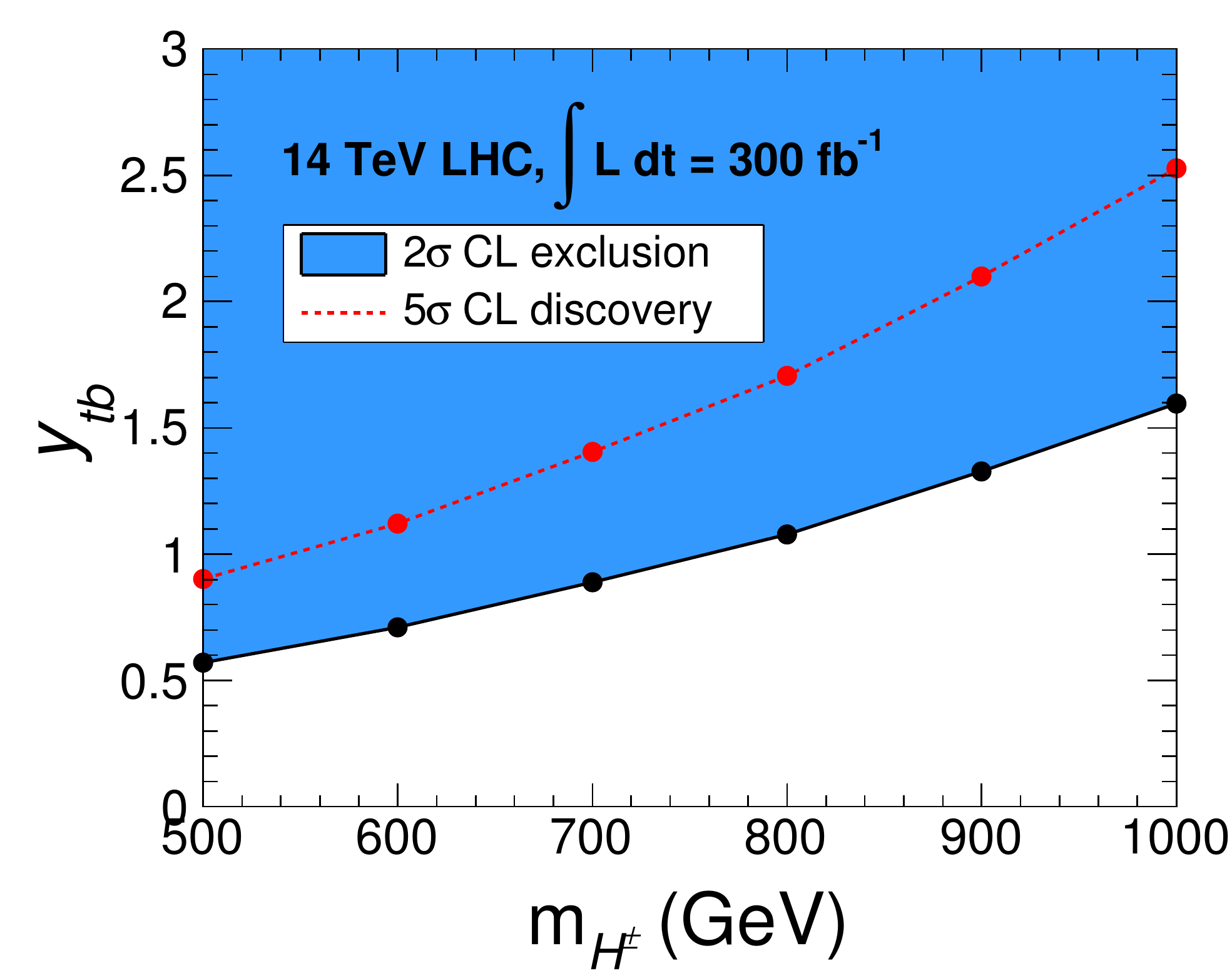}
\includegraphics[scale=0.35,clip]{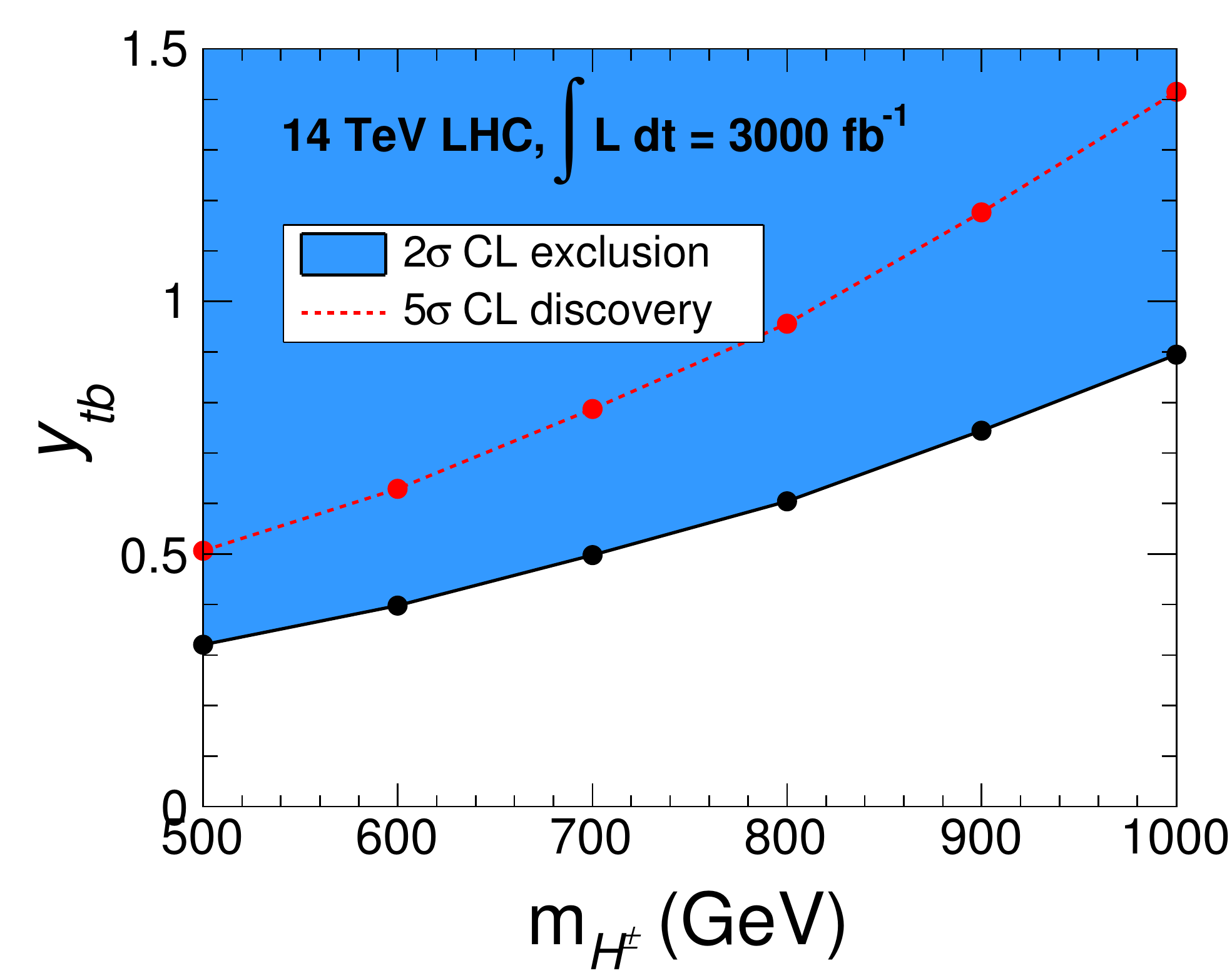}
\caption{The $2\sigma$ exclusion and the $5\sigma$ discovery 
bound of the charged Higgs via searching the $tb$ resonance
in the $ttbb$ channel at 14 TeV LHC with 300 fb$^{-1}$ (left panel) 
and 3000 fb$^{-1}$ (right panel) integrated luminosity. 
\label{fig:ex_dis} }
\end{figure}

\begin{figure}[!htb]
\includegraphics[scale=0.35,clip]{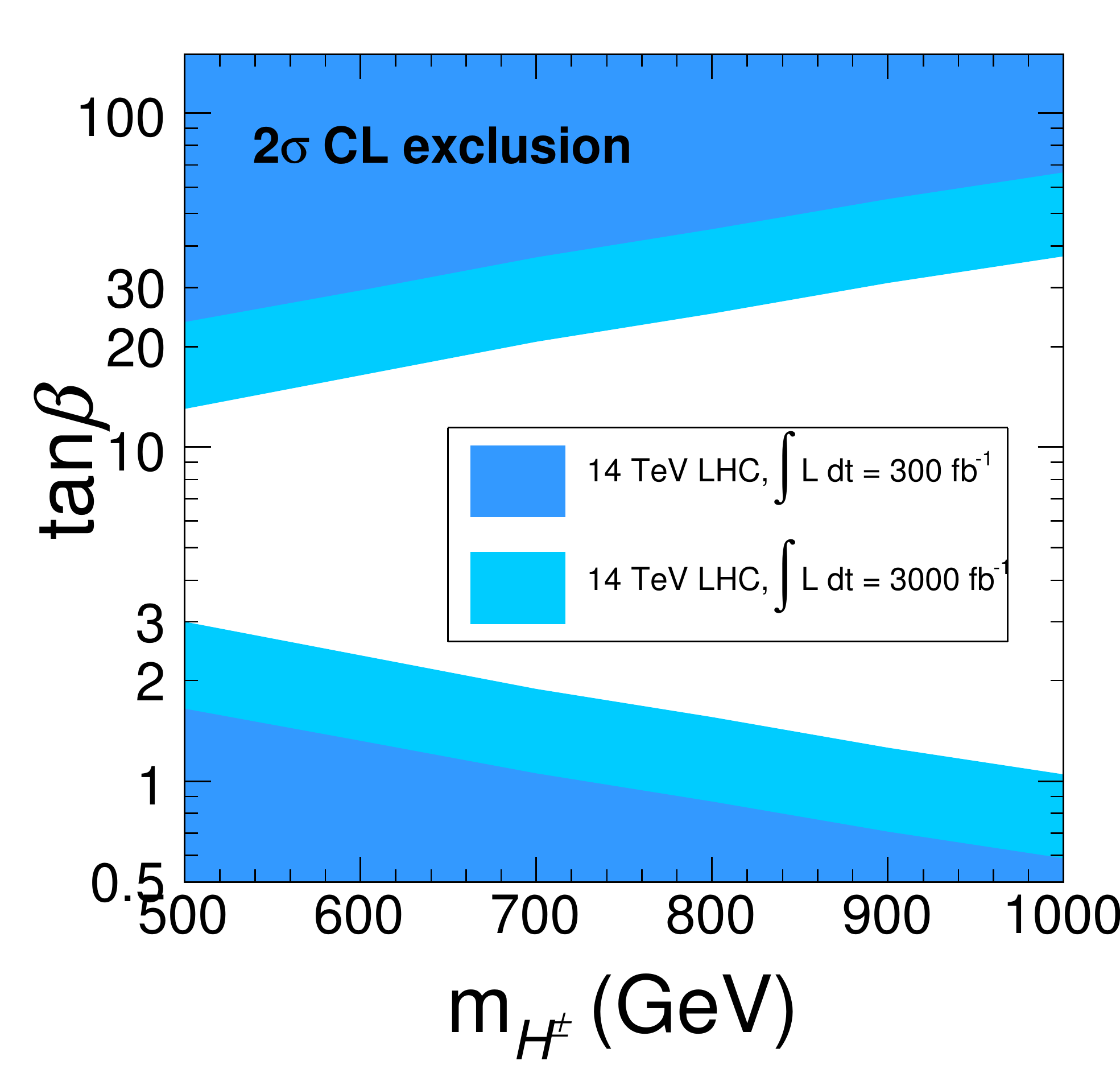}
\includegraphics[scale=0.35,clip]{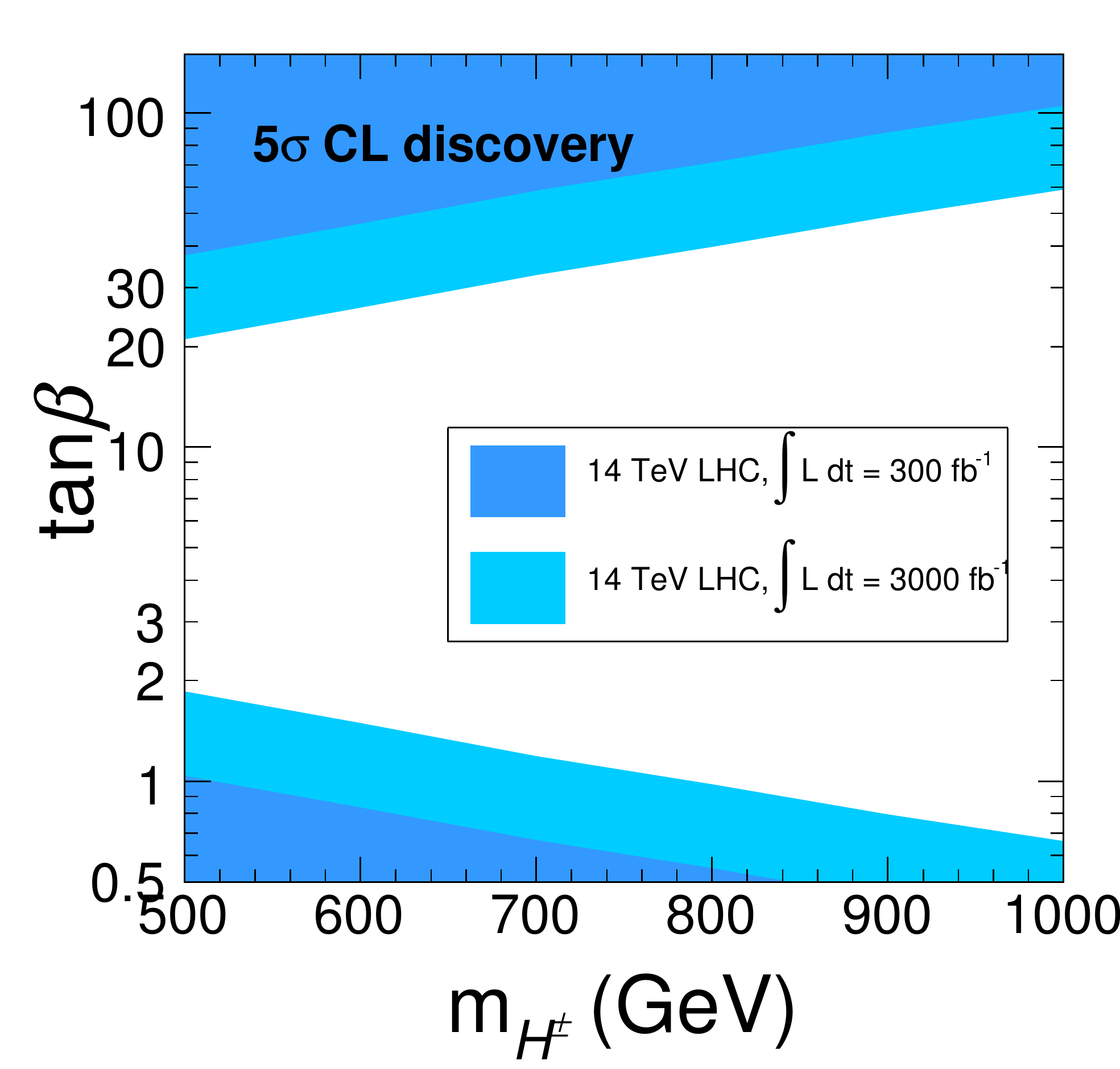}
\caption{The $2\sigma$ exclusion (left panel) and the $5\sigma$ discovery 
(right panel) bound of the charged Higgs via searching the $tb$ resonance
in the $ttbb$ channel at 14 TeV LHC with 300 fb$^{-1}$
and 3000 fb$^{-1}$ integrated luminosity. 
\label{fig:ex_dis_tb} }
\end{figure}

To obtain the exclusion and discovery bound, 
we again use the likelihood function given by Eq.~(\ref{eq:likelihood}), where now
where $x_j$ is the binned $m_{tb}$ distribution predicted by the 
model (with or without signal) and $n_j$ is the observed 
distribution. The $2\sigma$ exclusion bound is obtained as in 
Eq.~(\ref{eq:twosig}), while the
$5 \sigma$ discovery reach is obtained from
\be
\sqrt{-2\ln\left(\frac{L\left(b|\mu s+b\right)}{L\left(\mu s+b|\mu s+b\right)}\right)}=5.
\ee 
In figure \ref{fig:ex_dis}, we show discovery and exclusion curves for the coupling constant
$y_{tb}$ as a function of $m_H$. We have checked that these results are insensitive to the $\theta$ angle
in Eq.~(\ref{eq:interaction}). In the Type 2 2HDM, Eq. (\ref{eq:interaction})
can be written as
\be
\mathcal{L}_{\text{eff}}=\frac{\sqrt2}{v}H^+\bar t\left(P_Lm_t\cot\beta+P_Rm_b
\tan\beta\right)b
+{\text{h.c.}}
\label{eq:interaction2}
\ee
As such, the constraint on $y_{tb}$ shown in figure \ref{fig:ex_dis} is translated 
into a constraint to $\tan\beta$ in figure \ref{fig:ex_dis_tb}.

While there remains a hole in coverage at moderate values of $\tan \beta$, 
there is nonetheless considerable sensitivity for heavy charged Higgses in 
this channel at $\sqrt{s} = 14$ TeV. Note also that the reach of the semi-leptonic 
search at $\sqrt{s} = 14$ TeV is comparable to the naive extrapolation of the 
$\sqrt{s} = 8$ TeV CMS di-leptonic search, suggesting that an optimized 
search for charged Higgses can effectively employ both semi-leptonic and 
di-leptonic final states to constrain $pp\to H^{+}\bar{t}b (H^{-}t\bar{b})+X$ 
with $H^+\rightarrow t\bar{b}(H^-\to\bar tb)$.

\section{Conclusions}
\label{sec:conc}

The hunt for the rest of the Higgs bosons is entering a new phase, 
as an ever-broadening set of direct searches at the LHC begins 
to constrain the parameter space of extended Higgs sectors. In 
this work we have attempted to identify and analyze some of the most 
promising open channels in existing coverage of heavy Higgs 
bosons consistent with properties of the observed SM-like Higgs. 
These channels are the production of a heavy scalar or pseudoscalar 
with decay to $t \bar t$; $b \bar b$ and $t \bar t$ and associated production 
of a heavy scalar or pseudoscalar with decay to invisible final states; 
and $t \bar b$ associated production of a charged Higgs with decay to 
$\bar t b$. 

Heavy scalars or pseudoscalars decaying into $t \bar t$ constitute a significant gap in existing coverage of extended electroweak symmetry breaking scenarios. Taking into account the effects of detector resolution and $t \bar t$ 
reconstruction, we have found that searches for resonant production 
of heavy Higgses with decay into $t \bar t$ are likely to be systematics-limited 
at the LHC. We have correspondingly proposed several ancillary channels 
involving associated production that may provide complementary sensitivity. 
The most promising is $t\bar t H/A$ production with $H/A\rightarrow t\bar t$. 

Searches involving missing energy provide an effective probe of intriguing 
scenarios where a heavy Higgs decays invisibly. We demonstrated that $b \bar b H/A$ 
and $t\bar t H/A$ are valuable production channels in which to search for 
$H/A\rightarrow \rm{inv}$. Furthermore, their reach is expected to be competitive 
with -- or better than -- monojet searches in some models and mass ranges. 

Finally, heavy charged Higgses in the alignment limit decay dominantly to 
$tb$ if this channel is open, 
and the natural strong production channel for charged Higgses with $m_{H^\pm} \geq m_t + m_b$ is in association with $tb$. We studied the reach in the semileptonic channel, where the system can be 
completely reconstructed, and find considerable sensitivity to heavy charged Higgses that can complement existing searches in the dileptonic channel.

In conjunction with precision Higgs coupling measurements and existing direct searches for heavy Higgs bosons, these searches can maximize the LHC discovery potential for the most well-motivated 
extensions of the Higgs sector.

\begin{acknowledgments}
We thank Duane Dicus, John Paul Chou, 
Yuri Gershtein, and Sunil Somalwar for useful conversations. H. Zhang is supported 
by the U.S. DOE under contract No.~DE-SC0011702.
S. Thomas is supported by the U.S. DOE under grant DE-SC0010008.
\end{acknowledgments}

\appendix
\section{Higgs Couplings Fit} \label{app:fit}

We consider only a single SM-like Higgs boson of mass $m_h = 125$ GeV whose 
couplings to SM fermions and gauge bosons are modified relative to an SM Higgs 
boson of the same mass by coupling modifiers $\kappa_i \equiv g_i / g_{i,SM}$, 
where $i = t, b, \tau, W, Z, g, \gamma$. We treat the loop-induced couplings to 
gluons and photons independently from variations in the fermion couplings to 
allow for new degrees of freedom running in loops. For simplicity we assume 
custodial symmetry so that $\kappa_Z = \kappa_W$. We do not consider a 
potential invisible width or couplings with non-SM tensor structure.

We construct likelihoods for a Higgs coupling fit using data from Higgs analyses 
reported by the ATLAS and CMS collaborations. Single-channel likelihoods are 
constructed for each Higgs analysis using a two-sided Gaussian where the central 
value corresponds to the best-fit signal strength modifier $\hat \mu$ reported in 
the analysis and the variance on each side corresponds to the $1\sigma$ error 
on the signal strength modifier, i.e.
\begin{equation}
\mathcal{L}_i^\pm(\mu) \propto  \exp \left[ \frac{-(\mu - \hat \mu_i)^2}
{2 (\sigma_i^\pm)^2} \right].
\end{equation}
This two-sided likelihood accommodates the often sizable non-gaussianities 
found in low-statistics channels. 

The theory prediction for the signal strength modifier $\mu$ is constructed 
by summing over the production and decay modes considered in the analysis 
(each of which is a function of the coupling modifiers $\kappa_i$), weighted by 
the relative contribution $\epsilon$ of each production mode to the analysis. 
These relative contributions are extracted from experimental publications or 
inferred from the literature where appropriate. We neglect uncertainties on the 
values of $\epsilon$. We consider experimental analyses for which a single 
decay mode dominates the analysis, so that the signal strength modifier for 
a single experimental channel is given by
\begin{equation}
\mu = \left( \sum_a \epsilon_a \frac{\sigma_a}{\sigma_{a,SM}}\right)
\frac{BR}{BR_{SM}},
\end{equation}
where the index $a$ runs over the gluon fusion, vector boson fusion \& 
associated vector production, and associated $t \bar t$ production modes. 
The set of ATLAS and CMS Higgs analyses used to construct our coupling 
fit (with corresponding best-fit signal strength modifiers, $1\sigma$ errors, 
and relative efficiencies) are enumerated in tables \ref{tab:atlchannels} and 
\ref{tab:cmschannels}.

\begin{table}[h]
\small
\centering
\renewcommand{\arraystretch}{1.1}
\begin{tabular}{| l | c | c | c |}
\hline
Channel & $\sqrt{s}$ & $\hat \mu_a$ & $\epsilon_a{\rm (GGH, VBF/VH, TTH)}$ \\
\hline\hline 
$Vh \to b \bar b$ (0$\ell$) \cite{Aad:2014xzb} & 7/8 TeV & $-0.35^{+0.55}_{-0.52}$ & $( 0.0,1.0 ,0.0)$ \\ 
$Vh \to b \bar b$ (1$\ell$) \cite{Aad:2014xzb} & 7/8 TeV & $ 1.17^{+0.66 }_{-0.60 }$ & $(0.0 , 1.0, 0.0)$ \\
$Vh \to b \bar b$ (2$\ell$) \cite{Aad:2014xzb} & 7/8 TeV & $ 0.94^{+0.88 }_{-0.79 }$ & $( 0.0, 1.0, 0.0)$ \\
$tth \to b \bar b $ \cite{ATLtth} & 7/8 TeV & $ 1.7^{+1.4 }_{-1.4 }$ & $( 0.0, 0.0 ,1.0)$ \\
$h \to \tau \tau$ ($jj$) \cite{ATLtau} & 7/8 TeV & $ 3.6^{+2.0 }_{-1.6 }$ & $(0.60 ,0.4 , 0.0)$ \\
$h \to \tau \tau $ ($\ell j$) \cite{ATLtau} & 7/8 TeV & $ 0.9^{+1.0 }_{-0.9}$ & $(0.65 , 0.35, 0.0)$ \\
$h \to \tau \tau $ ($\ell \ell$) \cite{ATLtau} & 7/8 TeV & $ 3.0^{+1.9 }_{-1.7 }$ & $(0.65 ,0.35 ,0.0 )$ \\
$hjj \to \tau \tau $ ($jj$) \cite{ATLtau} & 7/8 TeV & $ 1.4^{+0.9 }_{-0.7 }$ & $( 0.15, 0.85, 0.0)$ \\
$hjj \to \tau \tau $ ($\ell j$) \cite{ATLtau} & 7/8 TeV & $ 1.0^{+0.6 }_{-0.5 }$ & $(0.12 , 0.88 ,0.0)$ \\
$hjj \to \tau \tau $ ($\ell \ell$) \cite{ATLtau} & 7/8 TeV & $ 1.8^{+1.1 }_{-0.9 }$ & $(0.10 ,0.90 , 0.0)$ \\
$h \to WW $ ($0j$) \cite{ATLWW} & 7/8 TeV & $1.14 ^{+0.34 }_{-0.30 }$ & $(0.98,0.02, 0.0)$ \\
$h \to WW $ ($1j$) \cite{ATLWW} & 7/8 TeV & $0.96 ^{+0.45 }_{-0.40 }$ & $(0.87,0.13,0.0)$ \\
$h \to WW $ ($2j$ ggH) \cite{ATLWW} & 7/8 TeV & $ 1.20^{+0.91 }_{-0.84 }$ & $(0.75,0.25,0.0)$ \\
$h \to WW $ ($2j$ VBF) \cite{ATLWW} & 7/8 TeV & $ 1.20^{+0.45 }_{-0.38 }$ &$(0.13,0.87,0.0)$ \\
$h \to ZZ $ (ggH) \cite{Aad:2014eva} & 7/8 TeV & $ 1.66^{+0.5 }_{-0.4 }$ & $(1.0,0.0,0.0)$ \\
$h \to ZZ $ (VBF+VH)\cite{Aad:2014eva} & 7/8 TeV & $ 0.26^{+1.6 }_{-0.9 }$ & $(0.0 ,1.0 , 0.0)$ \\
$h \to \gamma \gamma $ (ggH) \cite{Aad:2014eha} & 7/8 TeV & $1.32 ^{+ 0.38}_{-0.38 }$ & $(1.0 ,0.0 ,0.0 )$ \\
$h \to \gamma \gamma $ (VBF) \cite{Aad:2014eha} & 7/8 TeV & $0.8 ^{+0.7 }_{-0.7 }$ & $(0.0, 1.0,0.0 )$ \\
$h \to \gamma \gamma $ (WH) \cite{Aad:2014eha} & 7/8 TeV & $ 1.0^{+1.6 }_{-1.6 }$ & $( 0.0, 1.0, 0.0)$ \\
$h \to \gamma \gamma $ (ZH) \cite{Aad:2014eha} & 7/8 TeV & $ 0.1^{+3.7 }_{-0.1 }$ & $( 0.0, 1.0, 0.0)$ \\
$h \to \gamma \gamma $ (ttH) \cite{Aad:2014eha} & 7/8 TeV & $ 1.6^{+2.7 }_{-1.8 }$ & $(0.0 , 0.0, 1.0)$ \\

\hline
\end{tabular}
\caption{ATLAS Higgs analyses used in constructing coupling fits. 
The best-fit signal strength modifier is denoted by $\hat \mu$ with 
corresponding $\pm 1 \sigma$ errors. The relative contributions 
$\epsilon$ are reported for production initiated by gluons via gluon 
fusion (GGH), weak gauge bosons via vector boson fusion or vector 
associated production (VBF/VH), and top quarks via $t \bar t$ 
associated production (TTH).}
\label{tab:atlchannels}
\end{table}
\begin{table}[h]
\small
\centering
\renewcommand{\arraystretch}{1.1}
\begin{tabular}{| l | c | c | c |}
\hline
Channel & $\sqrt{s}$ & $\hat \mu_a$ & $\epsilon_a{\rm (GGH, VBF/VH, TTH)}$ \\
\hline\hline 
$h \to bb $ \cite{CMS:2014ega} & 7/8 TeV & $ 1.0^{+0.53 }_{-0.50 }$ & $( 0.0, 1.0, 0.0)$ \\
$tth \to bb $ \cite{CMS:2014jga} & 7/8 TeV & $ 0.67^{+1.35 }_{-1.33 }$ & $(0.0, 0.0, 1.0)$ \\
$h \to \tau \tau $ (0,1$j$) \cite{CMS:2014ega} & 7/8 TeV & $ 0.84^{+0.42 }_{-0.38 }$ & $(0.87,0.13,0.0)$ \\
$hjj \to \tau \tau $ (2$j$) \cite{CMS:2014ega} & 7/8 TeV & $ 0.95^{+0.43 }_{-0.38 }$ & $(.17, .83, 0.0)$ \\
$Vh \to \tau \tau $ \cite{CMS:2014ega} & 7/8 TeV & $ 0.87^{+1.00 }_{-0.88 }$ & $(0.0, 1.0, 0.0)$ \\
$h \to WW $ (0,1$j$) \cite{CMS:2014ega} & 7/8 TeV & $ 0.77^{+0.23 }_{-0.21 }$ & $(0.83, 0.17, 0.0)$ \\
$h \to WW $ (2$j$) \cite{CMS:2014ega} & 7/8 TeV & $ 0.62^{+0.59 }_{-0.48 }$ & $(0.17, 0.83, 0.0)$ \\
$Vh \to WW $ \cite{CMS:2014ega} & 7/8 TeV & $ 0.80^{+1.09 }_{-0.93 }$ & $(0.0, 1.00, 0.0)$ \\
$h \to ZZ $ \cite{CMS:2014ega} & 7/8 TeV & $ 0.88^{+0.34 }_{-0.27 }$ & $(0.9, 0.1, 0.0)$ \\
$h \to ZZ $ ($2j$) \cite{CMS:2014ega} & 7/8 TeV & $ 1.55^{+0.95 }_{-0.66 }$ & $(0.58, 0.42, 0.0)$ \\
$h \to \gamma \gamma $ (ggH) \cite{Khachatryan:2014ira} & 7/8 TeV & $ 1.12^{+0.37 }_{-0.32 }$ & $(1.0, 0.0, 0.0)$ \\
$h \to \gamma \gamma $ (VBF) \cite{Khachatryan:2014ira}& 7/8 TeV & $ 1.58^{+0.77 }_{-0.68 }$ & $(0.0, 1.0, 0.0)$ \\
$h \to \gamma \gamma $ (VH) \cite{Khachatryan:2014ira}  & 7/8 TeV & $ -0.16^{+1.16 }_{-0.79 }$ & $( 0.0, 1.0, 0.0)$ \\
$h \to \gamma \gamma $ (ttH) \cite{Khachatryan:2014ira} & 7/8 TeV & $ 2.69^{+2.51}_{-1.81 }$ & $(0.0, 0.0, 1.0)$ \\
\hline
\end{tabular}
\caption{CMS Higgs analyses used in constructing coupling fits. 
The best-fit signal strength modifier is denoted by $\hat \mu$ with 
corresponding $\pm 1 \sigma$ errors. The relative contributions 
$\epsilon$ are reported for production initiated by gluons via gluon 
fusion (GGH), weak gauge bosons via vector boson fusion or vector 
associated production (VBF/VH), and top quarks via $t \bar t$ 
associated production (TTH).}
\label{tab:cmschannels}
\end{table}

We construct a combined likelihood from the product of all 
single-channel likelihoods, 
\begin{equation}
\mathcal{L}(\mu) = \prod_i \mathcal{L}_i(\mu).
\end{equation}
This approach does not take into account correlations among systematic 
uncertainties in different Higgs searches, as such information is not 
publicly available. However, this is a reasonable approximation since 
uncertainties in Higgs measurements are not yet dominated by systematics. 
We are often interested in treating some inputs as nuisance parameters 
$\theta$, in which case the combined likelihood may be expressed as a 
function of both $\mu$ and $\theta$. 

We construct coupling fits using the profile likelihood approach
\cite{Cowan:2010js}. In this approach, the best-fit signal strength modifier 
$\hat \mu$ and corresponding uncertainty $\Delta \hat \mu$ of the combined 
likelihood are calculated using the likelihood ratio $\lambda(\mu) = \mathcal{L}
(\mu,\hat{\hat{\theta}})/\mathcal{L}(\hat \mu, \hat \theta)$. This is the ratio of 
a likelihood function with nuisance parameters $\hat{\hat{\theta}}$ optimized 
for a given value of $\mu$ to a likelihood function where $\hat \mu$ and 
$\hat \theta$ are optimized simultaneously. Optimizing nuisance parameters 
$\hat{\hat{\theta}}$ for a given value of $\mu$ amounts to profiling these 
nuisance parameters. Given this likelihood ratio, the uncertainty 
$\Delta \hat \mu$ is computed using the test statistic $-2 \ln \lambda(\mu)$, 
which converges to a $\chi^2$ distribution in one degree of freedom as the 
data sample size is taken to be large. 

\section{The $gg \to H/A \to t \bar{t}$ Differential Cross Section } \label{appx:angular}

The kinematics of spin averaged 
two to two on-shell  scattering processes $12 \to 34$ 
are completely specified by the masses of the particles, energy-momentum 
conservation, and the 
Mandelstam variables
$s = (p_1 \! + \! p_2)^2$
and  $t = (p_1 \! - \! p_3)^2$. 
In order to characterize the invariant phase space distribution for these processes 
it is useful to define a shifted dimensionless version of the 
Mandelstam variable $t$
that has a flat metric with respect to the squared amplitude.  
For $m_1 = m_2=0$ and $m_3 = m_4 = m$ this variable is 
\begin{equation}
\varpi = 1 + {2 \over s}  \Big( t - m^2 \Big) =
\beta \cos \theta = 
\tanh \Big(  { y_3 - y_4 \over 2} \Big)
\end{equation}
where $\beta = \sqrt{1 - 4 m^2 / s}$ is the velocity of either final state 
particle in the center of mass frame, 
$\cos \theta$ is the cosine of the center of mass scattering angle, 
and 
$ y = \tanh^{-1} (p_z / E)$ are the individual rapidities of the final 
state particles in any longitudinal frame. 
The scattering variable lies in the range 
$- \beta \leq \varpi \leq \beta$. 
At fixed $s$ 
the differential cross section with respect to $\varpi$ and 
$t$ and $\cos \theta$ are related by 
\begin{equation}
{d \sigma \over d \varpi} = {s \over 2} ~ \! { d \sigma \over dt} = 
{1 \over \beta} ~ \! { d \sigma \over d  \cos \theta} 
\end{equation}
Experimentally the variable $\varpi$ is more robust than 
$\cos \theta$ near threshold since $\beta \to 0$ and 
$\cos \theta$ becomes ill-defined at threshold, but 
$\varpi \to 1$ in this limit. 
In addition, $s~ d \sigma / d \varpi$ is proportional to the dimensionless 
scattering amplitude squared everywhere in phase space. 
So the dimensionless coordinates $\varpi$ and $x^2 = s/(4 m^2)$ provide a  flat metric 
for the probability density in phase space, with the physical phase space region given by 
 $x^2 \geq1$ and $ - \sqrt{ 1 - 1/x^2} \leq \varpi \leq  \sqrt{ 1 - 1/x^2}$. 
 Central scattering corresponds to $\varpi=0$. 
The phase space volume vanishes at threshold, $x^2 =1 $, $\varpi = 0$. 
These coordinates for two to two scattering are 
the analogs of Dalitz coordinates for three-body decay. 

For heavy scalar and pseudo-scalar Higgs bosons with masses 
greater than twice the top quark mass, $m_H, m_A > 2 m_t$, 
the scattering of two gluons to a top and anti-top quark, 
$gg \to t \bar {t}$, receives contributions both from QCD interactions, 
as well as from $s$-channel gluon fusion production and decay of $H$ and 
$A$. 
The differential cross section including all these processes is given by 
\cite{Dicus:1994bm} 
\begin{eqnarray}
s ~ \!  { d \sigma \over d \varpi} (gg \to t \bar{t}) 
   \! & = & \! {7 \pi \alpha_s^2 \over 96} 
  \bigg[  ~ \! 
    f(s/4 m_t^2 ,\varpi)_{\rm QCD}
 + f(s/4 m_t^2 ,  s/4 m_b^2 , m_H^2 / m_t^2,\varpi)_{\rm H-QCD}
  \nonumber 
  \\
& & 
  ~ + f(s/4 m_t^2 ,  s/4 m_b^2 , m_A^2 / m_t^2,\varpi)_{\rm A-QCD}
  +     f(s/4 m_t^2 ,   s/4 m_b^2 , m_H^2 / m_t^2,\varpi)_{\rm H}
\nonumber 
 \\
& & 
  ~ +  f(s/4 m_t^2 ,  s/4 m_b^2 , m_A^2 / m_t^2, \varpi)_{\rm A} 
 ~  \bigg] 
\end{eqnarray}
where 
\begin{eqnarray}
f(x_t^2 , \varpi)_{\rm QCD} \! &  = &  \! {7 + 9 \varpi^2 \over 7( 1 - \varpi^2 )  } 
\bigg( {1 \over 1 + \varpi^2 } + {2 \over x_t^2} - {2 \over  x_t^4 (1 - \varpi^2 ) }
\bigg) 
  \nonumber \\
 f(x_t^2 , x_b^2 , r^2 , \varpi)_{\rm H-QCD} \! &  =  & \!  -  
  {24   \over  7 ( 1 - \varpi^2) }
   \Big(1 - {1  \over x_t^2} \Big) ~ \! {\rm Re}
\Bigg[ 
{  [ 4x_t^2 - r^2 - 4 i x_t^2 \gamma_H(x_t)] \tilde{N}(x_t^2 , x_b^2) 
     \over (4 x_t^2 - r^2)^2 + 16 x_t^4 \gamma^2_H (x_t) } 
 \Bigg]
 \nonumber \\
 f(x_t^2 , x_b^2 , r^2 , \varpi)_{\rm A-QCD} \! & =& \! - 
 { 6 \over  7 ( 1 - \varpi)^2 } 
~ \! {\rm Re}
\Bigg[ 
{  [ 4x_t^2 - r^2 - 4 i x_t^2 \gamma_A(x_t)] \tilde{P}(x_t^2,x_b^2) 
     \over (4 x_t^2 - r^2)^2 + 16 x_t^4 \gamma^2_A (x_t) } 
 \Bigg]
 \nonumber \\
 f(x_t^2 , x_b^2 , r^2 , \varpi)_{\rm H} \! &=& \! 
    { 72  x_t^2\over 7 } 
   \Big(1 - {1  \over x_t^2} \Big) 
  ~ \! { \Big|  [4x_t^2 - r^2 - 4 i x_t^2 \gamma_H(x_t) ] \tilde{N}(x_t^2,x_b^2) \Big|^2 \over 
 \Big[ (4 x_t^2 - r^2)^2 + 16 x_t^4 \gamma^2_H (x_t) \Big]^2} 
 \nonumber \\
 f(x_t^2 , x_b^2 , r^2 , \varpi)_{\rm A} \! &=& \! 
 {  72 x_t^2\over 112 }  
  ~ \! { \Big| [4x_t^2 - r^2 - 4 i x_t^2 \gamma_H(x_t) ] \tilde{P}(x_t^2,x_b^2) \Big|^2 \over 
 \Big[ (4 x_t^2 - r^2)^2 + 16 x_t^4 \gamma^2_A (x_t) \Big]^2} 
\end{eqnarray}
The first term for the
QCD scattering amplitude squared function 
is unity for central scattering both at and well above
threshold
where $f(1,0)_{\rm QCD}=f(\infty,0)_{\rm QCD}=1$. 
The second and third terms arise from interference 
between the heavy scalar and pseudo-scalar Higgs $s$-channel 
amplitudes and the spin and color singlet component of the QCD amplitude. 
Near threshold the scalar Higgs amplitude interferes with the $P$-wave
component of the QCD amplitude, while 
the pseudo-scalar amplitude interferes with the $S$-wave component.    
These terms include one-loop functions for the gluon fusion production 
amplitudes generated by top and bottom quark loops
\begin{eqnarray}
\tilde{N}(x_t^2,x_b^2) 
 &=& 
 {\sqrt{2} ~ \! G_F \over 4 \pi^2}  ~ \! m_t  \kappa^H_t
  \sum_{f = t,b} m_f \kappa_f^H  \Bigg[ 
 1 - {1 \over 4} \Big(1 - {1  \over x_f^2} \Big) 
\bigg( \ln  { 1 + \sqrt{ 1 - 1/x_f^2} \over 1 - \sqrt{ 1 - 1/x_f^2}  } 
- i \pi \bigg)^2  ~ \! \Bigg] 
  \nonumber 
  \\
\tilde{P}(x_t^2,x_b^2) &=& -
  {\sqrt{2} ~ \! G_F \over 4 \pi^2} ~ \!  m_t  \kappa^A_t
  \sum_{f = t,b} m_f \kappa_f^A 
 \bigg( \ln  { 1 + \sqrt{ 1 - 1/x_f^2} \over 1 - \sqrt{ 1 - 1/x_f^2}  } 
- i \pi \bigg)^2
\end{eqnarray}
The $i \pi$ terms come from absorptive branch cuts 
in the one-loop amplitudes. 
For the top quark loop contributions to the full amplitude, 
these absorptive pieces represent QCD production 
of an on-shell top and anti-top quark with final state re-scattering  
through intermediate $s$-channel heavy Higgs bosons. 
For $m_H = m_A \gg m_t$ the absorptive contributions to the interference 
of the heavy scalar and pseudo-scalar Higgs bosons 
with QCD are equal in both magnitude and sign for all center of mass scattering energies. 
Away from the heavy Higgs resonances, 
interference of the QCD and non-absorptive parts of the 
heavy Higgs 
amplitudes 
are suppressed compared with the QCD amplitude squared 
by ${\cal O}(\lambda_t \lambda_f / 4 \pi^2)$
where $f = t,b$.
The functional dependence of the interference terms on the angular scattering 
variable $\varpi$ is universal and independent of the center of mass scattering 
energy. 
The real part of the interference terms changes sign 
across the resonances, leading to a distinctive excess below, 
and deficit above, the resonances \cite{Dicus:1994bm}. 
The fourth and fifth terms arise from the square of the amplitudes for 
$s$-channel production and decay of the heavy Higgs bosons.  
Away from the heavy Higgs resonances, 
the non-absorptive parts of the 
heavy Higgs 
amplitudes squared  
are suppressed compared with the QCD amplitude squared 
by ${\cal O}(\lambda_t^2 \lambda_f^2 / 16 \pi^4)$
where $f = t,b$.
The running dimensionless widths of the heavy Higgs bosons are 
\begin{equation}
\gamma_X(x)  = {\Gamma_X(s) \over \sqrt{s}}
\end{equation}
where $\Gamma_X = \Gamma(X \to {\rm All})$ are the running total 
widths for $X = H,A$. 
These terms in the $s$-channel heavy Higgs 
propagators represent
absorptive final state re-scattering  
of the heavy Higgs bosons 
through all intermediate on-shell states that contribute 
to the Higgs boson decay widths. 

The scalar and pseudo-scalar Higgs amplitudes do not interfere
in the spin averaged 
top and anti-top quark production differential cross section.  
Interference could arise first in a spin averaged 
production differential cross section with 
at least two gluons radiated from the final state.   
It also arises in the spin averaged differential 
cross section of the full phase of the top and anti-top quark decay products.

\begin{figure}
\begin{center}
\includegraphics[scale=0.37]{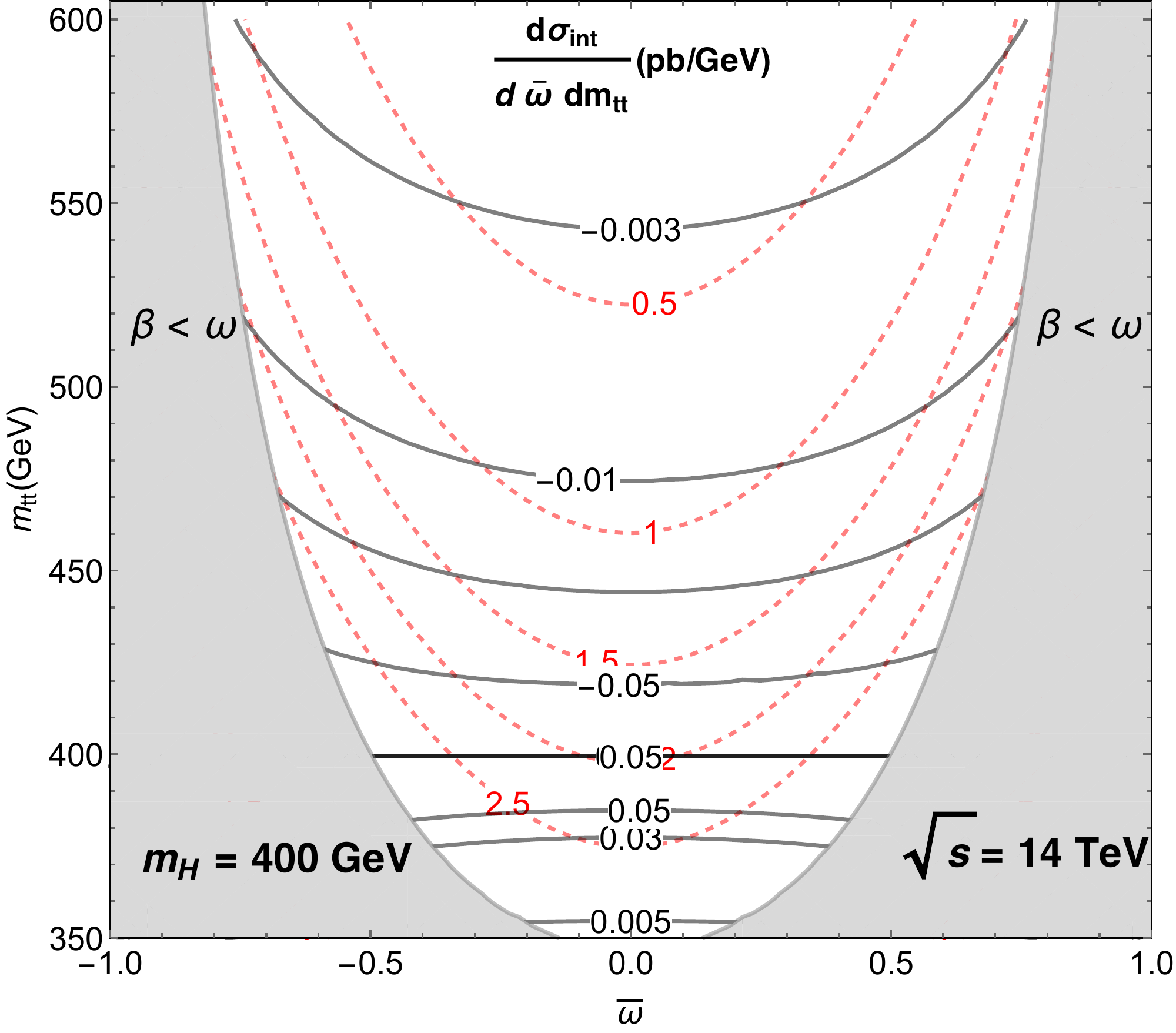} 
\includegraphics[scale=0.37]{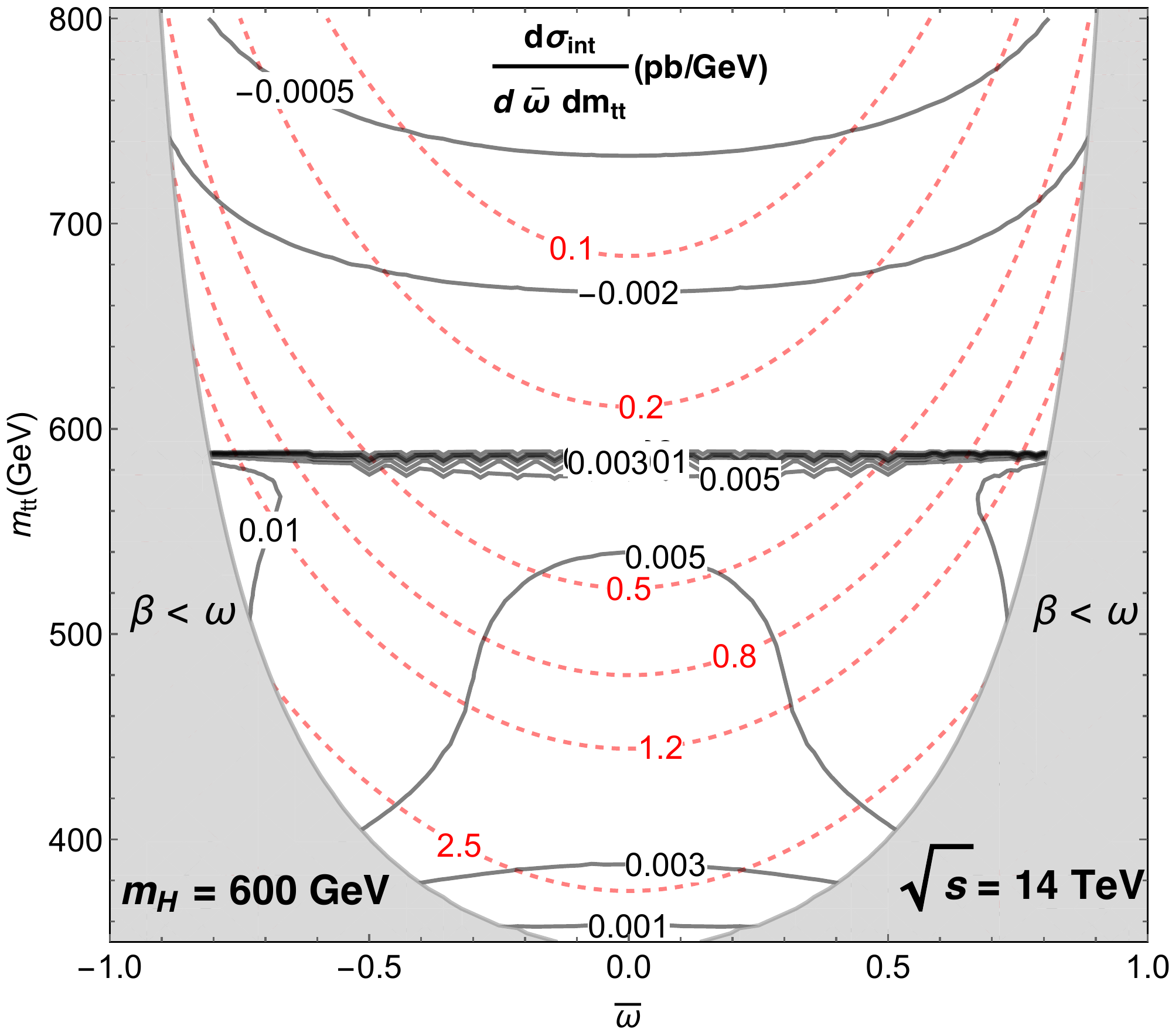}
\end{center}
\caption{Isocontours for the differential cross sections $d^2\sigma/d\varpi dm_{t \bar{t}}$ in QCD (red dotted lines) and for the interference terms in QCD+H (black solid lines) at 14 TeV with $\kappa^H_t =1$. The left (right) panel shows the results for $m_H=400$ (600) GeV.}\label{fig:ddsigmaH}
\end{figure}

In Fig.~\ref{fig:ddsigmaH} we plot contours of the differential $gg \to t\bar t$ cross section 
$d^2\sigma/d\varpi dm_{t \bar{t}}$ in QCD and of the interference contribution in QCD with a heavy scalar Higgs boson. 
We see that although the angular scattering variable offers some distinctive 
discrimination, there is no region where magnitude of this 
discrimination is able to effectively overcome the prodigious QCD background.

\section{Top Quark Reconstruction} \label{appx:reconstr}

In the $H/A \to t \bar t$, $b \bar b H/A \to b \bar b t \bar t$, and  charged Higgs analyses, we reconstruct the $W$ bosons and top quarks with the following algorithm.

The hadronically-decaying $W$ boson is reconstructed using
the non-$b$-tagged jets in the events. We choose the pair
of the jets $j_1j_2$ which minimizes $|m_{j_1j_2}-m_{Wh}|$, where
$m_{j_1j_2}$ is the invariant mass of the dijet system and 
$m_{Wh}=77.5$ GeV.\footnote{$m_{Wh}$ is the center of the $m_{j_1j_2}$ distribution obtained from the simulation of a pure sample of hadronic-$W$ events.} The reconstructed hadronic decaying 
$W$ 4-momentum is rescaled by a small correction factor of $m_W/m_{Wh}$,
where $m_W=80.4$ GeV. 

To reconstruct the leptonically-decaying 
top quark, we solve for the 4-momentum of the neutrino in the 
final state using neutrino and $W$ mass shell conditions. 
The solution for the $z$-component of the 
neutrino momentum is
\be
p_{\nu L}^\pm=\frac{A\left(m_W\right)p_{\ell L}
\pm E_\ell\sqrt{A\left(m_W\right)^2-4p_{\ell T}^2\met^2}}
{2p_{\ell T}^2}
\ee
where $A\left(m_W\right)\equiv m_W^2+2\overrightarrow{p}
_{\ell T}\cdot\overrightarrow{\met}$.
When there are two real solutions, the mass-shell condition of the 
top quark may be used to select the ``true" solution. Due to detector smearing effects and the finite width, there may be no real solution for $p_{\nu L}$. For such events with negative discriminant, we enforce 
\be
A\left(m_W\right)=\pm2p_{\ell T}\met
\ee
and look for a modified $\overrightarrow\met^\prime$ which 
minimizes $\left(\overrightarrow\met^\prime-\overrightarrow\met\right)^2$. 
It is clear that when $\overrightarrow{p}_{\ell T}$ is parallel to
$\overrightarrow\met$, the discriminant must be positive. So 
we can expand
\be
\overrightarrow\met^\prime=x\overrightarrow{p}_{\ell T}+(1+y)
\overrightarrow\met.
\ee
Then we have a constrained minimum value problem, which can be solved with Lagrange multipliers.
We obtain a cubic 
surface and with a unique real solution
\bea
y&=&-\frac{\left[d^2-\left(5m_W^2+4c\right)d+
\left(m_W^2-4c\right)^2\right]}
{\left[d^2+\left(m_W^2-4c\right)d+
\left(m_W^2-4c\right)^2\right]},\\
x&=&-\frac{\left[d^2+\left(m_W^2+8c\right)d+
\left(m_W^2-4c\right)^2\right]y}
{12ad},
\eea
where
\bea
a&=&\overrightarrow{p}_{\ell T}^2,\\
b&=&\overrightarrow{\met}^2,\\
c&=&\overrightarrow{p}_{\ell T}\cdot
\overrightarrow{\met},\\
d&=&\bigl\{216abm_W^2+\left(m_W^2+8c\right)
\left(m_W^4-20m_W^2c-8c^2\right)\nonumber\\
&+&12\sqrt3m_W\left[108abm_W^2+\left(m_W^2+2c\right)^2
\left(m_W^2-16c\right)\right]^{1/2}\nonumber\\
&\times&(ab-c^2)^{1/2}\bigr\}^{1/3}.
\eea
\begin{figure}[!htb]
\includegraphics[scale=0.35,clip]{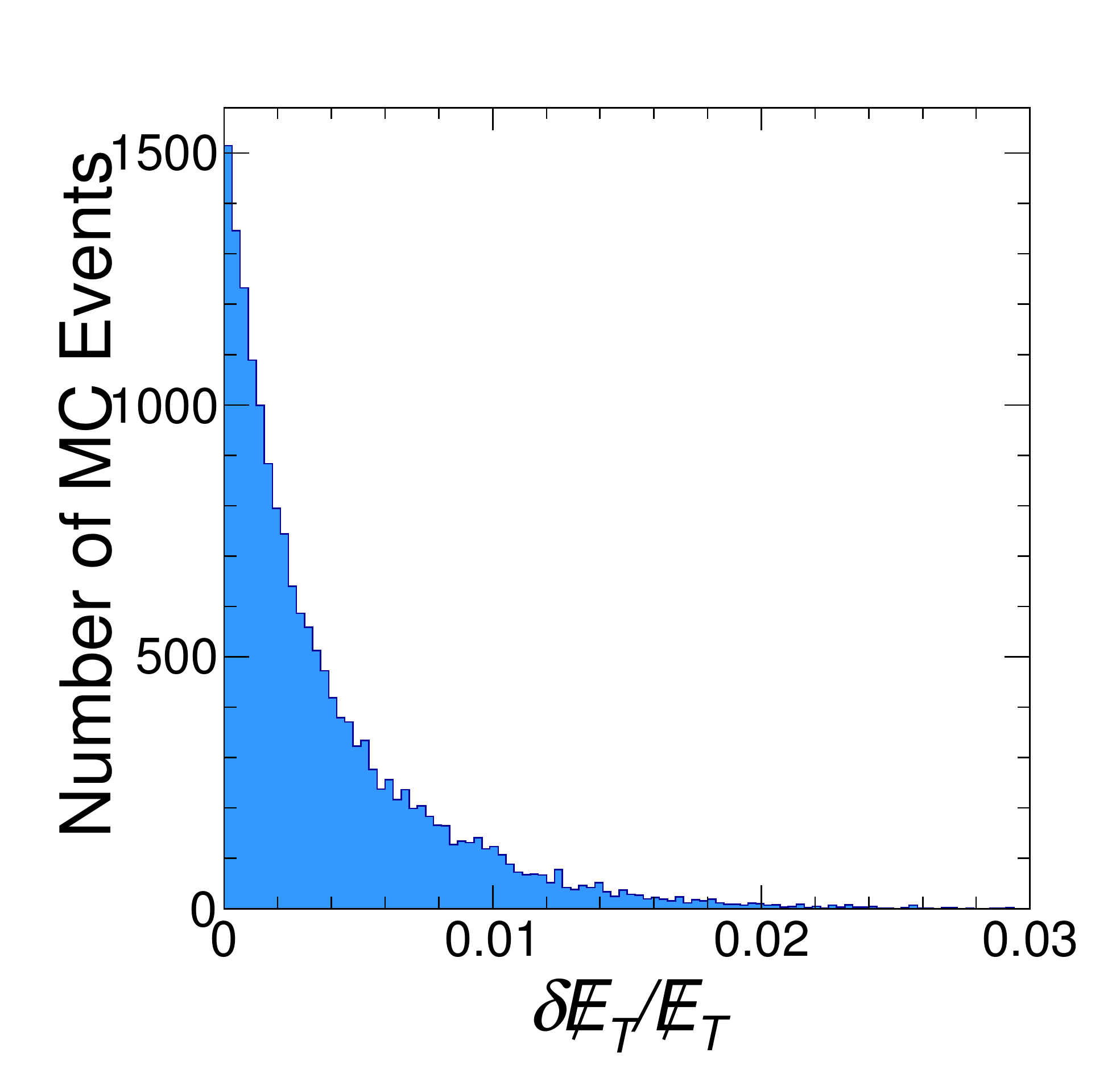}
\caption{The correction to the missing transverse momentum used to obtain a
real solution to the mass shell equation for the leptonically-decaying
$W$ boson. 
\label{fig:correction_met} }
\end{figure}
We show the correction to the missing transverse momentum in figure
\ref{fig:correction_met}. It is evident that for most of the events
with negative discriminant we only need to shift the missing 
transverse momentum by a factor of order $1$\%.

\begin{figure}[!htb]
\includegraphics[scale=0.35,clip]{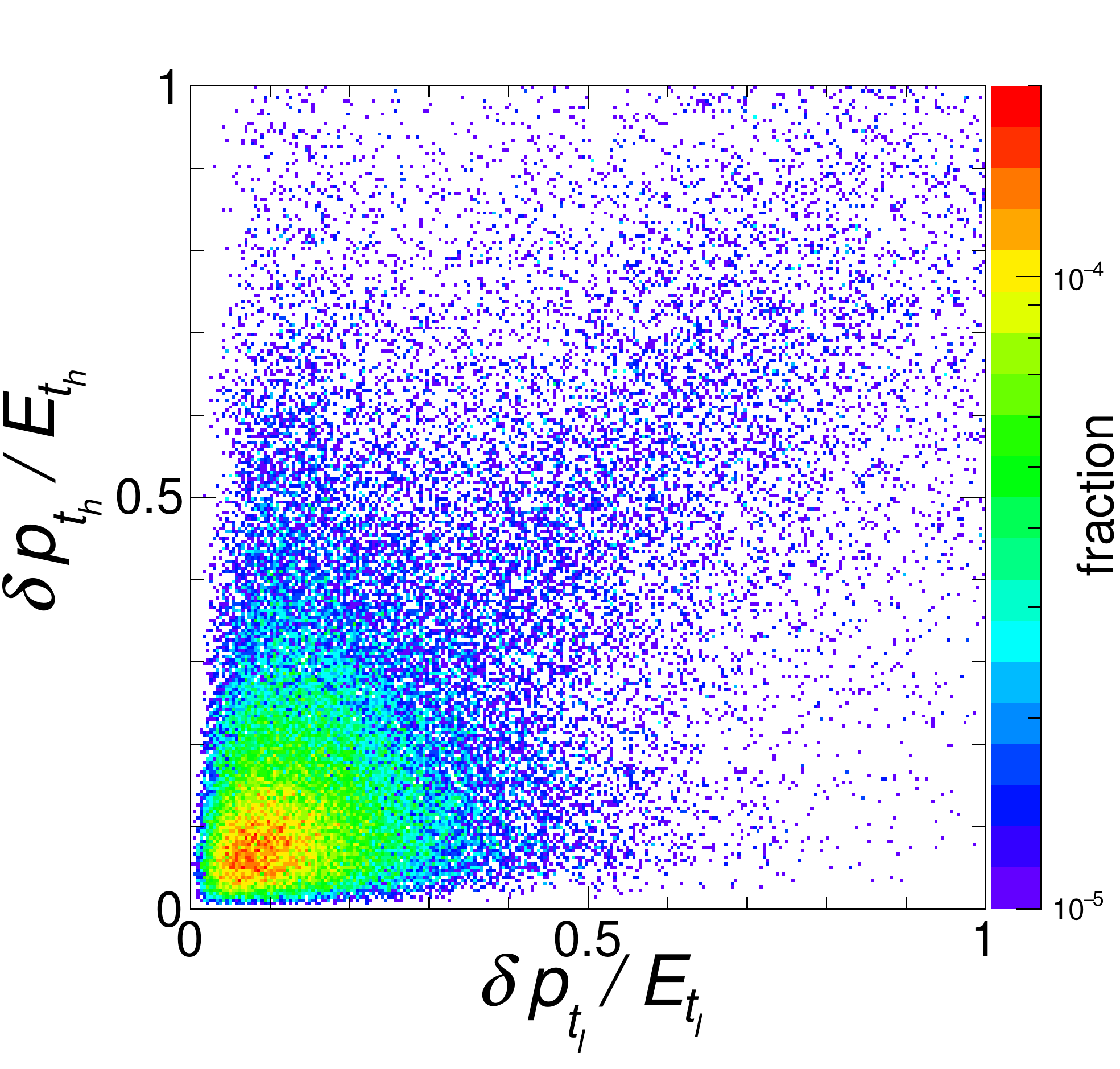}
\caption{The 2-dimensional distribution of the reconstruction 
efficiency of the hadronic and leptonic decaying top quarks 
in the events. 
\label{fig:topeff} }
\end{figure}

To reconstruct the top quarks in the final state, we try all of the combinations
$(W_hb_h)(W_\ell b_\ell)$, where $W_h$ and $W_\ell$ are the reconstructed 
hadronic and leptonic $W$ bosons and $b_h$ and $b_\ell$ are 
$b$-tagged jets in the event. When there are two different real solutions
of the neutrino longitudinal momentum, both of them are
used. We select the combination which minimizes 
\be
\chi^2=\frac{\left(m_{W_hb_h}-m_t\right)^2}{\sigma_h^2}+
\frac{\left(m_{W_\ell b_\ell}-m_t\right)^2}{\sigma_\ell^2},
\ee
where $m_t=173.2$ GeV is the pole mass of the top quark,
$\sigma_h=50$ GeV and $\sigma_\ell=25$ GeV. 
To check the reconstruction
efficiency, we compare the reconstructed  top quark 4-momenta with the real parton-level
momenta in the corresponding event. We 
calculate the ratio of the modulus of the
reconstructed top quark 3-momentum in the corresponding 
parton-level top quark rest frame to the energy of the parton-level
top quark in the laboratory frame, $\delta p_{t_{h(l)}}/E_{t_{h(l)}}$.
The result is shown in figure \ref{fig:topeff} (note the logarithmic z-axis). It is clear that
most of the reconstructed top quarks fall in the $\delta p/E<0.15$
region, meaning the top quarks in the events are well-reconstructed. 

\bibliography{2ndhiggs}
\bibliographystyle{jhep}

\end{document}